\RequirePackage{fix-cm}
\RequirePackage[log]{snapshot}
\documentclass[smallextended]{svjour3}
\smartqed  
\usepackage{graphicx}
\usepackage{url}
\usepackage[round, authoryear]{natbib}
\usepackage{booktabs}
\usepackage{array,multirow}
\usepackage[dvipsnames,table]{xcolor}
\usepackage{amssymb}
\usepackage{nth}
\usepackage{longtable}
\usepackage{float}
\usepackage{subfigure}
\usepackage[linesnumbered,lined,boxed,commentsnumbered]{algorithm2e}

\SetCommentSty{mycommfont}
\let\oldnl\nl
\newcommand{\nonl}{\renewcommand{\nl}{\let\nl\oldnl}}

\begin{document}
\title{Data-driven generation of spatio-temporal routines in human mobility}

\author{Luca Pappalardo and Filippo Simini}

\institute{Luca Pappalardo \at
              Department of Computer Science, University of Pisa, Italy\\
              Institute of Information Sciences and Technologies (ISTI), CNR, Italy \\
              \email{lpappalardo@di.unipi.it, luca.pappalardo@isti.cnr.it}           
           \and
           Filippo Simini \at
           Department of Engineering Mathematics, University of Bristol, UK\\
           Institute of Information Sciences and Technologies (ISTI), CNR, Italy \\
           \email{f.simini@bristol.ac.uk}  
}

\date{}

\maketitle

\newcommand{\ditras}{{\scshape{Ditras}} }

\begin{abstract}
The generation of realistic spatio-temporal trajectories of human mobility is of fundamental importance in a wide range of applications, such as the developing of protocols for mobile ad-hoc networks or what-if analysis in urban ecosystems. Current generative algorithms fail in accurately reproducing the individuals' recurrent schedules and at the same time in accounting for the possibility that individuals may break the routine during periods of variable duration. In this article we present {\scshape Ditras} (DIary-based TRAjectory Simulator), a framework to simulate the spatio-temporal patterns of human mobility. {\scshape Ditras} operates in two steps: the generation of a mobility diary and the translation of the mobility diary into a mobility trajectory. We propose a data-driven algorithm which constructs a diary generator from real data, capturing the tendency of individuals to follow or break their routine. We also propose a trajectory generator based on the concept of preferential exploration and preferential return. We instantiate {\scshape Ditras} with the proposed diary and trajectory generators and compare the resulting algorithm with real data and synthetic data produced by other generative algorithms, built by instantiating {\scshape Ditras} with several combinations of diary and trajectory generators. We show that the proposed algorithm reproduces the statistical properties of real trajectories in the most accurate way, making a step forward the understanding of the origin of the spatio-temporal patterns of human mobility.
\end{abstract}

\section{Introduction}
Understanding the complex mechanisms governing human mobility is of fundamental importance in different contexts, from public health \citep{colizza2007,Lenormand20150473} to official statistics \citep{jos2015,pappalardo2016}, urban planning \citep{Wang:1502355,DeNadai:2016:DLG:2872427.2883084} and transportation engineering \citep{Janssens:2013:DSS:2613577}. 
In particular, human mobility modelling has attracted a lot of interest in recent years for two main reasons. On one side, it is crucial in the performance analysis of networking protocols such as mobile ad hoc networks, where the displacements of network users are exploited to route and deliver the messages \citep{karamshuk2011human,Hess:2015:DHM:2856149.2840722}. On the other side human mobility modelling is crucial for urban simulation and what-if analysis \citep{meloni2011,koppetal2014}, e.g., simulating changes in urban mobility after the construction of a new infrastructure or when traumatic events occur like epidemic diffusion, terrorist attacks or international events. In both scenarios the developing of generative algorithms that reproduce human mobility patterns in an accurate way is fundamental to design more efficient and suitable protocols, as well as to design smarter and more sustainable infrastructures, economies, services and cities \citep{Batty2012,Kitchin2013}. 

Clearly, the first step in human mobility modelling is to understand how people move. The availability of big mobility data, such as massive traces from GPS devices \citep{pappalardo2013}, mobile phone networks \citep{gonzalez08} and social media records \citep{spinsanti2013mobility}, offers nowadays the possibility to observe human movements at large scales and in great detail \citep{barbosa2017survey}. Many studies relied on this opportunity to provide a series of novel insights on the quantitative spatio-temporal patterns characterizing human mobility. These studies observe that human mobility is characterized by a stunning heterogeneity of travel patterns, i.e., a heavy tail distribution in trip distances \citep{brockmann,gonzalez08} and the characteristic distance traveled by individuals, the so-called radius of gyration \citep{gonzalez08,pappalardo2015}. Moreover human mobility is characterized by a high degree of predictability \citep{pentland,song2010limits}, a strong tendency to spend most of the time in a few locations \citep{SongNaturePhysics2010}, and a propensity to visit specific locations at specific times \citep{gonzalez_clustering,rinzi14}. 

Building upon the above findings, many generative algorithms of human mobility have been proposed which try to reproduce the characteristic properties of human mobility trajectories \citep{karamshuk2011human, barbosa2017survey}. The goal of generative algorithms of human mobility is to create a population of agents whose mobility patterns are statistically indistinguishable from those of real individuals.
Typically each generative algorithm focuses on just a few properties of human mobility. A class of algorithms aims to realistically represent spatial properties: they are mainly concerned with reproducing the trip distance distribution \citep{brockmann,gonzalez08} or the visitation frequency to a set of preferred locations \citep{SongNaturePhysics2010,pappalardo2015}. Another class of algorithms focus on the accurate representation of the time-varying behavior of individuals, relying on detailed schedules of human activities \citep{gonzalez_clustering,rinzi14}. 
However, the major challenge for generative algorithms lies in the creation of realistic temporal patterns, in which various temporal statistics observed empirically are simultaneously reproduced, including the number and sequence of visited locations together with the time and duration of the visits. In particular, the biggest hurdle consists in the simultaneous description of an individual's routine and sporadic mobility patterns. Currently there is no algorithm able to reproduce the individuals' recurrent or quasi-periodic daily schedules, and at the same time to allow for the possibility that individuals may break the routine and modify their habits during periods of unpredictability of variable duration.

In this work we present {\scshape Ditras} (DIary-based TRAjectory Simulator), a framework to simulate the spatio-temporal patterns of human mobility. The key idea of {\scshape Ditras} is to separate the temporal characteristics of human mobility from its spatial characteristics. In order to do that, {\scshape Ditras} operates in two steps. First, it generates a mobility diary using a diary generator. A mobility diary captures the temporal patterns of human mobility by specifying the arrival time and the time spent in each location visited by the individual. A diary generator is an algorithm which generates a mobility diary for an individual given a diary length. In this paper we propose a data-driven algorithm called $\mbox{MDL}$ (Mobility Diary Learner) which is able to infer from real mobility data a diary generator, MD, represented as a Markov model. The Markov model captures the propensity of individuals to follow quasi-periodic daily schedules as well as to break the routine and modify their mobility habits.

Second, {\scshape Ditras} transforms the mobility diary into a mobility trajectory by using proper mechanisms for the exploration of locations on the mobility space, so capturing the spatial patterns of human movements. The trajectory generator we propose, $d$-EPR, is based on previous research by the authors \citep{pappalardo2015,Pappalardo2016934} and embeds mechanisms to explore new locations and return to already visited locations. The exploration phase takes into account both the distance between locations and their relevance on the mobility space, though taking into account the underlying urban structure and the distribution of population density. 

We instantiate {\scshape Ditras} with the proposed diary and trajectory generators and compare it with nation-wide mobile phone data, region-wide GPS vehicular data and synthetic trajectories produced by other generative algorithms on a set of nine different standard mobility measures. We show that $d$-EPR$_{\mbox{\small MD}}$, a generative algorithm created by combining diary generator MD with trajectory generator $d$-EPR, simulates the spatio-temporal properties of human mobility in a realistic manner, typically reproducing the mobility patterns of real individuals better than the other considered algorithms. Moreover, we show that the distribution of standard mobility measures can be accurately reproduced only by modelling both the spatial and the temporal aspects of human mobility. In other words, the spatial mechanisms and the temporal mechanisms have to be modeled together by proper diary and trajectory generators in order to reproduce the observed human mobility patterns in an accurate way. The generative algorithm we propose, $d$-EPR$_{\mbox{\small MD}}$, captures both the spatial and the temporal dimensions of human mobility and is a useful tool to develop more reliable protocols for ad hoc networks as well as to perform realistic simulation and what-if scenarios in urban contexts. In summary this paper provides the following novel contributions: 
\begin{itemize}
\item the modeling framework {\scshape Ditras} which allows for the combinations of different spatial and temporal mechanisms of human mobility and whose code is freely available (\url{https://github.com/jonpappalord/DITRAS}); 
\item the data-driven algorithm MDL to construct from real mobility data a diary generator (MD) which is realistic in reproducing the temporal patterns of human mobility; 
\item a comparison of existing algorithms as well as algorithms resulting from novel combinations of temporal and spatial mechanisms, on a set of nine mobility measures and two large-scale mobility datasets. 
\end{itemize}
Our modeling framework goes towards a comprehensive approach which combines a network science perspective and a data mining perspective to improve the accuracy and the realism of human mobility models.

This paper is organized as follows. Section \ref{sec:related} revises the relevant literature on human mobility modelling. In Section \ref{sec:DITRAS} we present the structure of the {\scshape Ditras} framework. Section \ref{sec:mobility_diary} describes the first step of {\scshape Ditras}, the generation of the mobility diary, and in Section \ref{sec:markov_gen} we describe the mobility diary learner MDL and the Markov model. Section \ref{sec:mobility_traj} describes the second step of {\scshape Ditras}, the generation of the mobility trajectory, and in Section \ref{sec:dEPR} we propose a trajectory generator called $d$-EPR. Section \ref{sec:results} shows the comparison between an instantiation of {\scshape Ditras} with the proposed diary and trajectory generators with real trajectory data and the trajectories produced by other generative algorithms. In Section \ref{sec:discussion} we discuss the obtained results and, finally, Section \ref{sec:conclusion} concludes the paper.

\section{Related Work}
\label{sec:related}
All the main studies in human mobility document a stunning heterogeneity of human travel patterns that coexists with a high degree of predictability: individuals exhibit a broad spectrum of mobility ranges while repeating daily schedules dictated by routine \citep{giannotti2013complexity}. Brockmann et al.\ study the scaling laws of human mobility by observing the circulation of bank notes in United States, finding that travel distances of bank notes follow a power-law behavior \citep{brockmann}. Gonz\'{a}lez et al. analyze a nation-wide mobile phone dataset and find a large heterogeneity in human mobility ranges \citep{gonzalez08}: (i) travel distances of individuals follow a power-law behavior, confirming the results by Brockmann et al.; (ii) the radius of gyration of individuals, i.e., their characteristic traveled distance, follows a power-law behavior with an exponential cutoff. Song et al.\ observe on mobile phone data that individuals are characterized by a power-law behavior in waiting times, i.e., the time between a displacement and the next displacement by an individual \citep{SongNaturePhysics2010}.
Pappalardo et al.\ find the same mobility patterns on a dataset storing the GPS traces of 150,000 private vehicles traveling during one month in Tuscany, Italy \citep{pappalardo2013}.
Song et al.\ study the entropy of individuals' movements and find a high predictability in human mobility, with a distribution of users' predictability peaked at approximately 93\% and having a lower cutoff at 80\% \citep{song2010limits}. Pappalardo et al.\ analyze mobile phone data and GPS tracks from private vehicles and discover that individuals split into two profiles, returners and explorers, with distinct mobility and geographical patterns \citep{pappalardo2015}. Several studies focus on the prediction of the kind of activity associated to individuals' trips on the only basis of the observed displacements \citep{Liao2007,gonzalez_clustering,rinzi14}, and to discover geographic borders according to recurrent trips of private vehicles \citep{borders,thiemann2010structure}, or to predict the formation of social ties \citep{cho,wang2011}. Other works demonstrate the connection between human mobility and social networks, highlighting that friendships and other types of social relations are significant drivers of human movements \citep{brown2013place,hristova2016multilayer,wang2011,volkovich2012length,brown2013social,hossmann2011complex,hossmann2011putting}.

How to combine the discovered patterns to create a generative algorithm that reproduces the salient aspects of human mobility is an open task. This task is particularly challenging because generative algorithms should be as simple, scalable and flexible as possible, since they are generally purposed to large-scale simulation and what-if analysis.
In the literature many generative algorithms have been proposed so far to model individual human mobility patterns \citep{karamshuk2011human,barbosa2017survey}. 

Some algorithms try to reproduce the heterogeneity of individual human mobility and simulate how individuals visits locations. 
ORBIT \citep{orbit} is an example of such algorithms. It 
splits into two phases: (i) at the beginning of the simulation it generates a predefined set of locations on a bi-dimensional space; (ii) then every synthetic individual selects a subset of these locations and moves between them according to a Markov chain. In the Markov chain every state represents a specific location in the scenario and proper probability of transitions guarantee a realistic distribution of location frequencies.
SLAW (Self-similar Least-Action Walk) produces mobility traces having specific statistical features observed on human mobility data, namely power-law waiting times and travel distances with a heavy-tail distribution \citep{slaw1,slaw2}. In a first step SLAW generates a set of locations on a bi-dimensional space so that the distance among them features a heavy-tailed distribution. Then, a synthetic individual starts a trip by randomly choosing a location as starting point and making movement decisions based on the LATP (Least-Action Trip Planning) algorithm. In LATP every location has a probability to be chosen as next location that decreases with the power-law of the distance to the synthetic individual's current location. SLAW is used in several studies of networking and human mobility modelling and is the base for other generative algorithms for human mobility, such as SMOOTH \citep{smooth}, MSLAW \citep{mslaw} and TP \citep{tp1,tp2}. 

SWIM (Small World In Motion) is based on the concept of location preference \citep{swim}. First, each synthetic individual is assigned to a home location, which is chosen uniformly at random on a bi-dimensional space. Then the synthetic individual selects a destination for the next move depending of the weight of each location, which grows with the popularity of the location and decreases with the distance from the home location. The popularity of a location depends on a collective preference calculated as the number of other people encountered the last time the synthetic individual visited the location. Another category of generative algorithms combine notions about the sociality of individuals with mobility patterns to define socio-mobility models, demonstrating how they can be exploited to design more realistic protocols for ad hoc and opportunistic networks \citep{borrel2009simps,yang2010using,fischer2010gesomo,boldrini2010hcmm,musolesi2007designing}.

In contrast with many generative algorithms of human mobility, the Exploration and Preferential Return (EPR) model does not fix in advance the number of visited locations on a bi-dimensional space but let them emerge spontaneously  \citep{SongNaturePhysics2010}. The model exploits two basic mechanisms that together describe human mobility: exploration and preferential return. Exploration is a random walk process with a truncated power-law jump size distribution \citep{SongNaturePhysics2010}. Preferential return reproduces the propensity of humans to return to the locations they visited frequently before \citep{gonzalez08}. A synthetic individual in the model selects between these two mechanisms: with a given probability the synthetic individual returns to one of the previously visited places, with the preference for a location proportional to the frequency of the individual's previous visits. With complementary probability the synthetic individual moves to a new location, whose distance from the current one is chosen from the truncated power-law distribution of travel distances as measured on empirical data \citep{gonzalez08}. The probability to explore decreases as the number of visited locations increases and, as a result, the model has a warmup period of greedy exploration, while in the long run individuals mainly move around a set of previously visited places. Recently the EPR model has been improved in different directions, such as by adding information about the recency of location visits during the preferential return step \citep{barbosa2015}, or adding a preferential exploration step to account for the collective preference for locations and the returners and explorers dichotomy, as the authors of this paper have done in previous research by defining the $d$-EPR model \citep{pappalardo2015,Pappalardo2016934}. It is worth noting that although the algorithms described above are able to reproduce accurately the heterogeneity of mobility patterns, none of them can reproduce realistic temporal patterns of human movements.

Recent research on human mobility show that individuals are characterized by a high regularity and the tendency to come back to the same few locations over and over at specific times \citep{gonzalez08,pappalardo2013}. Temporal models focus on these temporal patterns and try to reproduce accurately human daily activities, schedules and regularities. Zheng et al.\ \citep{zheng_agenda} use data from a national survey in the US to extract realistic distribution of address type, activity type, visiting time and population heterogeneity in terms of occupation. They first describe streets and avenues on a bi-dimensional space as horizontal and vertical lines with random length, and then use the Dijkstra's algorithm to find the shortest path between two activities taking into account different speed limits assigned to each street. WDM (Working Day Movement) distinguishes between inter-building and intra-building movements \citep{wdm}. It consists of several submodels to describe mobility in home, office, evening and different transportation means. For example a home model reproduces a sojourn in a particular point of a home location while an office model reproduces a star-like trajectory pattern around the desk of an individual at specific coordinates inside an office building.
Although Zheng et al.'s algorithm and WDM provide an extremely thorough representations of human movements in particular scenarios, they suffer two main drawbacks: (i) they represent specific scenarios and their applicability to other scenarios is not guaranteed; (ii) they are too complex for analytical tractability; (iii) they generally fail in capturing some global mobility patterns observed in individual human mobility, e.g., the distribution of radius of gyration.
A recent study \citep{mcinerney2013breaking} proposes methods to identify and predict departures from routine in individual mobility using information-theoretic metrics, such as the instantaneous entropy, and developing a Bayesian framework that explicitly models the tendency of individuals to break from routine. 

\paragraph{Position of our work.}
From the literature it clearly emerges that existing generative algorithms for human mobility  are not able to accurately capture at the same time the heterogeneity of human travel patterns and the temporal regularity of human movements.  On the one hand exploration models accurately reproduce the heterogeneity of human mobility but do not account for regularities in human temporal patterns. On the other hand temporal models accurately reproduce human mobility schedules paying the price in complexity, but fail in capturing some important global mobility patterns observed in human mobility. In this paper we try to fill this gap and propose $d$-EPR$_{\mbox{\small MD}}$, a scalable generative algorithm that creates synthetic individual trajectories able to capture both the heterogeneity of human mobility and the regularity of human movements. 
Despite its great flexibility, $d$-EPR$_{\mbox{\small MD}}$ is to a large extent analytically tractable and several statistics about the visits to routine and non-routine locations can be derived mathematically.
In fact, since the temporal mechanism of $d$-EPR$_{\mbox{\small MD}}$ is based on a Markov chain, using standard results in probability theory one can compute various quantities, including the probability to go between any two states in a given number of steps, the average number of visits to a state before visiting another state, the average time to go from one state to another and the probability to visit one state before another. 
Moreover the spatial mechanism of $d$-EPR$_{\mbox{\small MD}}$ is based on the EPR model for which various analytical results, such as the distributions of the radii of gyration and of the location frequencies, have been derived \citep{SongNaturePhysics2010}. The data-driven algorithm MDL (Mobility Diary Learner), is another novel contribution of this paper. MDL infers from real mobility data a diary generator for realistic mobility diaries. It is highly adaptive and can be applied to different geographic areas and different types of mobility data. 

The modelling framework we propose, {\scshape Ditras}, can generate synthetic mobility trajectories and can be easily integrated in transportation forecast models to infer trip demand. Our approach has some similarity with activity-based models \citep{bellemans2010implementation}, as they both aim to estimate trip demand by reproducing realistic individual temporal patterns, however there are important differences between the two approaches. In fact, while the goal of activity-based models is to produce detailed agendas filled with activities performed by the agents and are calibrated on surveys with a limited number of participants, our framework produces mobility diaries containing the time and duration of the visits in the various locations without explicitly specifying the type of activity performed there, and is calibrated on a large population of mobile phone users.

A recent paper introduces TimeGeo, a modelling framework to generate a population of synthetic agents with realistic spatio-temporal trajectories \citep{jiang2016}. 
Similarly to the modelling framework presented here, TimeGeo combines a Markov model to generate temporal patterns with the correct periodicity and duration of visits, with a model to reproduce spatial patterns with the characteristic number of visits and distribution of distances. 
Albeit having similar aims, there are important differences between our modelling approach and TimeGeo's. 
In fact, while TimeGeo proposes a parsimonious model which is based on few tunable parameters and is to some extent analytically tractable, the approach proposed in this paper is markedly data driven and parameter-free, with a greater level of complexity which ensures the necessary flexibility to reproduce realistic temporal patterns.

\section{The DITRAS modelling framework}
\label{sec:DITRAS}

\begin{figure}[htb]\centering
\includegraphics[scale=0.4]{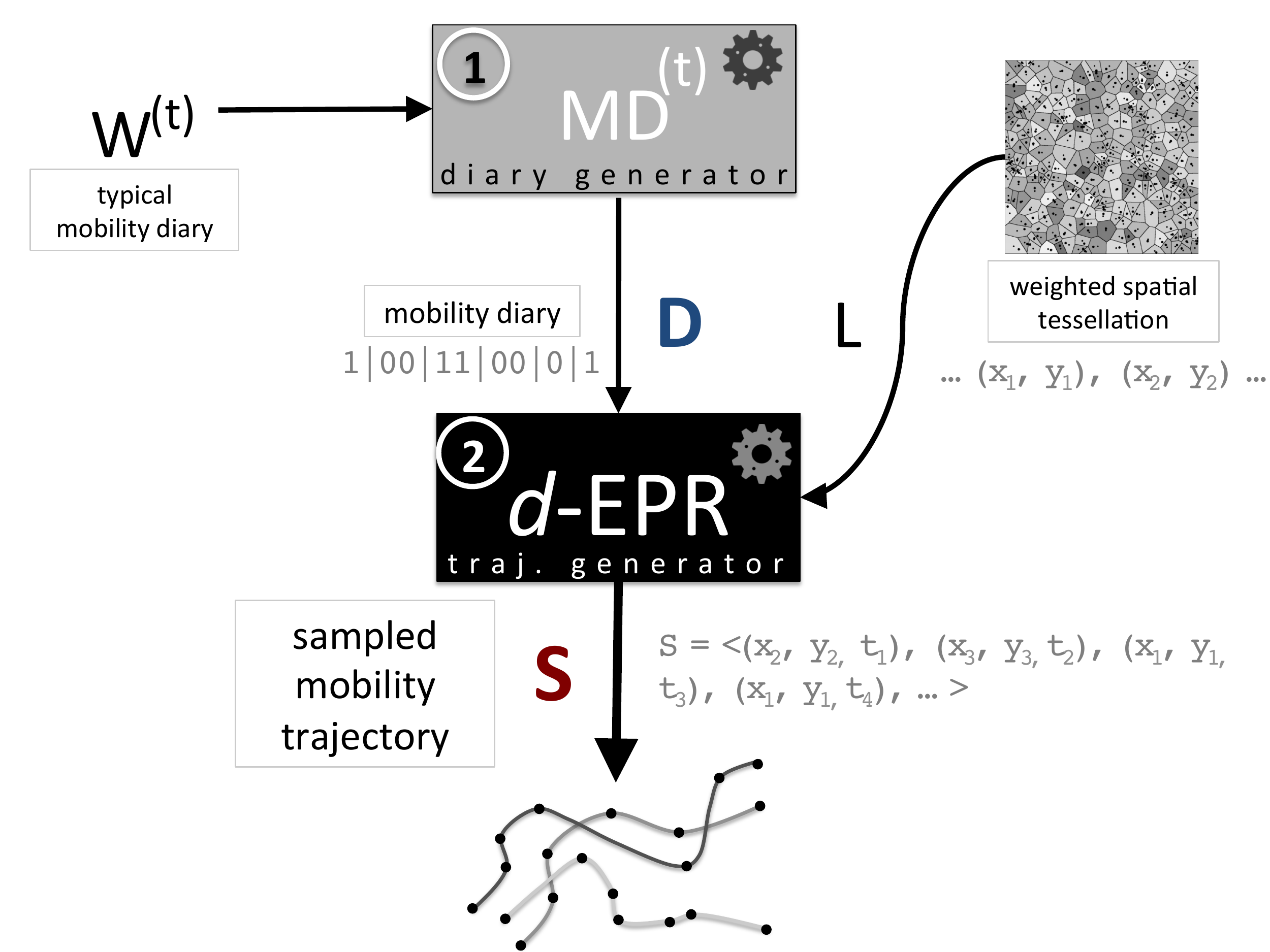}
\caption{\textbf{Outline of the DITRAS framework.} {\scshape Ditras} combines two probabilistic models: a diary generator (e.g., MD$^{(t)}$) and trajectory generator (e.g., $d$-EPR). The diary generator uses a typical diary $W^{(t)}$ to produce a mobility diary $D$. The mobility diary $D$ is the input of the trajectory generator together with a weighted spatial tessellation of the territory $L$. From $D$ and $L$ the trajectory generator produces a sampled mobility trajectory $S$.}
\label{fig:DITRAS_schema}
\end{figure}

{\scshape Ditras} is a modelling framework to simulate the spatio-temporal patterns of human mobility in a realistic way.\footnote{The Python code of {\scshape Ditras} is freely available for download on a public GitHub repository: \url{https://github.com/jonpappalord/DITRAS}} The key idea of {\scshape Ditras} is to separate the temporal characteristics of human mobility from its spatial characteristics. For this reason, {\scshape Ditras} consists of two main phases (Fig.\ \ref{fig:DITRAS_schema}): first, it generates a mobility diary which captures the temporal patterns of human mobility; second it transforms the mobility diary into a sampled mobility trajectory which captures the spatial patterns of human movements. In this section we define the main concepts which constitute the mechanism of {\scshape Ditras}. 

The output of a {\scshape Ditras} simulation is a sampled mobility trajectory for a synthetic individual. A mobility trajectory describes the movement of an object as a sequence of time-stamped locations. The location is described by two coordinates, usually a latitude-longitude pair or ordinary Cartesian coordinates, as formally stated by the following definition:

\begin{definition}[Mobility trajectory]
\label{def:mob_traj}
A mobility trajectory is a sequence of triples $T = \langle (x_1, y_1, t_1), \dots, (x_n, y_n, t_n) \rangle$, where $t_i$ $(i = 1, \dots, n)$ is a timestamp, $\forall_{1 \le i < n}$ $t_i < t_{i + 1}$ and 
$x_i, y_i$ 
are coordinates on a bi-dimensional space. 
\end{definition}

For modelling purposes it is convenient to define a sampled mobility trajectory, $S^{(t)}$, which can be obtained by sampling the mobility trajectory at regular time intervals of length $t$ seconds: 

\begin{definition}[Sampled mobility trajectory]
\label{def:mob_traj}
A sampled mobility trajectory is a sequence $S^{(t)} = \langle l_1, \dots, l_N \rangle$, where $l_i$ $(i = 1, \dots, N)$ is the geographic location where the individual spent the majority of time during time slot $i$, i.e., between $(i-1) t$ and $t i$ seconds from the first observation. 
$N$ is the total number of time slots considered. A location $l_i$ is described by coordinates on a bi-dimensional space. 
\end{definition}

To generate a sampled mobility trajectory {\scshape Ditras} exploits two probabilistic models: a diary generator and a trajectory generator (see Fig.\ \ref{fig:DITRAS_schema}). 
In this paper we propose as diary generator MD$^{(t)}$, a Markov model responsible for reproducing realistic temporal mobility patterns, such as the distribution of the number of trips per day and the tendency of individuals to change location at specific hours of the day \citep{gonzalez08,gonzalez_clustering}. Essentially, MD$^{(t)}$ captures the tendency of individuals to follow or break a temporal routine at specific times. As trajectory generator we propose the $d$-EPR generative model \citep{pappalardo2015,Pappalardo2016934}, which is able to reproduce realistic spatial mobility patterns, such as the heavy-tail distributions of trip distances \citep{brockmann,gonzalez08,pappalardo2013} and radii of gyration \citep{gonzalez08,pappalardo2013,pappalardo2015}, as well as the characteristic visitation patterns, such as the uneven distribution of time spent in the various locations \citep{SongNaturePhysics2010,pappalardo2013}. $d$-EPR embeds a mechanism to choose a location to visit on a bi-dimensional space given the current location, the spatial distances between locations and the relevance of each location. 

Fig.\ \ref{fig:DITRAS_schema} provides an outline of {\scshape Ditras} and Algorithm \ref{alg:DITRAS} describes its pseudocode. {\scshape Ditras} is composed of two main steps. During the first step, the diary generator builds a mobility diary $D$ of $N$ time slots, each of duration $t$. The operation of this step is described in detail in Section \ref{sec:mobility_diary}. During the second step, {\scshape Ditras} uses the trajectory generator and a given spatial tessellation $L$ to transform the mobility diary into a sampled mobility trajectory. We describe in detail the second step of {\scshape Ditras} in Section \ref{sec:mobility_traj}. Note that the two-step process described above is a general framework common to many generative models of human mobility, which are often composed by two sequential parts, the first generating temporal patterns and the second determining the spatial trajectory. 
However, in some models the division between the temporal and the spatial mechanisms is present but not explicitly acknowledged. 

In Section \ref{sec:results} we will instantiate {\scshape Ditras} by using MD$^{(t)}$ and $d$-EPR and compare it with other generative models obtained combining diary generators (first step) with trajectory generators (second step).

\SetKw{Continue}{continue}
\begin{algorithm}[htb]
\nonl {\large \textbf{The DITRAS framework}}\\
 \BlankLine
 \BlankLine
 \small
\DontPrintSemicolon
\SetKwInOut{Input}{input}
    \SetKwInOut{Output}{output}
    \Input{$L = \{(l_1, r_1), \dots, (l_n, r_n)\}$, weighted spatial tessellation\\
    G, diary generator\\
    $N$, length of trajectory to generate\\
    $W$, typical diary 
    }
    \Output{$S = \langle (x_1, y_1, t_1), \dots, (x_n, y_n, t_n) \rangle$, sampled mobility trajectory of length $N$}
 \SetKwFunction{executeMarkovChain}{generateMobilityDiary}
 \SetKwFunction{append}{append}
 \SetKwFunction{update}{update}
 \SetKwFunction{weightedRandom}{weightedRandom}
  \SetKwFunction{hasCoordinates}{hasCoordinates}
  \SetKwFunction{dEPR}{dEPR}
  \SetKwFunction{TG}{TG}
  \SetKwFunction{assignLocations}{generateMobilityTrajectory}
  \SetKwFunction{assignW}{assignLocationsTo}
\normalsize
 \BlankLine
 \BlankLine
\setcounter{AlgoLine}{0}
\nl $D = \executeMarkovChain(\mbox{G}, N)$ \tcp{use the diary generator DG to create a mobility diary $D$ of length $N$} 
\BlankLine
$S = \assignLocations(D, L, W)$ \tcp{scan the mobility diary $D$ and create a sample mobility trajectory $S$ of length $N$}
\BlankLine
\BlankLine
\KwRet $S$\;
\small

\BlankLine
\BlankLine

\setcounter{AlgoLine}{0}
\SetKwProg{myalg}{Function}{}{}
  \myalg{\assignLocations{D, L, W}}{
 \BlankLine
$S = newList()$\;
$t = 1$\;
\BlankLine
$W_m = \assignW(W)$ \tcp{assign a physical location to every abstract location in typical diary $W$}
\BlankLine
\While{$d < length(D)$}{
        
{\tcp{scan the mobility diary $D$}
\BlankLine
	\If{$D[d] = |$} 
       {
       \tcp{when it sees a separator `|'}
       		$d = d + 1$\;
			\Continue
		}
		\BlankLine
	\If{$D[d] = 0$} 
       {
       		\tcp{the individual follows the routine (i.e., she visits a typical location)}
			$S.\append((W_m[t], t))$\;
			$t = t + 1$\;
		}
       
       \Else
       {
       \tcp{the individual breaks the routine}
       	$l = \TG(S, P)$ \tcp{call the trajectory generator TG to obtain the next location to visit}
		$S.\append((l, t))$\;
		$t = t + 1$\;
		\BlankLine
		$j = d + 1$\;
		\While{$D[d] = D[j]$}
		{ \tcp{stay in location $l$ until the next separator appears}
			$S.\append((l, t))$\;
			$t = t + 1$\;
			\BlankLine
			$j = j + 1$\;
			
		}
		\BlankLine
		$d = j - 1$\;
       }
       \BlankLine
       }
       $d = d + 1$\;
}
}{\KwRet $S$\;}

\caption{The algorithm describing the {\scshape Ditras} framework.}
\label{alg:DITRAS}
\end{algorithm}

\section{Step 1: Generation of mobility diary}
\label{sec:mobility_diary}

A diary generator G produces a mobility diary, $D^{(t)}$, containing the sequence of trips made by a synthetic individual during a time period divided in time slots of $t$ seconds.
For example, G$^{(3600)}$ and G$^{(60)}$ produce mobility diaries with temporal resolutions of one hour and one minute, respectively. In Section \ref{sec:markov_gen} we illustrate a data-driven algorithm to construct a diary generator, MD$^{(t)}$, using real mobility trajectory data such as mobile phone data. 

To separate the temporal patterns from the spatial ones, we define the abstract mobility trajectory, $A^{(t)}$, which contains the time ordered list of the ``abstract locations'' visited by a synthetic individual during a period divided in time slots of $t$ seconds.
An abstract location uniquely identifies a place where the individual is stationary, like home or the workplace, but it does not contain any information on the specific geographic position of the location (i.e., its coordinates). 
The abstract mobility trajectory is thus equivalent to the sampled mobility trajectory where the geographic locations, $l_k$, are substituted by placeholders, $a_k$, called abstract locations:

\begin{definition}[Abstract mobility trajectory]
\label{def:abs_mob_traj}
An abstract mobility trajectory is a sequence $A^{(t)} = \langle a_1, \dots, a_N \rangle$, where $a_i$ $(i = 1, \dots, N)$ is the abstract location where the individual spent the majority of time during time slot $i$, i.e., between $(i-1) t$ and $i t$ seconds from the first observation. 
\end{definition}

The mobility diary, $D^{(t)}$, is generated with respect to a typical mobility diary, $W^{(t)}$, which represents the individual's routine.
$W^{(t)}$ is a sequence of time slots of duration $t$ seconds and specifies the typical and most likely abstract location the individual visits in every time slot. 
Here we consider the simplest choice of typical mobility diary, in which the most likely location where a synthetic individual can be found at any time is her home location. 
It is possible to relax this simplifying assumption and estimate an individual's typical mobility diary from the data by computing her mobility \emph{regularity}, which is the time series of the most visited location in each time slot \citep{song2010limits}.
Computing the weekly mobility regularity of individuals on real large-scale mobile phone data and GPS vehicular data and performing a clustering of their typical diaries we find that there is one dominant cluster containing $\approx$90\% of the individuals and whose representative typical diary has a single location (see Appendix \ref{app:clustering}). This result supports the validity of the simplifying assumption to consider one typical diary with a single location for all agents. The proposed generative model does not change if there are two or more typical mobility diaries which have more than one typical location. When a synthetic individual is generated it can be randomly assigned to one of the typical diaries in proportion to the overall frequency of the various diaries among real users. Then, the rest of the algorithm remains the same.

\begin{definition}[Typical mobility diary]
A typical mobility diary is a sequence $W^{(t)} = \langle w_1, \dots, w_N \rangle$ where $w_k = w \quad \forall k=1, \dots, N$ denotes the home location of the synthetic individual. $N$ is the total number of time slots considered.
\end{definition} 

The mobility diary, $D^{(t)}$, specifies whether an individual's abstract mobility trajectory, $A^{(t)}$, follows her typical mobility diary, $W^{(t)}$, or not. 
In particular, for every time slot $i$, $D^{(t)}(i)$ can assume two values:
\begin{itemize} 
\item $D^{(t)}(i) = 1$ if $A^{(t)}(i) = W^{(t)}(i)$, meaning that the individual visits the abstract location $W^{(t)}(i)$ following her routine, i.e., she is at home; 
\item $D^{(t)}(i) = 0$ if $A^{(t)}(i) \neq W^{(t)}(i)$, meaning that the individual visits a location other than the abstract location $W^{(t)}(i)$ being out of her routine. 
\end{itemize}

\begin{definition}[Mobility Diary]
A mobility diary is a sequence $D^{(t)}$ of time slots of duration $t$ seconds generated by the regular language $\mathbb{L} = ( 1^+| (0^+|)^* )^*$, where $1$ at time slot $i$ indicates that the individual visits the abstract location in her typical diary at time $i$, $W^{(t)}(i)$, and $0$ indicates a visit to a location different from the abstract location $W^{(t)}(i)$. The symbol ``$|$'' indicates a transition or trip between two different abstract locations. 
\end{definition}

An example of mobility diary generated by language $\mathbb{L}$ is $D^{(t)} = \langle 1 1 | 0 0 | 0 | 1 \rangle$. The first two entries indicate that $A^{(t)}(1) = W^{(t)}(1)$ and $A^{(t)}(2) = W^{(t)}(2)$, i.e., the individual follows her routine and she is at home. 
Next, the third, fourth and fifth entries indicate that $A^{(t)}(3) \ne W^{(t)}(3)$, $A^{(t)}(4) \ne W^{(t)}(4)$ and $A^{(t)}(5) \ne W^{(t)}(5)$, i.e., the individual breaks the routine and visits a non-typical location for two consecutive time slots, then she visits a different non-typical location for one time slot. Finally, the last time slot indicates that $A^{(t)}(6) = W^{(t)}(6)$, the individual follows the routine and returns home. 
We assume that the travel time between any two locations is of negligible duration.

\subsection{Mobility Diary Learner (MDL)}
\label{sec:markov_gen}

In this section we propose diary generator MD$^{(t)}$ and illustrate MDL (Mobility Diary Learner), a data-driven algorithm to compute MD from the abstract mobility trajectories of a set of real individuals (Algorithm \ref{alg:diary_generator_builder}).
We use a Markov model to describe the probability that an individual follows her routine and visits a typical location at the usual time, or she breaks the routine and visits another location. First, MDL translates mobility trajectory data of real individuals into abstract mobility trajectories (Section \ref{sec:mob_traj}). Second, it uses the obtained abstract trajectory data to compute the transition probabilities of the Markov model MD$^{(t)}$ (Section \ref{sec:markov_chain}).

\subsubsection{Mobility trajectory data}
\label{sec:mob_traj}

The construction of MD$^{(t)}$ is based on mobility trajectory data of real individuals. We assume that raw mobility trajectory data describing the movements of a set of individuals are in the form $ \langle (u_1, x_1, y_1, t_1), \dots, (u_n, x_n, y_n, t_n)\rangle$ where $u_i$ indicates the individual who visits location $(x_i, y_i)$ at time $t_i$, $\forall_{1 \le i < n}$ $t_i < t_{i + 1}$.

Mobility trajectory data can be obtained from various sources (e.g., mobile phones, GPS devices, geosocial networks) and describe the movements of individuals on a territory. Since the purpose of MD$^{(t)}$ is to capture the temporal patterns regardless the geographic position of locations, we translate raw mobility trajectory data into abstract mobility trajectories (see definition in Section \ref{def:abs_mob_traj}). 

Starting from the raw trajectory data, we assign an abstract location to every time slot in an individual's abstract mobility trajectory $A^{(t)}$ according to the following method. If the individual visits just one location during time slot $i$, we assign that location to $i$. If the individual visits multiple locations during slot $i$, we choose the most frequent location in $i$, i.e., the location where the individual spends most of the time during the time slot. If there are multiple locations with the same visitation frequency in time slot $i$, we choose the location with the highest overall frequency. 
If there is no information in the abstract trajectory data about the location visited in time slot $i$ (e.g., no calls during the time slot in the case of mobile phone data), we assume no movement and choose the location assigned to time slot $i-1$.

To clarify the method let us consider the following example. A mobile phone user has the following hourly time series of calls: $[A, A, \bullet, \bullet, B, (C, C, B, B)]$, where $A,B,C$ are placeholders for different cell phone towers (i.e., abstract locations). Here the symbol $\bullet$ indicates that there is no information in the data about the location visited during the 1-hour time slot, while all the locations in round brackets are visited during the same time slot. Using the method described above, the abstract mobility trajectory of the individual becomes $A^{(3600)} = \langle A, A, A, A, B, B \rangle$ because: (i) the two $\bullet$ symbols in the third and fourth time slots are substituted by location $A$ assuming no movement with respect to the second time slot; (ii) the location assigned to the last time slot is $B$ since $C$ and $B$ have the same visitation frequency in $(C, C, B, B)$ but $f(B) > f(C)$, i.e., $B$ has the highest overall visitation frequency.

It is worth noting that the choice of the duration of the time slot, $t$, is crucial and depends on the specific kind of mobility trajectory data used. GPS data from private vehicles, for example, generally provide accurate information about the location of the vehicle every few seconds. In this scenario, a time slot duration of one minute can be a reasonable choice. In contrast when dealing with mobile phone data a time slot duration of an hour or half an hour is a more reliable choice, since the majority of individuals have a low call frequency during the day \citep{pappalardo2015}. 

\subsubsection{Markov model transition probabilities}
\label{sec:markov_chain}

Let $A_u = \langle a^{(u)}_0, \dots, a^{(u)}_{n-1}\rangle$ and $W_u = \langle w^{(u)}_0, \dots, w^{(u)}_{n-1}\rangle$ be the abstract mobility trajectory and the typical mobility diary of individual $u \in U$, where $U$ is the set of all individuals in the data -- we omit the superscript ${(t)}$ for clarity. 
Elements $a^{(u)}_h \in A_u$ and $w^{(u)}_h \in W_u$ denote the abstract and the typical locations visited by individual $u$ at time slot $h$ with $h = 0, \dots, N{-}1$. 

A state in the Markov model MD is a tuple of two elements $s = (h, R)$. The state's first element, $h$, is the time slot of the time series denoted by an integer between 0 and $N{-}1$. The state's second element, $R$, is a boolean variable that is 1 (True) if at time slot $h$ the individual is in her typical location, $w_h^{(u)}$, and 0 (False) otherwise -- just like in the mobility diary. In total there are $N \times 2 = 2N$ possible states in the model. The transition matrix, MD, is a $2N \times 2N$ stochastic matrix whose element MD$_{s s'}$ corresponds to the conditional probability of a transition from state $s$ to state $s'$, MD$_{s s'} \equiv p(s' | s)$. The normalization condition imposes that the sum over all elements of any
row $s$ is equal to 1, $\sum_{s'}$ MD$_{s s'} = 1, \forall s$. We consider two types of transitions, $s \to s'$, depending on whether in state $s$ the individual is in typical location or not:
\begin{itemize} 
\item if the individual is in the typical location at time slot $h$, i.e., $s = (h, 1)$, then she can either go to the next typical location at time slot $h+1$, $s = (h, 1) \to s' = (h+1, 1)$, or go to a non-typical location and stay there for $\tau$ time slots, $s = (h, 1) \to s' = (h+\tau, 0)$;
\item if instead the individual is not in the typical location at time slot $h$, i.e., $s = (h, 0)$, then she can either go to the typical location at time slot $h+1$, $s = (h, 0) \to s' = (h+1, 1)$, or go to a different non-typical location and stay there for $\tau$ time slots, $s = (h, 0) \to s' = (h+\tau, 0)$.
\end{itemize}

The formulae to compute the empirical frequencies for the four types of transitions are shown in Table \ref{mmt}. In the table, $\delta_x^u(a) = \delta(a_x^{(u)}, w_x^{(u)})$, $\hat{\delta}_{x}^u(a) = \delta(a^{(u)}_{x} , a^{(u)}_{x+1}) $, where $\delta(i ,j) = 1$ if $i=j$ and $0$ otherwise, is the Kronecker delta.
By convention, the product $\prod_{i=1}^{\tau-1} \dots$ is equal to $1$ if
$\tau = 1$.

\begin{longtable}[c]{@{}lc@{}}
\toprule\addlinespace
\label{mmt} 
Transition, $s \to s'$ & Frequency, MD$_{s s'}$
\\\addlinespace
\midrule\endhead
$(h, 1) \to (h+1, 1)$ &
$\frac{\sum_{u \in U} \sum_{a \in A_u} \delta_h^u(a)  \delta_{h + 1}^u(a)}{\sum_{u \in U} \ \sum_{a \in A_u} \delta_h^u(a)}$

\\\addlinespace
$(h, 1) \to (h+\tau, 0)$ &
$\frac{\sum_{u \in U} \ \sum_{a \in A_u} \delta_h^u(a) [1- \delta_{h+1}^u(a)]  \prod_{i=1}^{\tau-1} \hat{\delta}_{h+i}^u(a) [1- \hat{\delta}_{h + \tau}^u(a)] }{\sum_{u \in U} \ \sum_{a \in A_u} \delta_h^u(a) }$

\\\addlinespace
$(h, 0) \to (h+1, 1)$ &
$\frac{\sum_{u \in U} \ \sum_{a \in A_u} [1 - \delta_h^u(a)]  \delta_{h+1}^u(a)}{\sum_{u \in U} \ \sum_{a \in A_u} [1 - \delta_h^u(a)] }$

\\\addlinespace
$(h, 0) \to (h+\tau, 0)$ &
$\frac{ \sum_{d \in D} [1 - \delta_h^u(a)] [1- \delta_{h+1}^u(a)]  [1- \hat{\delta}_h^u(a)]  \prod_{i=1}^{\tau-1} \hat{\delta}_{h + i}^u(a) [1- \hat{\delta}_{h + \tau}^u(a)] }{\sum_{u \in U} \ \sum_{a \in A_u} [1 - \delta_h^u(a)] }$
\\\addlinespace
\bottomrule
\caption{Formulae to compute the transition probabilities of the Markov chain MD from abstract mobility trajectories.} \label{t:markov}
\end{longtable}

\begin{algorithm}
\nonl {\large \textbf{MDL (Mobility Diary Learner)}}\\
 \BlankLine
 \BlankLine
 \small
\DontPrintSemicolon
\SetKwInOut{Input}{input}
\SetKwInOut{Output}{output}
    \Input{$D=\{T_1, \dots, T_n\}$, dataset of real trajectories of $n$ agents\\
    $t$, time slot length\\
    }
    \Output{$G$, a Markov chain}
\setcounter{AlgoLine}{0}    
    \SetKwFunction{createTimeSeries}{createTimeSeries}
    \SetKwFunction{updateMarkovChain}{updateMarkovChain}
    \SetKwFunction{emptyMarkovChain}{emptyMarkovChain}
    \SetKwFunction{normalizeMarkovChain}{normalizeMarkovChain}
    \DontPrintSemicolon
    \BlankLine
    \BlankLine
    \normalsize
    \nl $G = \emptyMarkovChain()$\;
    \ForAll{$i \in \{1, \dots, n\}$}{
     $A_i = \createTimeSeries(T_i)$ \tcp{create abstract trajectory of $i$}
    $G = \updateMarkovChain(A_i)$ \tcp{update the Markov chain using $A_i$}
     }
    \KwRet $G$
    \BlankLine
    \small
    \setcounter{AlgoLine}{0}
\SetKwProg{myalg}{Function}{}{}
  \myalg{\updateMarkovChain{A, G}}{
$slot = 0$\;
\While{$slot < len(A) - 1$} 
{
$h = slot \% 24$ \tcp{hour of the day}
$next_h = (h + 1) \% 24$  \tcp{next hour of the day}
$loc_h = A[slot]$  \tcp{abstract location at the slot}
$loc_{h + 1} = A[slot + 1]$  \tcp{abstract location at next slot}
\BlankLine
\If{$loc_h == 1$}
{
   \If{$loc_{h + 1} == 1$}
   {
      \tcp{Case 1: $loc_h$ is typical and $loc_{h + 1}$ is typical}
      $G[(h, 1), (next_h, 1)] =  G[(h, 1), (next_h, 1)] + 1$\;
      \BlankLine
   }
   \Else{
       \tcp{Case 2: $loc_h$ is typical and $loc_{h+1}$ is not typical}
       $\tau = 1$\;
          \For{$j=slot + 2$ to $len(A)$}{
             $loc2_{h} = A[j]$\;
             \If{$loc2_h == loc2_{h + 1}$}{
                $\tau = \tau + 1$\;
             }
             \Else{
                \textbf{break}\;
             }
          }
          $h_{\tau} = (h + \tau)\%24$\;
          $G[(h, 1), (h_{\tau}, 0)] =  G[(h, 1), (h_{\tau}, 0)] + 1$\;
          $slot = j - 1$\;
   }
}
\Else{
   \If{$loc_{h + 1} == 1$}
   {
   \tcp{Case 3: $loc_h$ is not typical and $loc_{h + 1}$ is typical}
       $G[(h, 0), (next_h, 1)] =  G[(h, 0), (next_h, 1)] + 1$\;
   }
   \Else{
   \tcp{Case 4: both $loc_h$ and $loc_{h + 1}$ are not typical}
   $\tau = 1$\;
   \For{$j=slot + 2$ to $len(A)$}{
             $loc2_{h} = A[j]$\;
             \If{$loc2_h == loc2_{h + 1}$}{
                $\tau = \tau + 1$\;
             }
             \Else{
                \textbf{break}\;
             }
          $h_{\tau} = (h + \tau)\%24$\;
          $G[(h, 0), (h_{\tau}, 0)] =  G[(h, 0), (h_{\tau}, 0)] + 1$\; 
          $slot = j - 1$\;
          }
   }
}
$slot = slot + 1$
}
$G = \normalizeMarkovChain(G)$\;
\KwRet $G$
}
\caption{Algorithm for the construction of MD generator.}
\label{alg:diary_generator_builder}
\end{algorithm}

\section{Step 2: Generation of sampled mobility trajectory}
\label{sec:mobility_traj}

Starting from the mobility diary $D^{(t)}$, the sampled mobility trajectory $S^{(t)}$ is generated to describe the movement of a synthetic individual between a set of discrete locations called weighted spatial tessellation. 
A weighted spatial tessellation is a partition of a bi-dimensional space into locations each having a weight corresponding to its relevance. 

\begin{definition}[Weighted spatial tessellation]
A weighted spatial tessellation is a set of tuples $L = \{(l_1, r_1), \dots, (l_m, r_m)\}$, where $r_j \in \mathbb{N}$ $(j = 1, \dots, m)$ is the relevance of a location and the $l_j$ are a set of non-overlapping polygons that cover the bi-dimensional space where individuals can move. 
The location of each polygon is identified by the coordinates of its centroid, $(x_{j}, y_{j})$.
\end{definition}

The weighted spatial tessellation indicates the possible physical locations on a finite bi-dimensional space a synthetic individual can visit during the simulation. 
The relevance of a location measures its popularity among real individuals: locations of high relevance are the ones most frequently visited by the individuals \citep{pappalardo2015,Pappalardo2016934}. 
The relevance is introduced to generate synthetic trajectories that take into account the underlying urban structure. 
An example of weighted spatial tessellation is the one defined by a set of mobile phone towers, where the relevance of a tower can be estimated as the number of calls performed by mobile phone users during a period of observation, and the polygons correspond to the regions obtained from the Voronoi partition induced by the towers. If information about location relevance is not available to the user of the simulator, the distribution of population can be used to estimate the relevance of the locations. For example, the websites \url{http://sedac.ciesin.columbia.edu/} and \url{http://www.worldpop.org.uk/} provide a fine-grained spatial tessellation for the entire globe, together with an estimate of population density in every location.

First, {\scshape Ditras} assigns to every abstract location in the typical mobility diary $W^{(t)}$ a physical location on the weighted spatial tessellation $L$, creating $W_m^{(t)}$, a typical mobility diary where each abstract location has a specific geographic position (Algorithm \ref{alg:DITRAS}, line 4, procedure \texttt{assignLocationsTo}). The geographic position of an abstract location is chosen according to the distribution of location relevance specified in the spatial tessellation, i.e., the more relevant a location is the more likely it is chosen as a geographic position of an abstract location. This choice ensures the generation of synthetic data with a realistic distribution of locations across the territory \citep{Pappalardo2016934}. 
Next, {\scshape Ditras} scans $D^{(t)}$ to assign a physical location to every entry. For every entry $D^{(t)}(i) \in D^{(t)}$ we have two possible scenarios: 
\begin{itemize}
\item $D^{(t)}(i) = 1$, the entry indicates a visit to a typical location, i.e., the abstract location in $W^{(t)}(i)$ (Algorithm \ref{alg:DITRAS}, line 12). In this scenario the synthetic individual visits location $l = W_m^{(t)}(i)$ which is added to the sampled trajectory at time slot $i$, i.e. $S^{(t)}(i) = W_m^{(t)}(i)$ (Algorithm \ref{alg:DITRAS}, lines 14); \\
\item $D^{(t)}(i) = 0$, the entry indicates a visit to a non-typical location (Algorithm \ref{alg:DITRAS}, line 17). In this second scenario {\scshape Ditras} calls the trajectory generator to choose a location $l$ to visit, where $l \ne W_m^{(t)}(i)$ (Algorithm \ref{alg:DITRAS}, lines 19). The chosen location $l$ is added to the sampled mobility trajectory $k$ times, where $k$ is the number of consecutive $0$ characters before the next separator character `$|$' appears in $D^{(t)}$, i.e., the total number of time slots spent in location $l$ (Algorithm \ref{alg:DITRAS}, lines 23-27).
\end{itemize}

\paragraph{Example of trajectory generation.} To clarify how the second step of {\scshape Ditras} works let us consider the following example. A synthetic individual is assigned a mobility diary $D^{(t)} = \langle 1|00|1\rangle$ and the chosen typical diary is $W^{(t)} = \langle w,w,w,w\rangle$, where $w$ denotes the individual's home. To generate a synthetic sampled mobility trajectory $S$, {\scshape Ditras} operates as follows. First, {\scshape Ditras} assigns a physical location to the individual's home $w$, generating $W_m^{(t)} = \langle (x_1, y_1), (x_1, y_1), (x_1, y_1), (x_1, y_1)\rangle$. Next, {\scshape Ditras} starts from the first entry $D^{(t)}(1)$. Since $D^{(t)}(1) = 1$ the synthetic individual is at home. Therefore, tuple $(x_1, y_1, 1)$ is added to trajectory $S$. Next, {\scshape Ditras} processes the second entry $D^{(t)}(2)$, sees a separator and then proceeds to entry $D^{(t)}(3)$. Since $D^{(t)}(3) = 0$, the synthetic individual is not at home in the third time slot. Hence, {\scshape Ditras} calls a trajectory generator (e.g., $d$-EPR) which chooses to visit physical location $(x_2, y_2)$. {\scshape Ditras} hence adds the tuples $(x_2, y_2, 2)$ and $(x_2, y_2, 3)$ to trajectory $S$, since there two $0$ characters until the next separator in $D^{(t)}$. The last entry $D^{(t)}(6) = 1$ indicates that the synthetic individual returns home in the fourth time slot. So, {\scshape Ditras} adds tuple $(x_1, y_1, 4)$ to trajectory $S$. At the end of the execution, the sampled mobility trajectory generated by {\scshape Ditras} is $S = \langle (x_1, y_1, 1), (x_2, y_2, 2), (x_2, y_2, 3), (x_1, y_1, 4) \rangle$.

\subsection{The $d$-EPR model}
\label{sec:dEPR}

As trajectory generator we propose the $d$-EPR individual mobility model \citep{pappalardo2015,Pappalardo2016934} that assigns a location on the bi-dimensional space to an entry in mobility diary $D^{(t)}$. The $d$-EPR (density-Exploration and Preferential Return) is based on the evidence that an individual is more likely to visit relevant locations than non-relevant locations \citep{pappalardo2015,Pappalardo2016934}. For this reason $d$-EPR incorporates two competing mechanisms, one driven by an individual force (preferential return) and the other driven by a collective force (preferential exploration). The intuition underlying the model can be easily understood: when an individual returns, she is attracted to previously visited places with a force that depends on the relevance of such places at an individual level. In contrast, when an individual explores she is attracted to new places with a force that depends on the relevance of such places at a collective level. In the preferential exploration phase a synthetic individual selects a new location to visit depending on both its distance from the current location, as well as its relevance measured as the collective location's relevance in the bi-dimensional space. In the model, hence, the synthetic individual follows a personal preference when returning and a collective preference when exploring. The $d$-EPR uses the gravity model \citep{zipf1946,jung2008,Lenormand2016158} to assign the probability of a trip between any two locations in $L$, which automatically constrains individuals within a territory's boundaries. The usage of the gravity model is justified by the accuracy of the gravity model to estimate origin-destination matrices even at the country level \citep{RefWorks:134,RefWorks:40,RefWorks:192,RefWorks:121,Lenormand2016158}. 

Algorithm \ref{alg:dEPR} describes how $d$-EPR assigns a location on the bi-dimensional space defined by a spatial tessellation $L$ for an entry in mobility diary $D^{(t)}$. The $d$-EPR takes in input two variables: (i) the current sampled mobility trajectory of the synthetic individual $S = \langle (x_1, y_1, t_1), \dots, (x_n, y_n, t_n) \rangle$; (ii) a probability matrix $P$ indicating, for every pair of locations $i, j \in L, i \ne j$ the probability of moving from $i$ to $j$. Every probability $p_{ij}$ is computed as: $$p_{ij} = {1\over Z} \frac{r_i r_j}{d_{ij}^2},$$ where $r_{i(j)}$ is the relevance of location $i (j)$ as specified in the weighted spatial tessellation $L$, $d_{ij}$ is the geographic distance between $i$ and $j$, and $Z = \sum_{i,j\neq i} p_{ij}$ is a normalization constant. The matrix $P$ is computed before the execution of the {\scshape Ditras} model by using the spatial tessellation $L$. 

With probability $p_{new} = \rho N^{-\gamma}$ where $N$ is the number of distinct locations in $S$ and $\rho=0.6$, $\gamma=0.21$ are constants \citep{pappalardo2015,Pappalardo2016934,SongNaturePhysics2010}, the individual chooses to explore a new location (Algorithm \ref{alg:dEPR}, line 5), otherwise she returns to a previously visited location (Algorithm \ref{alg:dEPR}, line 10). If the individual explores and is in location $i$, the new location $j \neq i$ is selected according to the probability $p_{ij} \in P$ (Algorithm \ref{alg:dEPR}, function \texttt{PreferentialExploration}). If the individual returns to a previously visited location, it is chosen with probability proportional to the number of her previous visits to that location (Algorithm \ref{alg:dEPR}, function \texttt{preferentialReturn}). The $d$-EPR model hence returns the chosen location $j$.

It is worth highlighting the difference between typical locations and preferred locations. Typical locations indicate places where individuals repeatedly return as part of their mobility routine. Examples of typical locations are home and work locations, where individuals regularly return in their everyday routine. Besides typical locations, individuals can also return to preferred locations, i.e., places which are not part of a schematic routine but where people return occasionally, such as cinemas or restaurants. The preferential return mechanism of $d$-EPR models the existence of such preferred locations, allowing the agents to return to previously visited locations with a probability depending of the past visitation frequency.

\begin{algorithm}[htb!]
\nonl {\large \textbf{The $d$-EPR model}}\\
 \BlankLine
 \BlankLine
\small
\DontPrintSemicolon
\SetKwInOut{Input}{input}
    \SetKwInOut{Output}{output}
    \Input{
    $S= \langle (x_1, y_1, t_1), \dots, (x_n, y_n, t_n) \rangle$, the current sample mobility trajectory of the synthetic individual\\
    $P$, the gravity-probability matrix
    }
    \Output{$j$, the next location to visit}
 \SetKwFunction{getReturnProbability}{getReturnProbability}
 \SetKwFunction{explore}{PreferentialExploration}
 \SetKwFunction{return}{PreferentialReturn}
 \SetKwFunction{append}{append}
 \SetKwFunction{update}{update}
  \SetKwFunction{getRandomLocation}{getRandomLocation}
 \SetKwFunction{distance}{distance}
 \SetKwFunction{weightedRandom}{weightedRandom}
 \SetKwFunction{density}{density}
  \SetKwFunction{updateMatrix}{updateMatrix}
 \BlankLine
 $\rho = 0.6$, $\gamma = 0.21$ \tcp{distributions' constants \citep{pappalardo2015,Pappalardo2016934,SongNaturePhysics2010}}
 \BlankLine
 \BlankLine
 $N=|set(S)|$ \tcp{number of distinct visited locations}
 $i = last(S)$\tcp{the current location of the synthetic individual}
 
 \BlankLine
 \BlankLine
 \BlankLine
 
 $p_{new} = \getReturnProbability()$ \tcp{generate a probability to return or explore}
 \If{$p_{new} \le \rho N^{-\gamma}$}
 {
 	$j = \explore(i, P)$ \tcp{explore a new location}
	\nl \KwRet $j$\;
 }
 \Else{
	$j = \return(S)$ \tcp{return to a previously visited location}
	\nl \KwRet $j$\;
 }
 
 \BlankLine
 \BlankLine
 \BlankLine
 
\setcounter{AlgoLine}{0}
\SetKwProg{myalg}{Function}{}{}
  \myalg{\explore{i}}{
$j = \weightedRandom(P[i])$ \tcp{choose $j$ according to prob.s in $P[i]$}
  \nl \KwRet $j$\;}{}
  
\BlankLine
\BlankLine

\setcounter{AlgoLine}{0}
\SetKwProg{myalg}{Function}{}{}
  \myalg{\return{S}}{
  $j = \weightedRandom(S)$ \tcp{choose $j$ according to visitation frequency of locations in $S$}
  \nl \KwRet $j$\;}{}
  
\caption{The psuedo-code of the $d$-EPR trajectory generator. The function \texttt{weightedRandom} randomly chooses an element in a vector according to its probability.}
\label{alg:dEPR}
\end{algorithm}


\section{Results}
\label{sec:results}

In this section we show the results of simulation experiments where we instantiate {\scshape Ditras} by using $d$-EPR as trajectory generator and MD$^{(t)}$ as diary generator. We construct MD$^{(t)}$ from nation-wide mobile phone data covering a period of three month using MDL.
We refer to the spatio-temporal model as $d$-EPR$_{\mbox{\small MD}}^{\mbox{\tiny (CDR)}}$ and use it to generate sampled mobility trajectories of 10,000 agents. We compare the resulting sampled mobility trajectories with:
 
\begin{itemize} 
\item the trajectories of 10,000 mobile phone users whose mobility is tracked during 3 months in a European country; \\
\item the sampled mobility trajectories produced by other 8 spatio-temporal mobility models created through the {\scshape Ditras} framework by combining different diary and trajectory generators, whose parameters are fitted on the mobile phone data. 
\end{itemize}

Similarly we instantiate {\scshape Ditras} by using $d$-EPR and MD$^{(t)}$ constructed on GPS vehicular tracks covering a period of one month. We refer to the spatio-temporal model as $d$-EPR$_{\mbox{\small MD}}^{\mbox{\tiny (GPS)}}$. We use this model to generate sample mobility trajectories of 10,000 agents and compare the resulting sample mobility trajectories with:

\begin{itemize}
\item the trajectories of 10,000 private vehicles whose mobility is tracked through on-board GPS devices during 4 weeks in Tuscany; \\
\item the sampled mobility trajectories produced by other 8 spatio-temporal mobility models created through the {\scshape Ditras} framework by combining different diary and trajectory generators, whose parameters are fitted on the GPS vehicular data.
\end{itemize}

In Section \ref{sec:data} and in Section \ref{sec:gps} we describe respectively the mobile phone data and the GPS vehicular data we use in our experiments to describe the mobility of real individuals and the pre-processing operations we carry out on the data. 
In Section \ref{sec:comparison} we provide a comparison on a set of spatio-temporal mobility patterns of $d$-EPR$_{\mbox{\small MD}}^{\mbox{\tiny(CDR)}}$'s trajectories, mobile phone data's trajectories, and the trajectories produced by the other models. These simulations are performed by using a weighted spatial tessellation induced by the mobile phone towers. Analogously, we provide a comparison on a set of spatio-temporal mobility patterns of $d$-EPR$_{\mbox{\small MD}}^{\mbox{\tiny(GPS)}}$'s trajectories, GPS data's trajectories, and the trajectories produced by the other models. These simulations are performed by using a weighted spatial tessellation induced by the census cells in Tuscany. All the simulations are performed using a time slot duration $t = 3600$s $= 1$h.

\subsection{CDR data}
\label{sec:data}
We have access to a set of Call Detail Records (CDRs) gathered by a European carrier for billing and operational purposes. The dataset records all the calls made during 11 weeks by $\approx$1 million anonymized mobile phone users. CDRs collect geographical, temporal and interaction information on mobile phone use and show an enormous potential to empirically investigate the structure and dynamics of human mobility on a society wide scale \citep{reades2007,Hidalgo2008,gonzalez08,gonzalez_clustering,calabrese2011,pappalardo2015,pappalardo2015bigdata}. Each time an individual makes a call the mobile phone operator registers the connection between the caller and the callee, the duration of the call and the coordinates of the phone tower communicating with the phone, allowing to reconstruct the user's approximate position. Table \ref{tab:CDR_data} illustrates an example of the structure of CDRs.
\begin{table}[H]\centering
\def\arraystretch{1.0}
\subfigure[]{
\normalsize
\begin{tabular}{| c | c | c | c |}
\hline
\textbf{timestamp} & \textbf{tower} & \textbf{caller} & \textbf{callee} \\
\hline
2007/09/10 23:34 & 36 & 4F80460 & 4F80331\\
2007/10/10 01:12 & 36 & 2B01359 & 9H80125\\
2007/10/10 01:43 & 38 & 2B19935 & 6W1199\\
\vdots & \vdots & \vdots & \vdots \\
\hline
\end{tabular}
}
\subfigure[]{
\normalsize
\begin{tabular}{| r | r | r |}
\hline
\textbf{tower} & \textbf{latitude} & \textbf{longitude}\\
\hline
36 & 49.54 & 3.64\\
37 & 48.28 & 1.258\\
38 & 48.22 & -1.52\\
\vdots & \vdots & \vdots \\
\hline
\end{tabular}
}
\caption{Example of Call Detail Records (CDRs). Every time a user makes a call, a record is created with timestamp, the phone tower serving the call, the caller identifier and the callee identifier (a). For each tower, the latitude and longitude coordinates are available to map the tower on the territory (b).}
\label{tab:CDR_data}
\end{table}

CDRs have been extensively used in literature to study different aspects of human mobility, due to several advantages: they provide a means of sampling user locations at large population scales; they can be retrieved for different countries and geographic scales given their worldwide diffusion; they provide an objective concept of location, i.e., the phone tower. Nevertheless, CDR data suffer different types of bias \citep{Ranjan2012,iovan2013}, such as: (i) the position of an individual is known at the granularity level of phone towers; (ii) the position of an individual is known only when she makes a phone call; (iii) phone calls are sparse in time, i.e., the time between consecutive calls follows a heavy tail distribution \citep{gonzalez08,barabasi05}. In other words, since individuals are inactive most of their time, CDRs allow to reconstruct only a subset of an individual's mobility. Several works in literature study the bias in CDRs by comparing the mobility patterns observed on CDRs to the same patterns observed on GPS data \citep{pappalardo2013,pappalardo2015,pappalardo2012_sebd,pappalardo_comparing} or handover data (data capturing the location of mobile phone users recorded every hour or so) \citep{gonzalez08}. The studies agree that the bias in CDRs does not affect significantly the study of human mobility patterns.

\emph{Data preprocessing.} In order to cope with sparsity in time of CDRs and focus on individuals with reliable call statistics, we carry out some preprocessing steps. Firstly, for each individual $u$ we discard all the locations with a visitation frequency $f = n_i/N \le 0.005$, where $n_i$ is the number of calls performed by $u$ in location $i$ and $N$ the total number of calls performed by $u$ during the period of observation \citep{Schneider20130246,pappalardo2015}. This condition checks whether the location is relevant with respect to the specific call volume of the individual. Since it is meaningless to analyze the mobility of individuals who do not move, all the individuals with only one location after the previous filter are discarded. 
We select only active individuals with a call frequency threshold of $f = N/(h*d) \ge 0.5$ calls per hour, where $N$ is the total number of calls made by $u$, $h=24$ is the hours in a day and $d=77$ the days in our period of observation. Starting from $\approx$1 millions users, the filtering results in $50,000$ active mobile phone users. 

\emph{Weighted Spatial Tessellation.} The weighted spatial tessellation $L$ we use in the experiments is defined by the mobile phone towers in the CDR data. The relevance of a phone tower is estimated as the total number of calls served by that tower by the 50,000 active mobile phone users during the 3 months. Every location's position on the space is identified by the latitude and longitude coordinates of a phone tower.

\subsection{GPS data}
\label{sec:gps}

The GPS dataset stores information of approximately 9.8 Million different trips from
159,000 private vehicles tracked during one month (May 2011) which passed through Tuscany (central Italy). The GPS traces are provided by Octo Telematics Italia Srl,\footnote{\url{http:
//www.octotelematics.com/}} a company that provides a data collection service for insurance
companies. The GPS device is embedded in the private vehicles' engine and automatically
turns on when the vehicle starts. The sequence of GPS points that the device transmits
every 30 seconds to the server via a GPRS connection forms the global trajectory of a vehicle.
When the vehicle stops no points are logged nor sent. 

We exploit these stops to split
the global trajectory into several sub-trajectories, corresponding to the trips performed
by the vehicle. Clearly, the vehicle may have stops of different duration, corresponding
to different activities. To ignore small stops like gas stations, traffic lights, bring and get activities and so on, we choose a stop duration threshold of at least 20 minutes: if the time
interval between two consecutive observations of the vehicle is larger than 20 minutes, the
first observation is considered as the end of a trip and the second observation is considered
as the start of another trip. We also performed the extraction of the trips by using different
stop duration thresholds (5, 10, 15, 20, 30, 40 minutes), without finding significant differences
in the sample of short trips and in the statistical analysis we present in the paper. Since GPS data do not provide explicit information about visited locations, we assign
each origin and destination point of the obtained sub-trajectories to the corresponding census
cell, according to the information provided by the Italian National Institute of Statistics
(ISTAT).\footnote{\url{www.istat.it}} We hence obtain a data format similar to CDR data, where we describe the movements of a vehicle by the time-ordered list of census cells where the vehicle stopped. We filter the data by discarding all the vehicles with only one visited location or with less than one trip per day on average during the period of observation. This filtering results in a dataset of 46,121 vehicles.

\emph{Weighted Spatial Tessellation.}
The weighted spatial tessellation $L$ we use in the experiments is defined by the census cells in Tuscany. The relevance of a location is estimated as the total number of stops in the corresponding cell by the 159,000 private vehicles during the month of observation. Every location's position on the space is identified by the latitude and longitude coordinates of the census cell.

\subsection{Models comparison and validation}
\label{sec:comparison}
We use the {\scshape Ditras} framework to build 18 models (9 models fitted on CDRs and 9 models fitted on GPS data) which use different combinations for the diary generator and the trajectory generator. In particular, we consider three diary generators -- MD, RD and WT -- and three trajectory generators -- $d$-EPR, SWIM and LATP. For every model we simulate the mobility of 10,000 agents for a period of $N = 1,848$ hours (3 months) and $N = 744$ hours (1 month) for models fitted on CDRs and GPS data respectively. Table \ref{tab:models} and Table \ref{tab:models_GPS} show the ability of every model in reproducing a set of characteristic statistical distributions derived from the CDR and the GPS data respectively, quantified by two measures: \emph{(i)} the Root Mean Square Error, $\mbox{RMSE}(\vec{y}, \hat{\vec{y}}) = \sqrt{ {\sum_{i=1}^n (\hat{y}_i - y_i)^2 \over n} }$ where $\hat{y}_i \in \hat{\vec{y}}$ indicates a point of the synthetic distribution $\hat{\vec{y}}$, $y_i \in \vec{y}$ the corresponding point in the empirical distribution $\vec{y}$ and $n$ the number of observations; \emph{(ii)} the Kullback-Leibler divergence, $\mbox{KL}(\vec{y}||\hat{\vec{y}}) = H(\vec{y}, \hat{\vec{y}}) - H(\vec{y})$, where $H(\vec{y}, \hat{\vec{y}})$ is the cross entropy between the real distribution and the empirical distribution and $H(\vec{y})$ is the entropy of the real distribution.
Here we use the notation TG$_{\mbox{\small DG}}$ to specify that trajectory generator TG is used in combination with diary generator DG. For example, $d$-EPR$_{\mbox{\small MD}}$ indicates the model using diary generator MD in combination with trajectory generator $d$-EPR. Notation TG$_{\{\mbox{\small DG}_1, \dots, \mbox{\small DG}_k\}}$ indicates the set of models \{TG$_{\mbox{\small DG}_1}$, \dots, TG$_{\mbox{\small DG}_k}$\}. Similarly, notation \{TG$^1$, \dots, TG$^k$\}$_{\mbox{\small DG}}$ indicates the set of models \{TG$^1_{\mbox{\small DG}}$, \dots, TG$^k_{\mbox{\small DG}}$\}.
 
\newcommand{\specialcell}[2][c]{\begin{tabular}[#1]{@{}c@{}}#2\end{tabular}}
\begin{table}[H]\centering
\scriptsize
\begin{tabular}{|c | c | l  l  l  l  l  l  l  l  l |}
\cline{3-11}
\multicolumn{2}{c|}{\large \textbf{CDR}}  & \large \specialcell{$\Delta r$} &\large \specialcell{$r_g$}  & \large \specialcell{$S^{unc}$}  & \large \specialcell{$T$} & \large \specialcell{$D$} & \large \specialcell{$\Delta t$} & \large \specialcell{$V$} & \large \specialcell{$N$} & \large \specialcell{$f(L)$}\\
\hline
\multirow{6}{*}{\rotatebox[origin=c]{90}{\color{red}\textbf{MD}}} & \multirow{2}{*}{\color{red}$d$-EPR} & \cellcolor{blue!25}.0001 & .0026 & \cellcolor{blue!25}.9643 & \cellcolor{blue!25}.0061 & \cellcolor{blue!25}.0659 & \cellcolor{blue!25}.0014 & $2.6e^{-5}$ & \cellcolor{blue!25}.0218 & \cellcolor{blue!25}.0122\\
 &  & \cellcolor{blue!25}.0006 & .0247 & \cellcolor{blue!25}29.34 & \cellcolor{blue!25}.0101 & \cellcolor{blue!25}.0682 & \cellcolor{blue!25}.1915 & .0016 & \cellcolor{blue!25}.5449 & \cellcolor{blue!25}.1200\\
\cline{2-11}

& \multirow{2}{*}{SWIM} & .0005 & \multicolumn{1}{c}{\multirow{2}{*}{-}} & 3.6069 & .0062 & .0683 & .0029 & $5.6e^{-5}$ & \multicolumn{1}{c}{\multirow{2}{*}{-}} & .0669\\
 &  & .0067 &  & 60.97 & .0101 & .0808 & .4996 & .0451 &  & 1.2892\\
\cline{2-11}

& \multirow{2}{*}{LATP} & .0001 & .0061 & 3.2236 & .0062 & .0684 & .0027 & $6.3e^{-5}$ & \multicolumn{1}{c}{\multirow{2}{*}{-}} & .0625\\
 &  & .0008 & .3223 & 258.46 & .0101 & .0802 & .3282 & .0600 &  & .9353\\
\hline
\hline

\multirow{6}{*}{\rotatebox[origin=c]{90}{\textbf{RD}}} & \multirow{2}{*}{$d$-EPR} & .0004 & .0027 & 1.1745 & .0232 & .2098 & .0024 & $4.1e^{-5}$ & .0235 & .0521\\
 &  & .0029 & .0161 & 20.8015 & .197 & 4.3558 & .2048 & .0191 & 1.1773 & .3876\\
\cline{2-11}

& \multirow{2}{*}{SWIM} & .0041 & \multicolumn{1}{c}{\multirow{2}{*}{-}} & \multicolumn{1}{c}{\multirow{2}{*}{-}} & .0232 & \multicolumn{1}{c}{\multirow{2}{*}{-}} & .0033 & $7.2e^{-5}$ & \multicolumn{1}{c}{\multirow{2}{*}{-}} & .0947\\
 &  & .1501 & & & .1974 & & .3773 & .0460 & & 4.4057\\
\cline{2-11}

& \multirow{2}{*}{LATP} & .0002 & \multicolumn{1}{c}{\multirow{2}{*}{-}} & \multicolumn{1}{c}{\multirow{2}{*}{-}} & .0232 & \multicolumn{1}{c}{\multirow{2}{*}{-}} & .0033 & $4.6e^{-5}$ & \multicolumn{1}{c}{\multirow{2}{*}{-}} & .0874\\
 &  & .0014 &  & & .1974 & & .6967 & .0321 & & 2.2051\\
\hline
\hline

\multirow{6}{*}{\rotatebox[origin=c]{90}{\textbf{WT}}} & \multirow{2}{*}{$d$-EPR} & .0003 & \cellcolor{blue!25}.0024 & 1.1666 & .0232 & .1790 & .0023 & $4.0e^{-5}$ & .0224 & .0502\\
 &  & .0019 & \cellcolor{blue!25}.0130 & 20.00 & .1970 & 3.9769 & .1946 & .0189 & 1.0395 & .3537\\
\cline{2-11}

& \multirow{2}{*}{SWIM} & .0033 & \multicolumn{1}{c}{\multirow{2}{*}{-}} & \multicolumn{1}{c}{\multirow{2}{*}{-}} & .0232 & .2036 & .0033 & \cellcolor{blue!25}$1.9e^{-5}$ & \multicolumn{1}{c}{\multirow{2}{*}{-}} & .0943\\
 &  & .0601 &  &  & .1975 & 4.3806 & .1146 & \cellcolor{blue!25}.0070 & & 3.9605\\
\cline{2-11}

& \multirow{2}{*}{LATP} & .0001 & \multicolumn{1}{c}{\multirow{2}{*}{-}} & \multicolumn{1}{c}{\multirow{2}{*}{-}} & .0232 & .2037 & .0033 & $7.2e^{-5}$ & \multicolumn{1}{c}{\multirow{2}{*}{-}} & .0866\\
 &  & .0010 &  & & .1975 & 4.5672 & .6322 & .0309 & & 2.1015\\
\hline
\hline
\multicolumn{2}{c|}{\multirow{2}{*}{best model}} &\tiny $d$-EPR & \tiny$d$-EPR &\tiny $d$-EPR & \tiny$d$-EPR & \tiny$d$-EPR & \tiny$d$-EPR & \tiny SWIM & \tiny$d$-EPR & \tiny$d$-EPR\\
 \multicolumn{1}{c}{} & & \multicolumn{1}{c }{MD} & \multicolumn{1}{c}{WT} & \multicolumn{1}{c}{MD} & \multicolumn{1}{c}{MD} & \multicolumn{1}{c}{MD} & \multicolumn{1}{c}{MD} & \multicolumn{1}{c}{WT} & \multicolumn{1}{c}{MD} & \multicolumn{1}{c|}{MD}\\
 \hline
\end{tabular}

\caption{\textbf{Error of fit between CDR data and synthetic data.} Every row $i$ is a model and every column $j$ a mobility measure. A cell $(i, j)$ indicates the RMSE (first row) and the KL divergence (second row) of a synthetic distribution w.r.t.\ the real distribution. The best RMSE values are in blue. Symbol - indicates that the synthetic distribution is not comparable with the real distribution. We highlight in blue the cells with the best values of RMSE and KL divergence. We color in red the combination of temporal and spatial model leading to the highest number of blue cells.}
\label{tab:models}
\end{table}

\paragraph{Diary generators.}
In the Random Diary (RD) generator a synthetic individual is in perpetuum motion: in every time slot of the simulation she chooses a new location to visit. We use RD to highlight the difference between the diary generator we propose, MD (Section~\ref{sec:markov_gen}), and the temporal patterns of a non-realistic diary generator.

In the Waiting Time (WT) diary generator a synthetic individual chooses a waiting time $\Delta t$ between a trip and the next one from the empirical distribution $P(\Delta t) \sim \Delta t^{-1 - \beta} \exp^{-\Delta t / \tau}$, with $\beta = 0.8$ and $\tau=17$ hours as measured on CDR data \citep{SongNaturePhysics2010}. WT is the temporal mechanism usually used in combination with mobility models like EPR \citep{SongNaturePhysics2010} and SWIM \citep{swim}. It reproduces in a realistic way the distribution of the time between two consecutive trips \citep{SongNaturePhysics2010,pappalardo2013} but does not model the circadian rhythm and the tendency of individuals to be in certain places and specific times. 

We construct two diary generators, MD$_{\mbox{\tiny (CDR)}}$ and MD$_{\mbox{\tiny (GPS)}}$, by applying algorithm MDL (Section \ref{sec:markov_gen}) on CDR data and GPS data respectively. These diary generators are based on Markov models and can reproduce the circadian rhythm of individuals and their tendency to follow or break the routine.

\paragraph{Trajectory generators.} The trajectory generator SWIM \citep{swim} is a modelling approach based on location preference. The model initially assigns to each synthetic individual a home location $L_h$ chosen randomly from the spatial tessellation. The synthetic individual then selects a destination for the next movements depending on the weight of each location \citep{swim}: 
\begin{equation}
w(L)_{\mbox{swim}} = \alpha * d(L_h, L) + (1 - \alpha) * r(L),\quad\alpha=0.75
\label{eq:swim}
\end{equation}
which grows with the relevance $r(L)$ of the location and decreases with the distance from the home \citep{swim}: $$d(L_h, L) = {1 \over (1 + distance(L_h, L))^2}.$$ SWIM tries to model both the preference for short trips and the preference for relevant locations, though it does not model the preferential return mechanism.

The trajectory generator LATP (Least Action Trip Planning) \citep{slaw1,slaw2} is a trip planning algorithm used as exploration mechanism in several mobility models, such as SLAW \citep{slaw1,slaw2}, SMOOTH \citep{smooth}, MSLAW \citep{mslaw} and TP \citep{tp1,tp2}. In LATP a synthetic individual selects the next location to visit according to a weight function \citep{slaw1,slaw2}:
\begin{equation}
w(L)_{\mbox{latp}} = {1\over distance(c, L)^{1.5}}.
\end{equation} 
LATP only models the preference for short distances and does not consider the relevance of a location nor model the preferential return mechanism.

We compare the synthetic mobility trajectories of the nine models with CDR trajectories and GPS trajectories on the distributions of several measures capturing salient characteristics of human mobility. Table \ref{tab:models} and Table \ref{tab:models_GPS} display the mobility measures we consider, which are: trip distance $\Delta r$ \citep{gonzalez08,pappalardo2013}, radius of gyration $r_g$ \citep{gonzalez08,pappalardo2013,pappalardo2015}, mobility entropy $S^{unc}$ \citep{song2010limits,pentland,pappalardo2016}, location frequency $f(L)$ \citep{SongNaturePhysics2010,hasan2013,pappalardo2013}, visits per location $V$ \citep{Pappalardo2016934}, locations per user $N$ \citep{Pappalardo2016934}, trips per hour $T$ \citep{gonzalez08,pappalardo2013}, time of stays $\Delta t$ \citep{SongNaturePhysics2010,hasan2013} and trips per day $D$. 

\begin{table}[H]\centering
\scriptsize
\begin{tabular}{|c | c | l  l  l  l  l  l  l  l  l |}
\cline{3-11}
\multicolumn{2}{c|}{\large \textbf{GPS}}  & \large \specialcell{$\Delta r$} &\large \specialcell{$r_g$}  & \large \specialcell{$S^{unc}$}  & \large \specialcell{$T$} & \large \specialcell{$D$} & \large \specialcell{$\Delta t$} & \large \specialcell{$V$} & \large \specialcell{$N$} & \large \specialcell{$f(L)$}\\
\hline
\multirow{6}{*}{\rotatebox[origin=c]{90}{\color{red}\textbf{MD}}} & \multirow{2}{*}{\color{red}$d$-EPR} & .0254 &  \cellcolor{blue!25}.0148 &  \cellcolor{blue!25}1.9855 &  \cellcolor{blue!25}.0053 & .1334 & .0738 & .0123 &  \cellcolor{blue!25}.0113 & \cellcolor{blue!25}.0323\\
 &  & .5346 & \cellcolor{blue!25}.2850 & \cellcolor{blue!25}156.92 &  \cellcolor{blue!25}.0156  & .2992 & .7567 & .1415 & \cellcolor{blue!25}.0411 & \cellcolor{blue!25}.2429\\
\cline{2-11}

& \multirow{2}{*}{SWIM} & .0229 & \multicolumn{1}{c}{\multirow{2}{*}{-}} & 3.8403 & .0054 & .1232 & .0589 & .0123 & .0319 & .0358\\
 &  & .8970 &   & 210.87 & .0156 & .2634 & .7321 & .1522 & 1.6923 & .4914\\
\cline{2-11}

& \multirow{2}{*}{LATP} & .0258 & .0225 & 3.7636 & .0054 & .1233 & .0655 & .0178 & .0315 &  .0324\\
 &  & .5968 & .9508 & 151.35 & .0157 & .2636 & .7148 & .4639 & 1.9085 & .3811\\
\hline
\hline
\multirow{6}{*}{\rotatebox[origin=c]{90}{\textbf{RD}}} & \multirow{2}{*}{$d$-EPR} &  \cellcolor{blue!25}.0031 & .0237 &  \multicolumn{1}{c}{\multirow{2}{*}{-}} & .0231 & .0923 & .0349 &  \cellcolor{blue!25}.0042 &  .0271 & .0560\\
 &  &  \cellcolor{blue!25}.0420 & .9939 &  & .1906 & 1.2493 & .4221 & \cellcolor{blue!25}.0360 &  3.3216 & .5258\\
\cline{2-11}

& \multirow{2}{*}{SWIM} & .0274 &  \multicolumn{1}{c}{\multirow{2}{*}{-}} &  \multicolumn{1}{c}{\multirow{2}{*}{-}} & .0231 & \multicolumn{1}{c}{\multirow{2}{*}{-}} & .2647 & .0102 &  \multicolumn{1}{c}{\multirow{2}{*}{-}} & .0915\\
 &  & 1.6628 &  &  & .1912 &  & 1.4443 & .0919 &  & 3.6641\\
\cline{2-11}

& \multirow{2}{*}{LATP} & .0169 &  \multicolumn{1}{c}{\multirow{2}{*}{-}} &  \multicolumn{1}{c}{\multirow{2}{*}{-}} & .0231 & \multicolumn{1}{c}{\multirow{2}{*}{-}} & .1599 & .0168 &  \multicolumn{1}{c}{\multirow{2}{*}{-}} & .0899\\
 &  & .1381 &  &  & .1912 &  & 1.1524 & .3609 &  & 2.9663\\
\hline
\hline

\multirow{6}{*}{\rotatebox[origin=c]{90}{\textbf{WT}}} & \multirow{2}{*}{$d$-EPR} & .0069 & .0223 &  \multicolumn{1}{c}{\multirow{2}{*}{-}} & .0231 &  .0923 &  \cellcolor{blue!25}.0291 & .0045 & .0270 & .0530\\
 &  & .0518 & .8217 &   & .1906 & 1.0593 & \cellcolor{blue!25}.4369 & .0394 & 2.132 & .4623\\
\cline{2-11}

& \multirow{2}{*}{SWIM} & .0180 &  \multicolumn{1}{c}{\multirow{2}{*}{-}} &  \multicolumn{1}{c}{\multirow{2}{*}{-}} & .0231 & \cellcolor{blue!25}.0923 & .1608 & .0095 &  \multicolumn{1}{c}{\multirow{2}{*}{-}} & .0908\\
 &  & .7278 &  &  & .1912 & \cellcolor{blue!25}.9510 & 1.0941 & .0823 &  & 3.2346\\
\cline{2-11}

& \multirow{2}{*}{LATP} & .0190 &  \multicolumn{1}{c}{\multirow{2}{*}{-}} &  \multicolumn{1}{c}{\multirow{2}{*}{-}} & .0231 & .0923 & .1027 & .0166 &  \multicolumn{1}{c}{\multirow{2}{*}{-}} & .0890\\
 &  & .1840 &  &  & .1913 & 1.0398 & .9187 & .4282 & & 2.6838\\
\hline
\hline
\hline
\multicolumn{2}{c|}{\multirow{2}{*}{best model}} &\tiny $d$-EPR & \tiny$d$-EPR &\tiny $d$-EPR & \tiny$d$-EPR & \tiny SWIM & \tiny$d$-EPR & \tiny SWIM & \tiny$d$-EPR & \tiny $d$-EPR\\

 \multicolumn{1}{c}{} & & \multicolumn{1}{c}{RD} & \multicolumn{1}{c}{MD} & \multicolumn{1}{c}{MD} & \multicolumn{1}{c}{MD} & \multicolumn{1}{c}{WT} & \multicolumn{1}{c}{WT} & \multicolumn{1}{c}{WT} & \multicolumn{1}{c}{MD} & \multicolumn{1}{c|}{MD}\\
 \hline

\end{tabular}
\caption{\textbf{Error of fit between GPS data and synthetic data.} Every row $i$ is a model and every column $j$ a mobility measure. A cell $(i, j)$ indicates the RMSE (first row) and the KL divergence (second row) of a synthetic distribution w.r.t.\ the real distribution. The best RMSE values are in blue. Symbol - indicates that the synthetic distribution is not comparable with the real distribution. We highlight in blue the cells with the best values of RMSE and KL divergence. We color in red the combination of temporal and spatial model leading to the highest number of blue cells.}
\label{tab:models_GPS}
\end{table}

\begin{figure}[!htb]\centering
\textbf{\LARGE CDR}\par\medskip
\subfigure[]
   {\includegraphics[scale=0.2555]{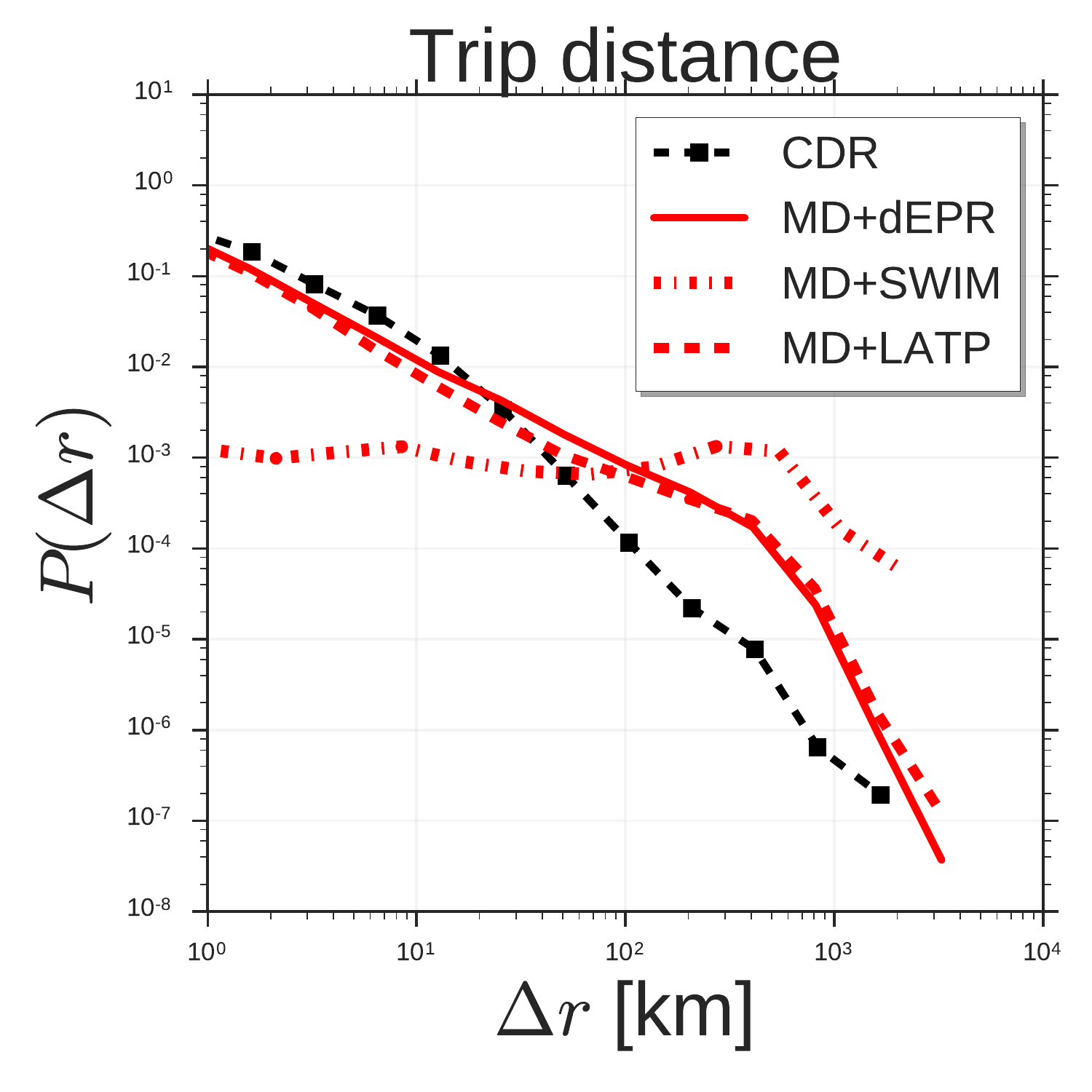}}
\subfigure[]
   {\includegraphics[scale=0.2555]{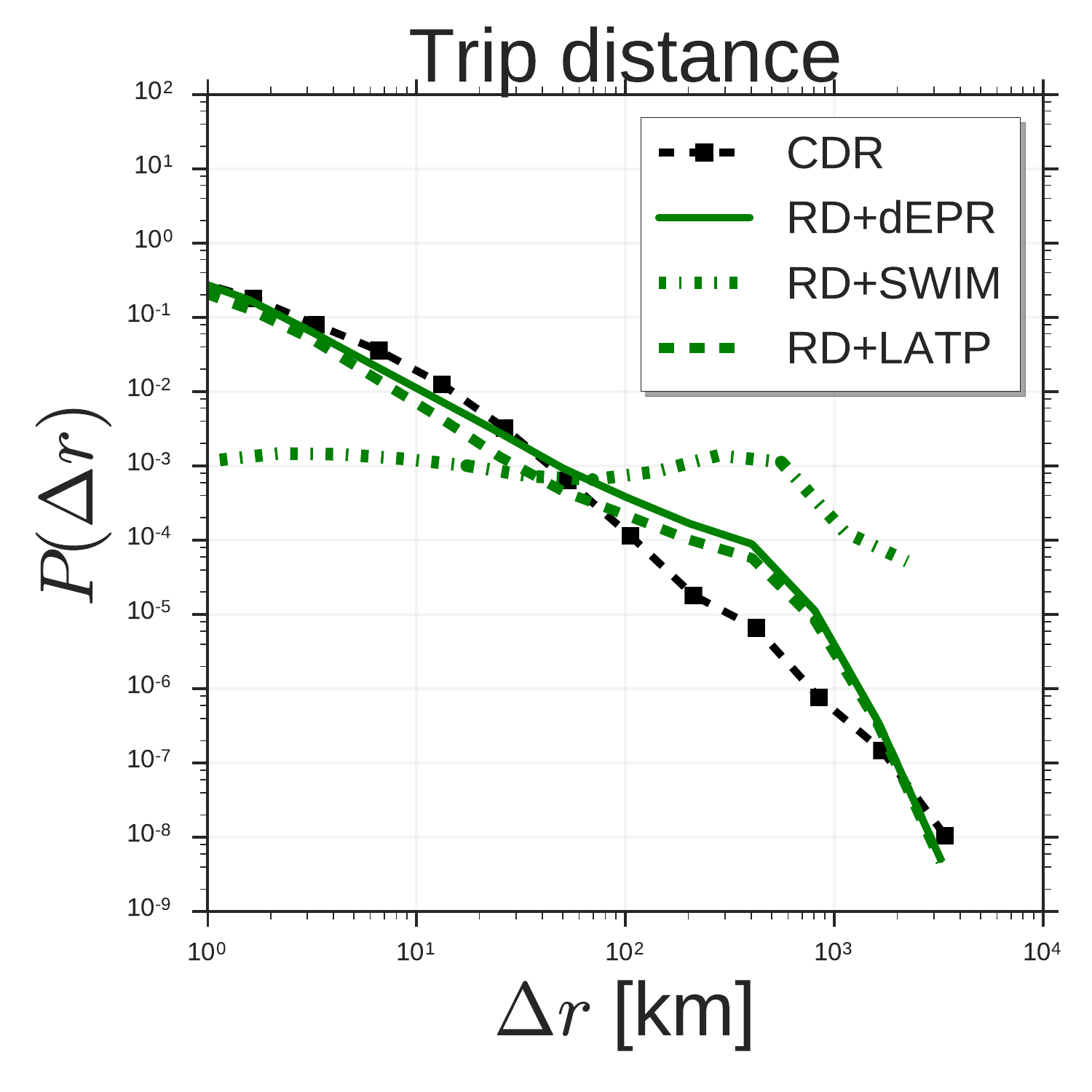}}
\subfigure[]
   {\includegraphics[scale=0.2555]{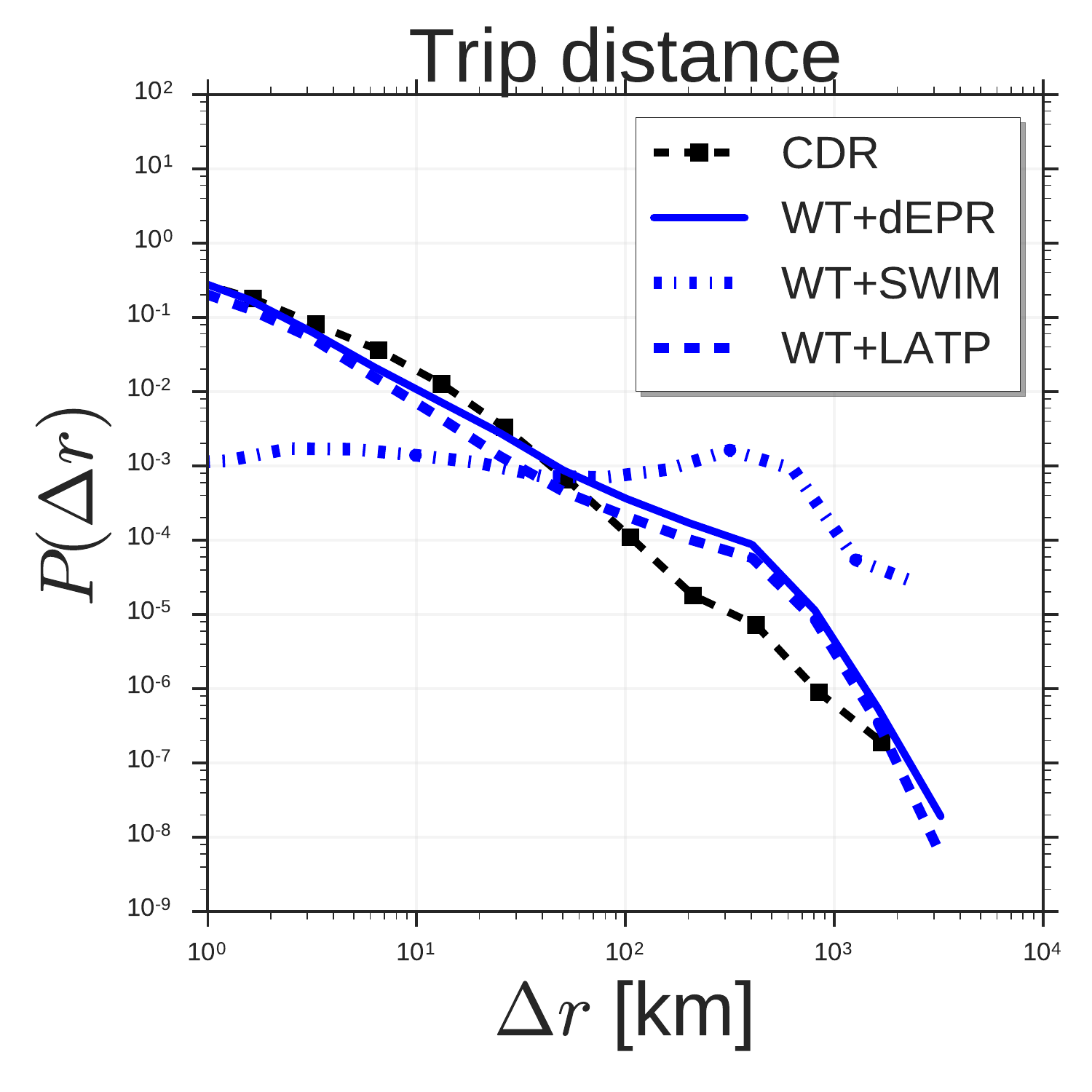}}
   
\subfigure[]
   {\includegraphics[scale=0.255]{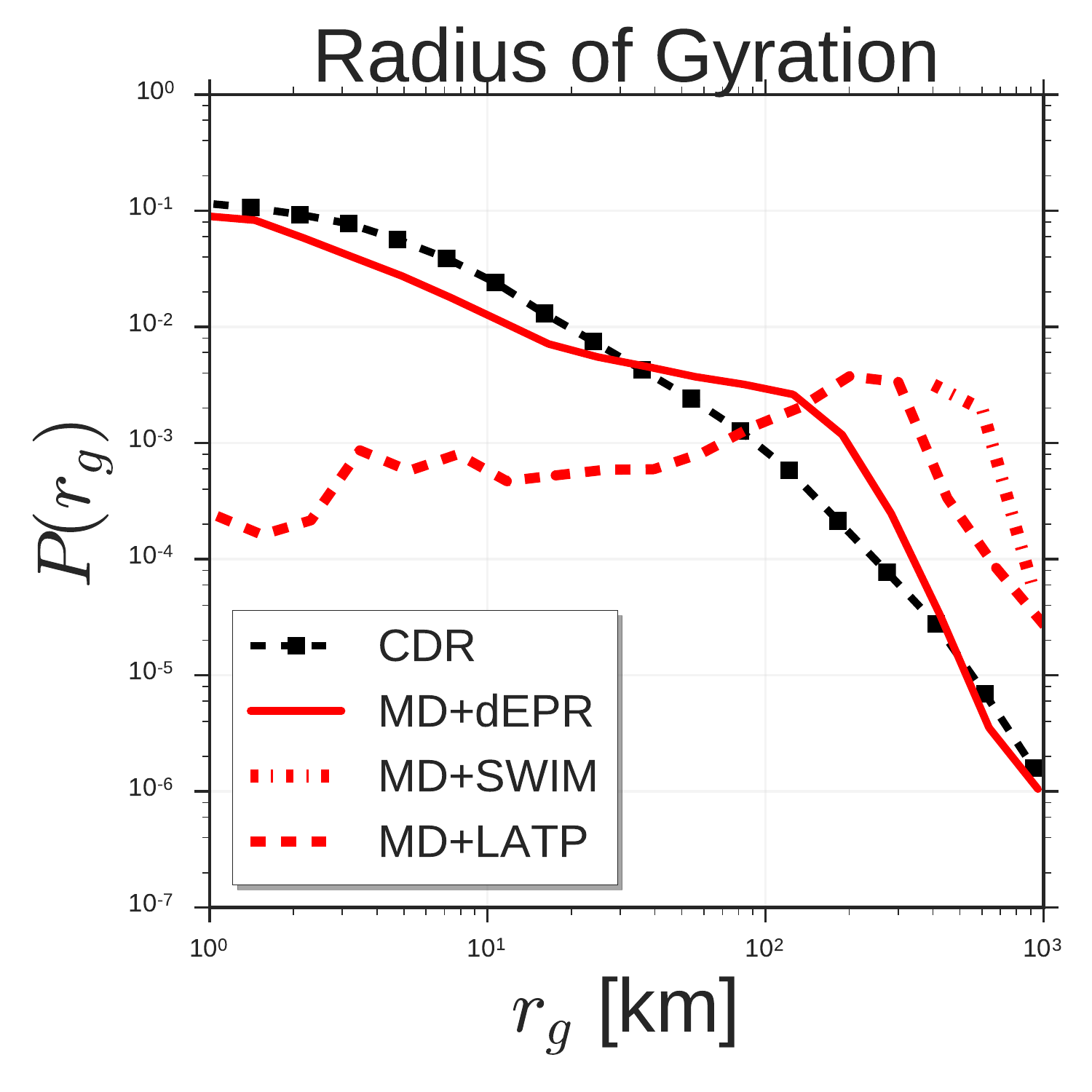}}
\subfigure[]
   {\includegraphics[scale=0.255]{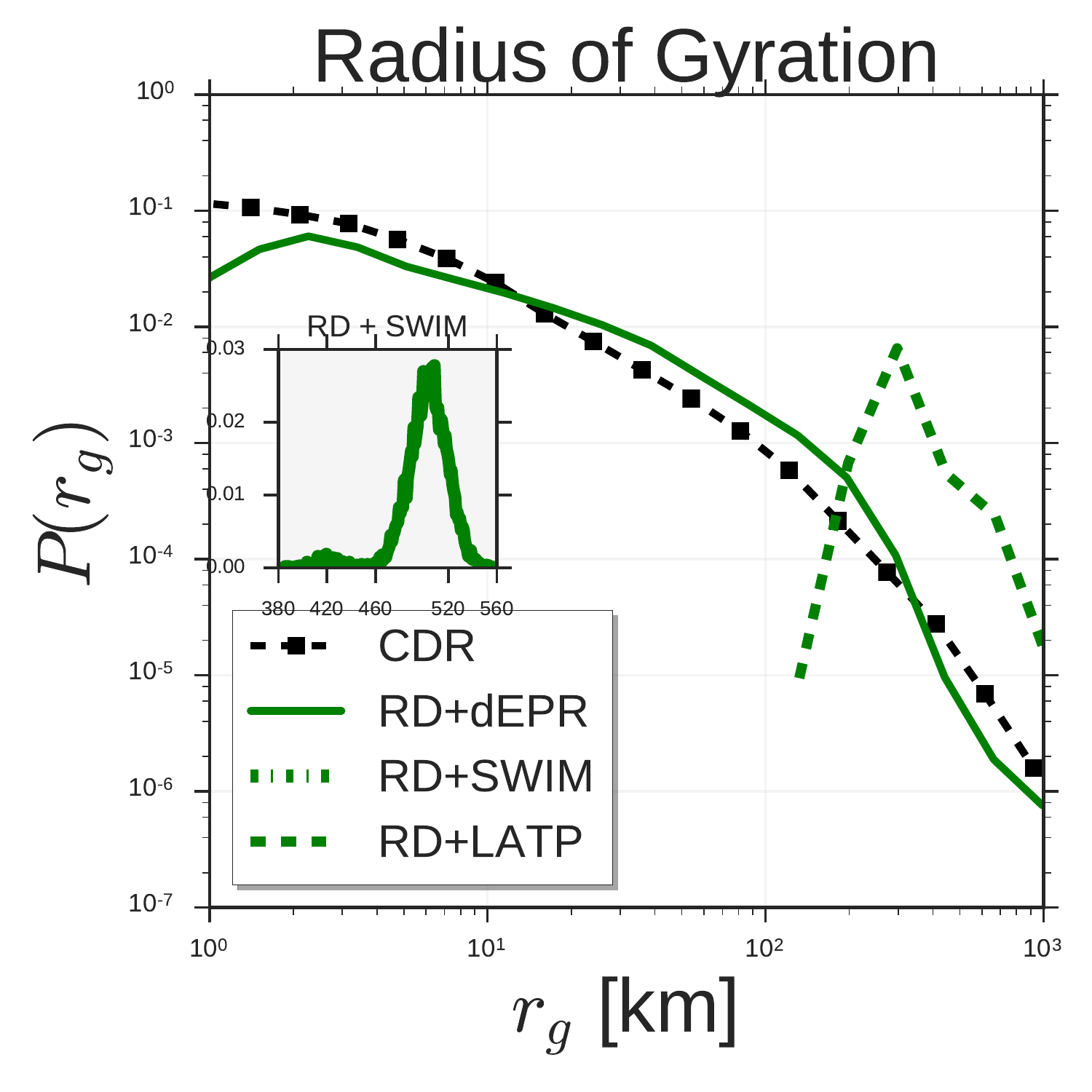}}
\subfigure[]
   {\includegraphics[scale=0.255]{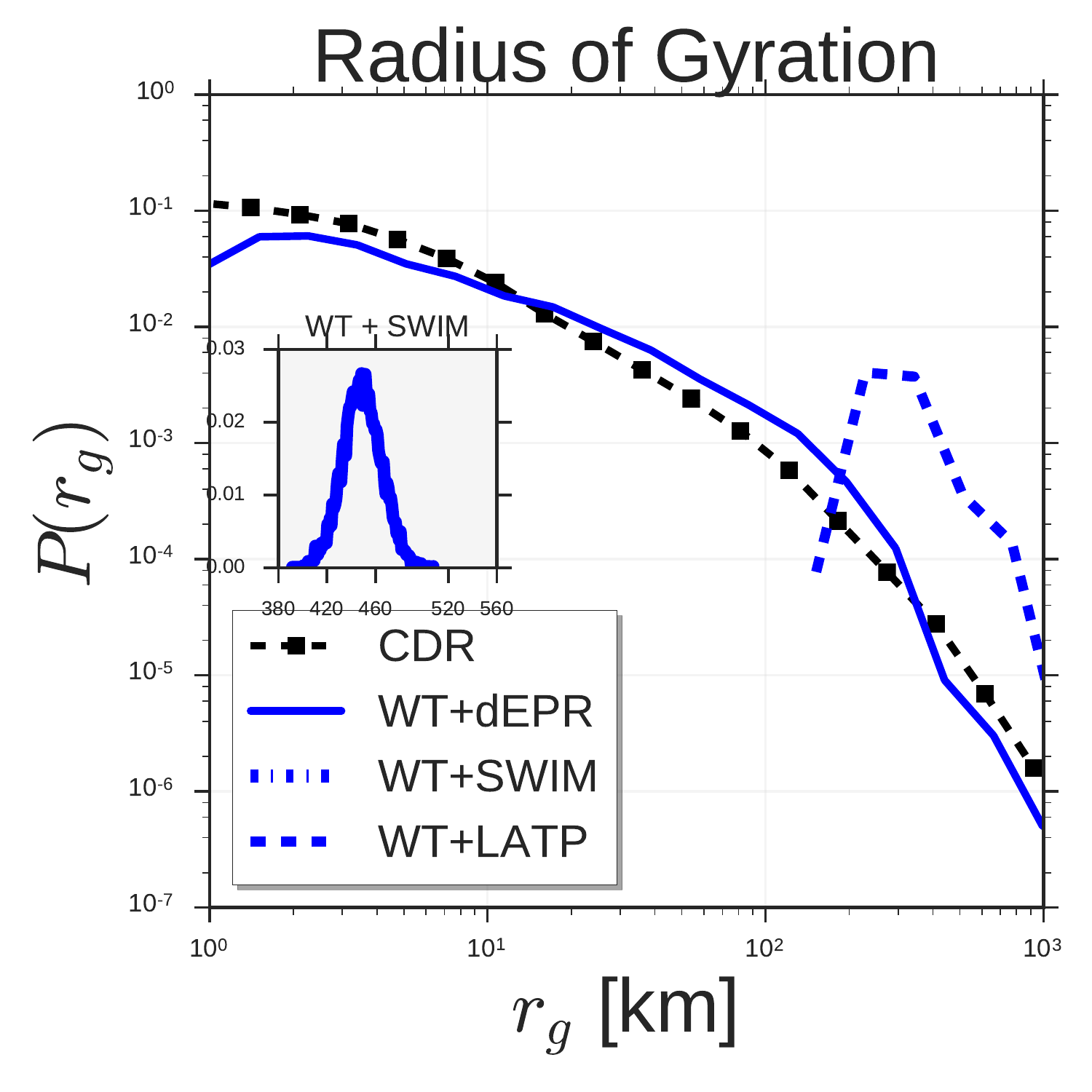}}

\subfigure[]
   {\includegraphics[scale=0.255]{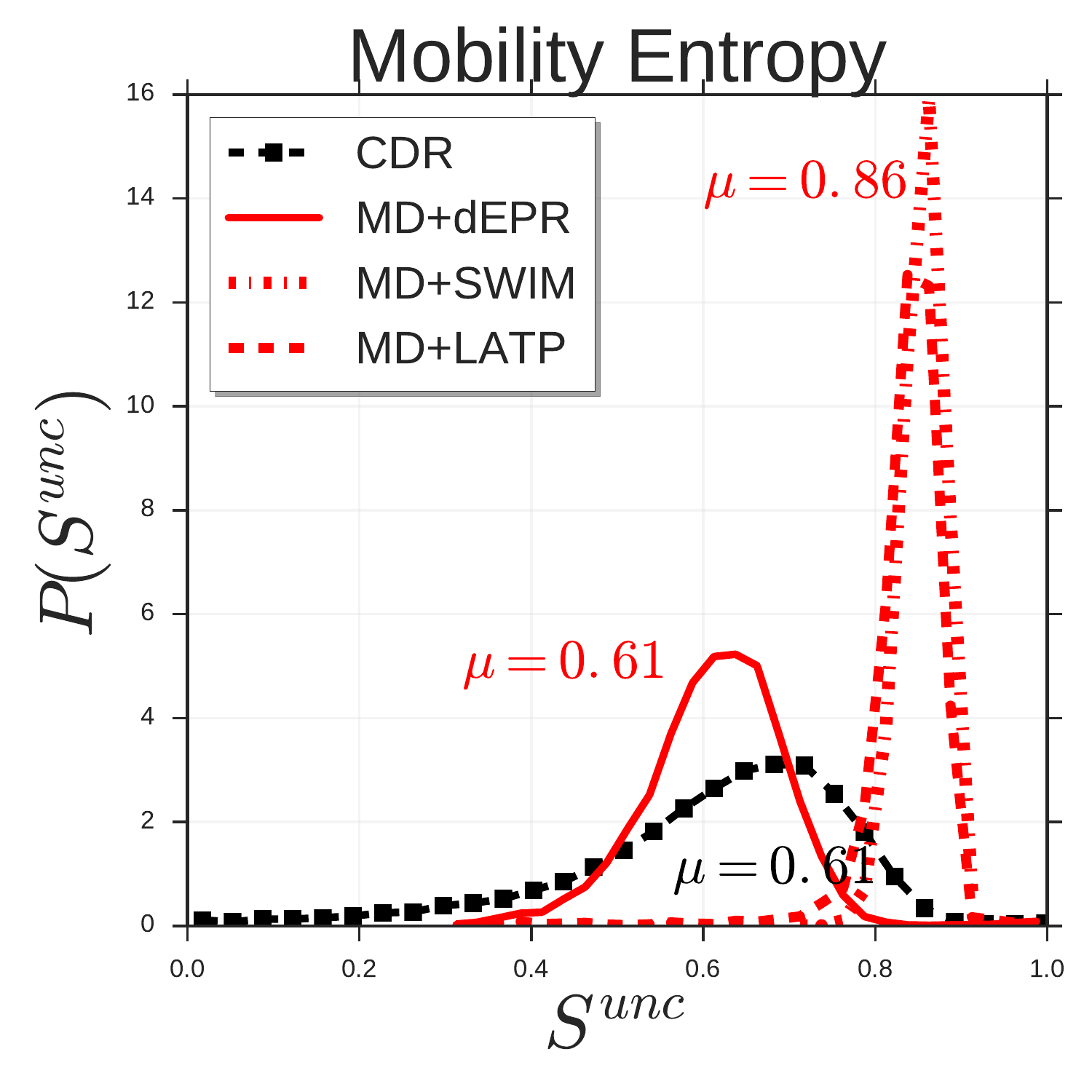}}
\subfigure[]
   {\includegraphics[scale=0.255]{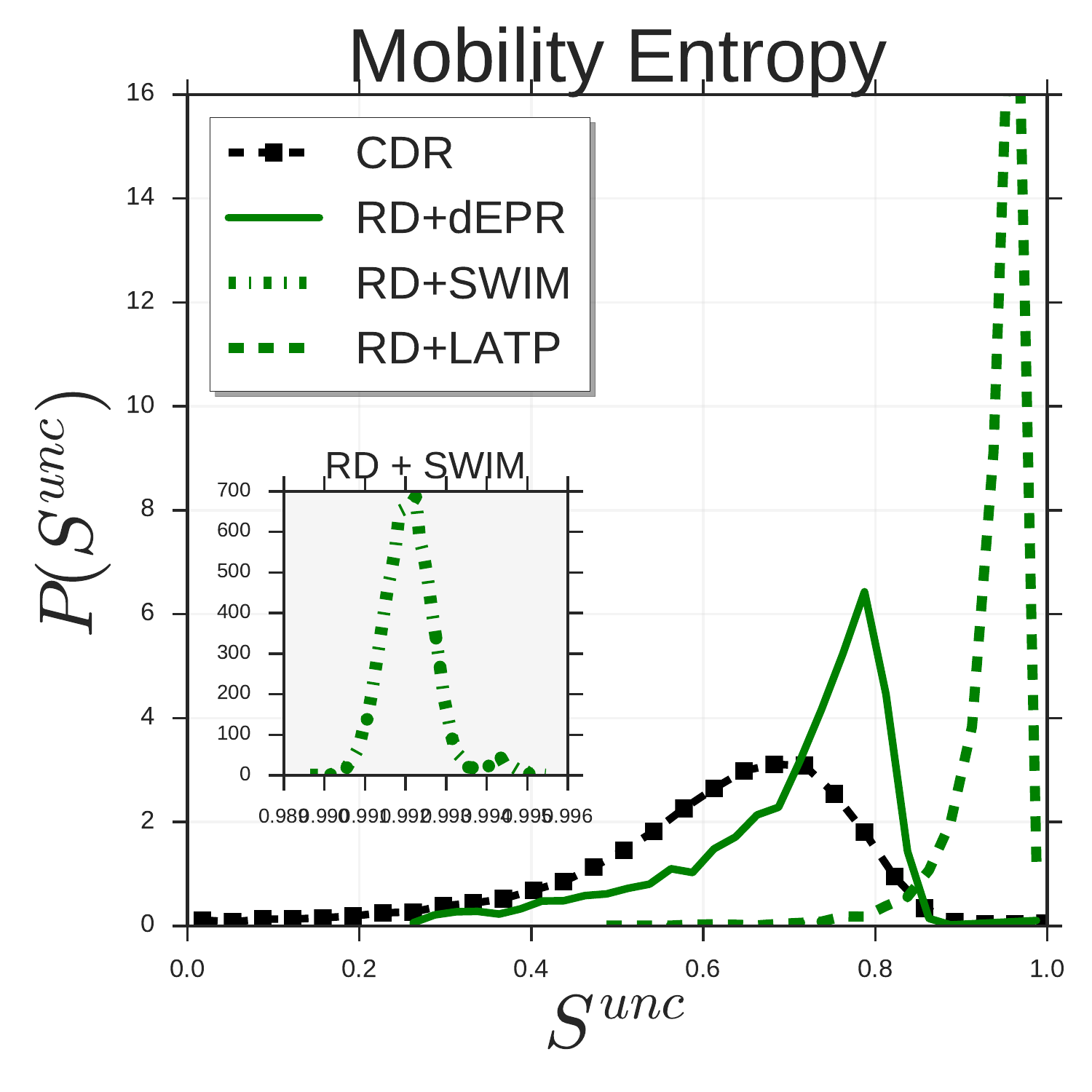}}
 \subfigure[]
   {\includegraphics[scale=0.255]{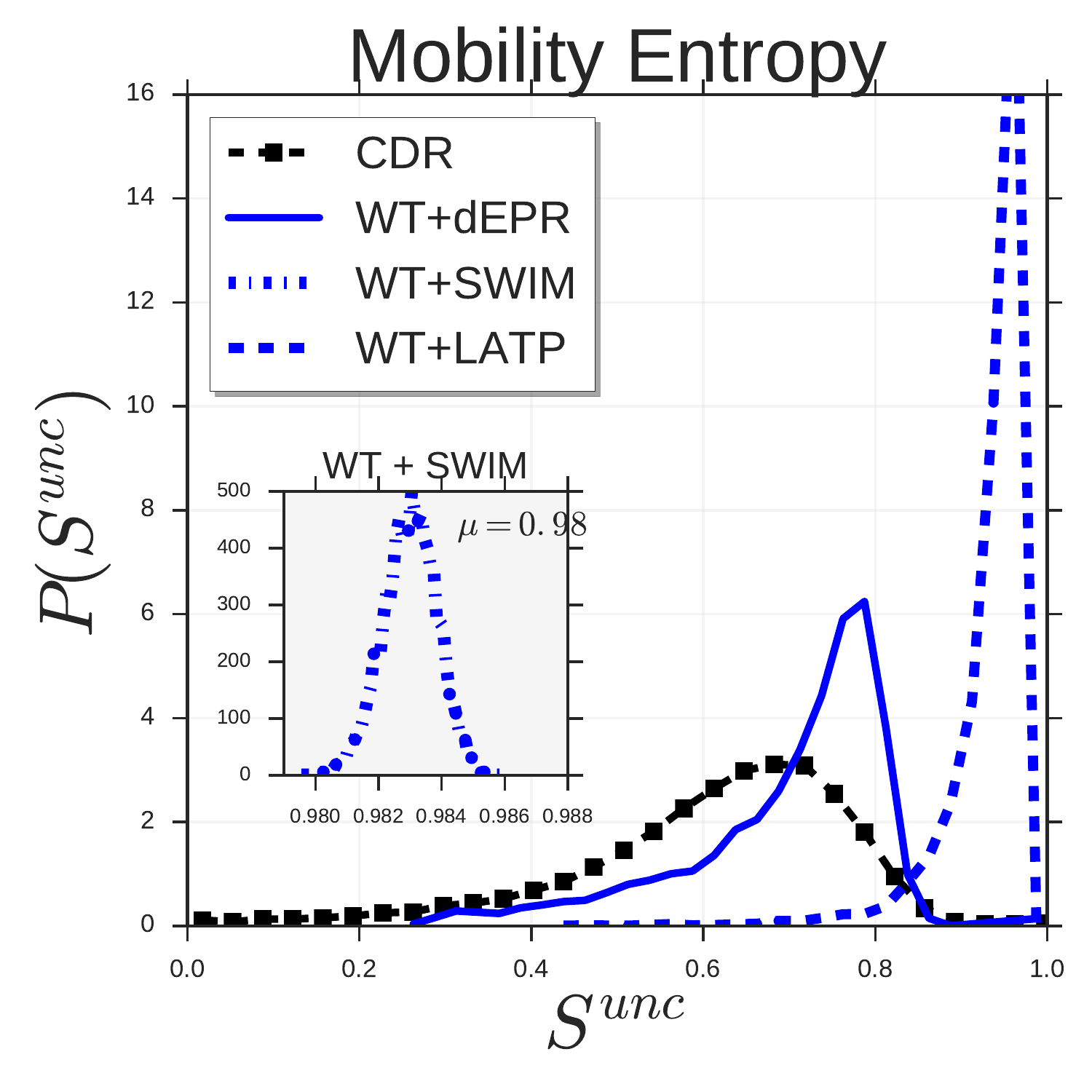}}

\caption{\textbf{Distributions of human mobility patterns (CDR).} The figure compares the models and CDR data on trip distance, radius of gyration and mobility entropy. Plots in (a), (b) and (c) show the distribution of trip distances $P(\Delta r)$ for real data (black squares) and data produced by three trajectory generators ($d$-EPR, SWIM and LATP) in combination with the MD generator (a), the RD generator (b) and the WT generator (c). Plots in (d), (e) and (f) show the distribution of radius of gyration $r_g$, while plots in (g), (h) and (i) show the distribution of mobility entropy $S^{unc}$. }
\label{fig:plots}
\end{figure}

\begin{figure}[!htb]\centering
\textbf{\LARGE GPS}\par\medskip
\subfigure[]
   {\includegraphics[scale=0.2555]{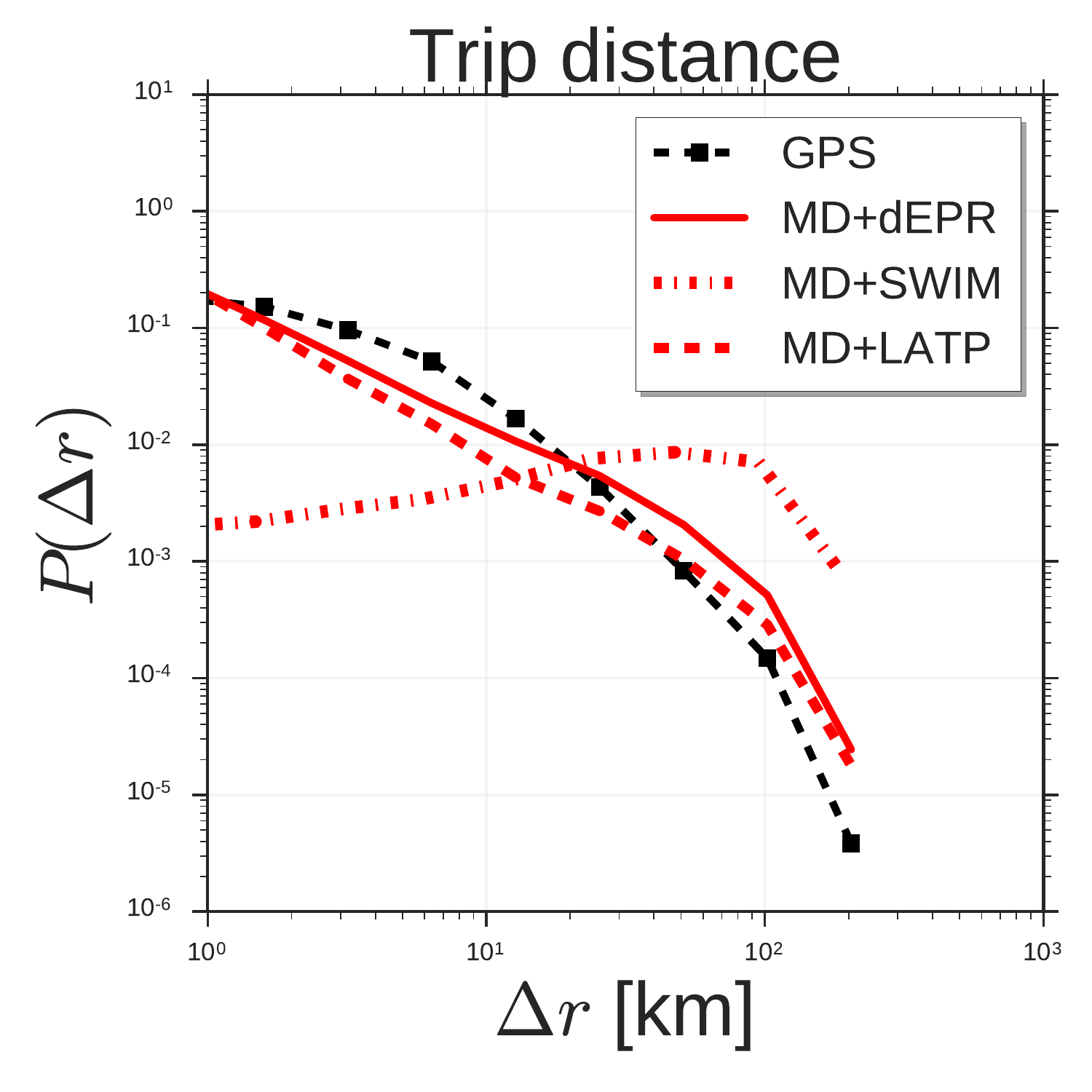}}
\subfigure[]
   {\includegraphics[scale=0.2555]{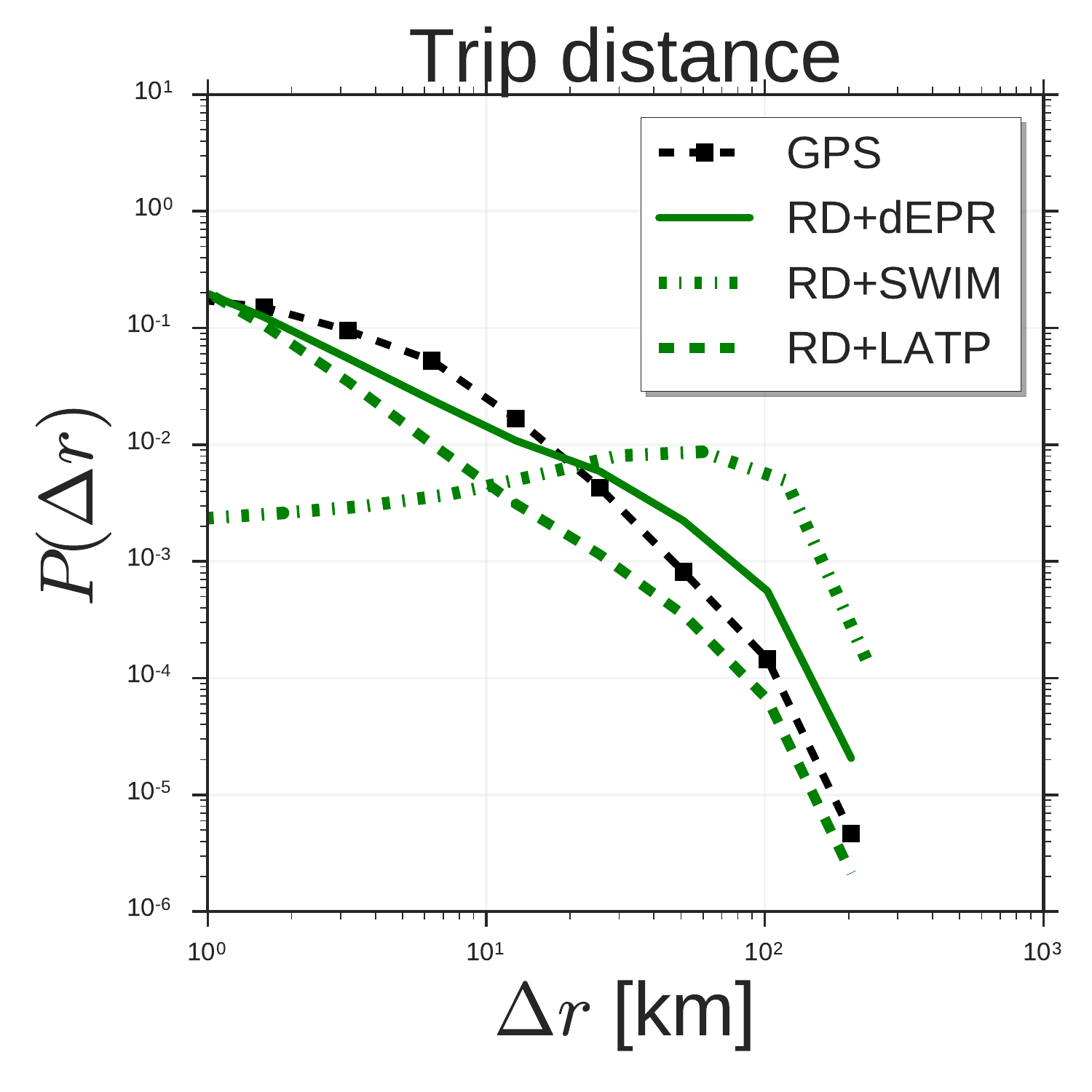}}
\subfigure[]
   {\includegraphics[scale=0.2555]{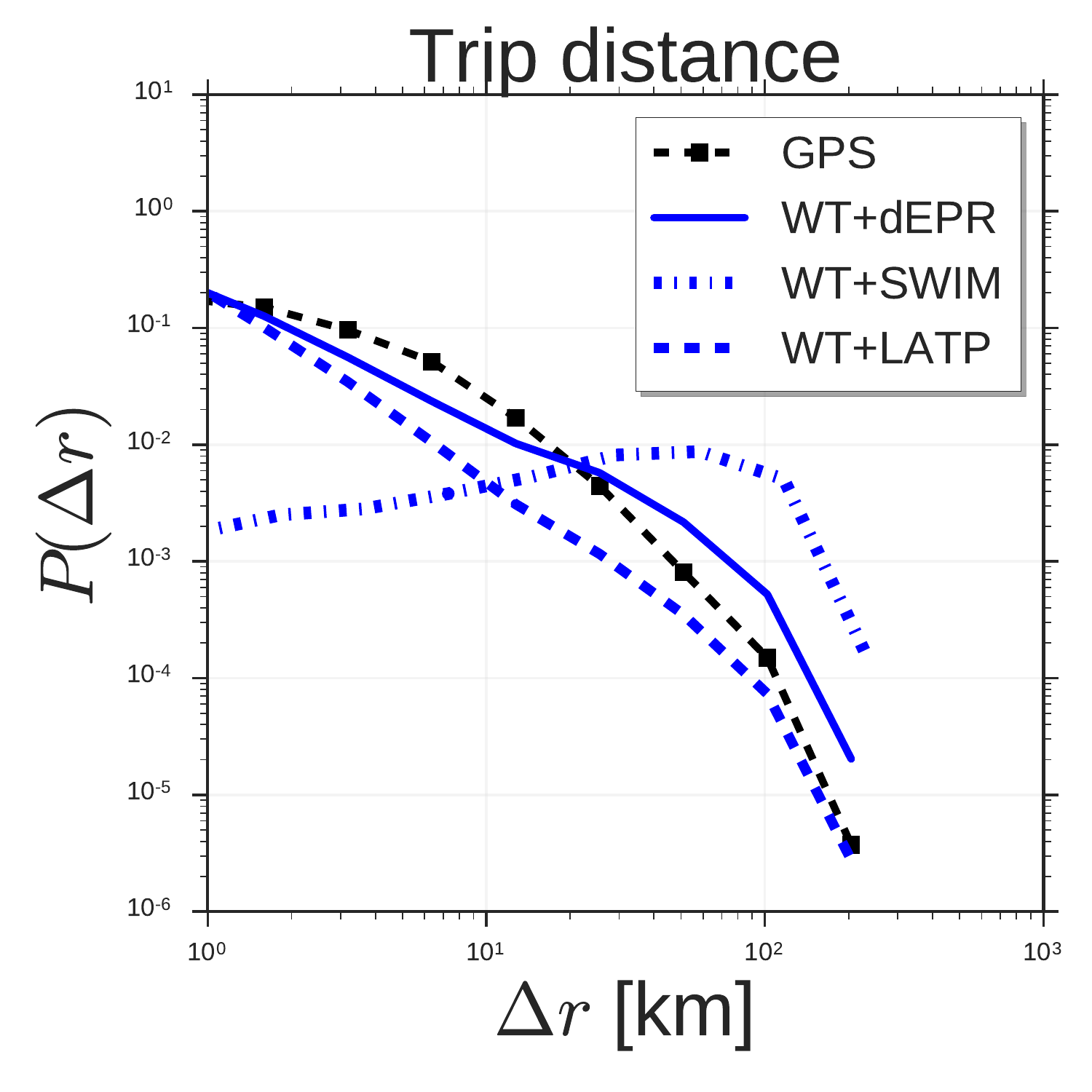}}
   
\subfigure[]
   {\includegraphics[scale=0.255]{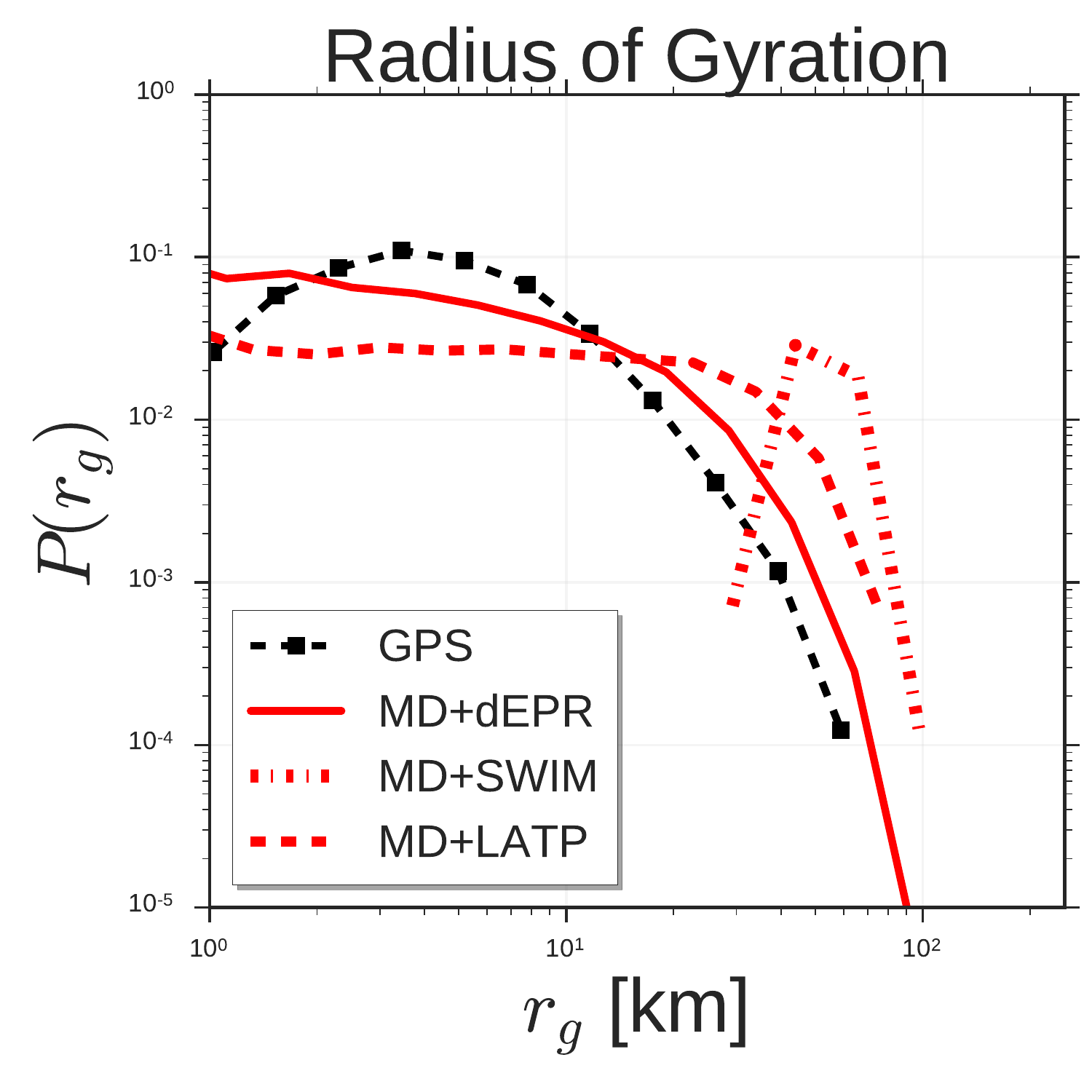}}
\subfigure[]
   {\includegraphics[scale=0.255]{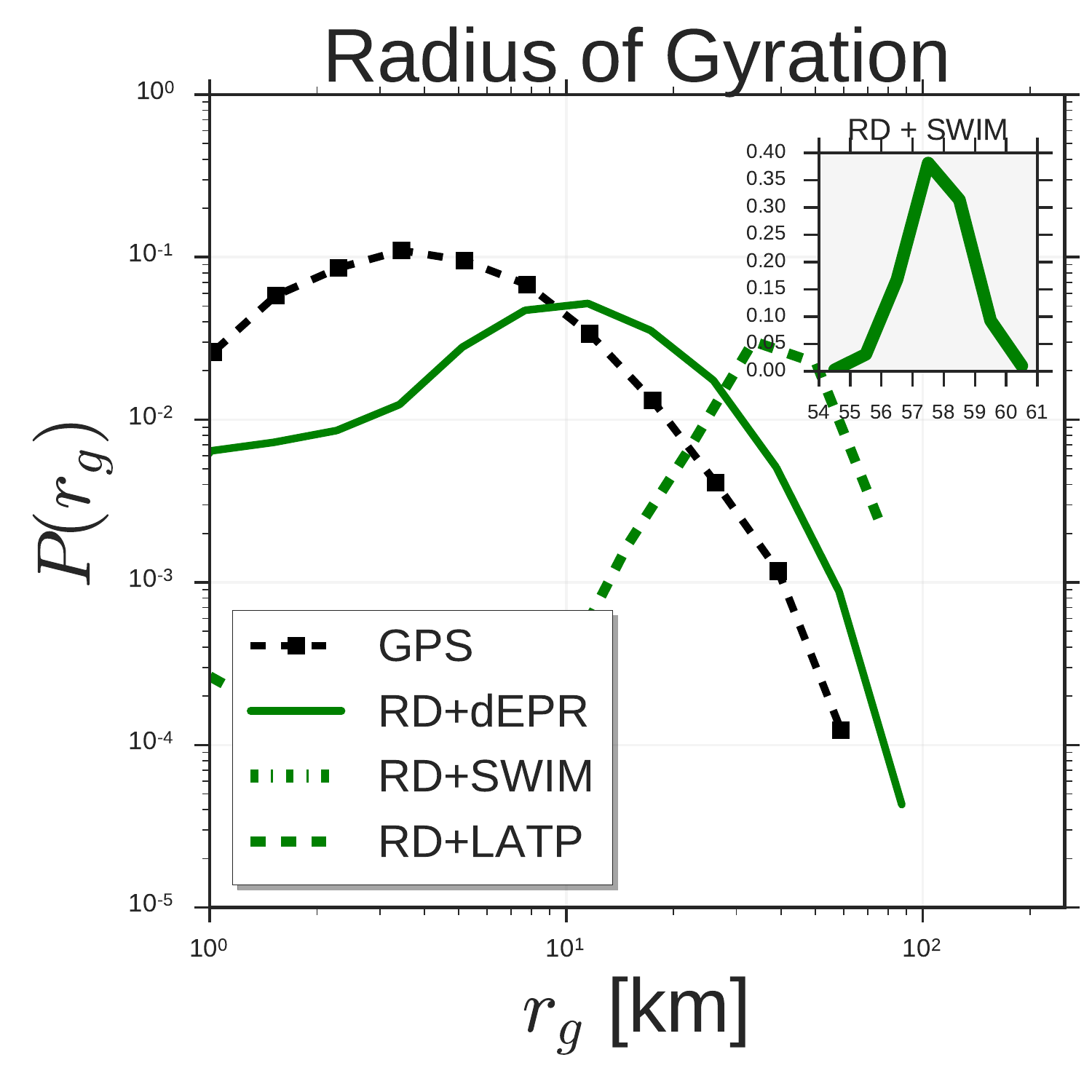}}
\subfigure[]
   {\includegraphics[scale=0.255]{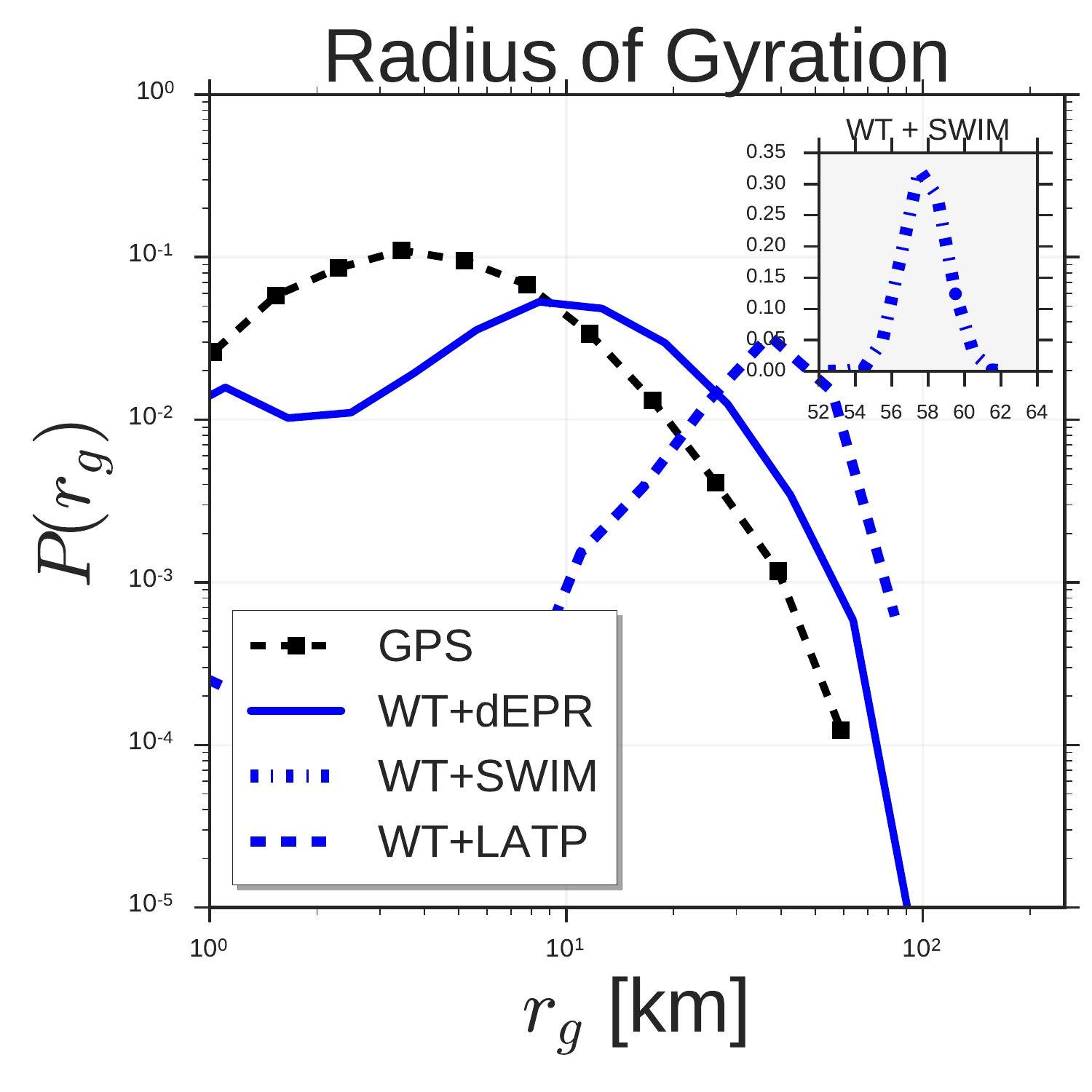}}

\subfigure[]
   {\includegraphics[scale=0.255]{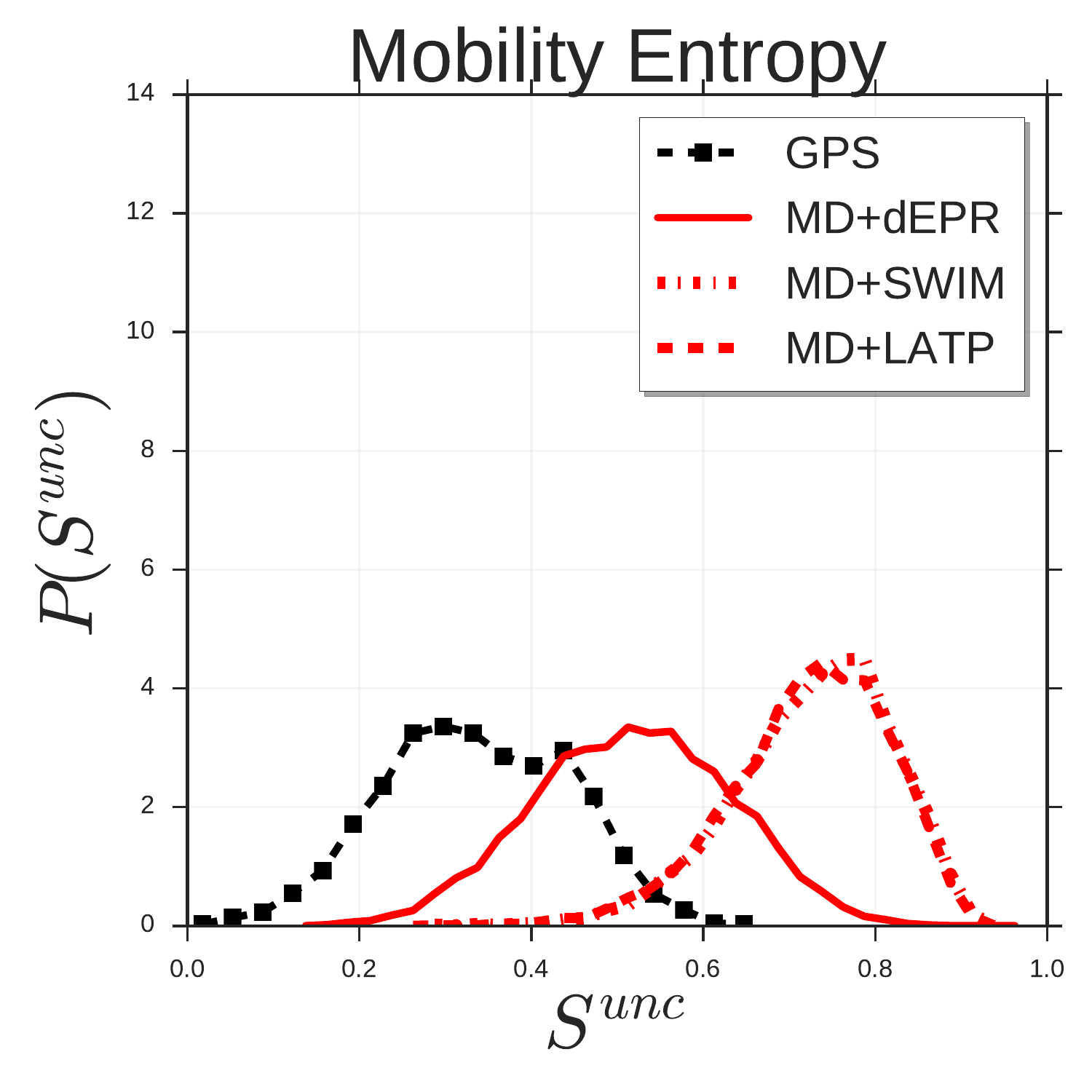}}
\subfigure[]
   {\includegraphics[scale=0.255]{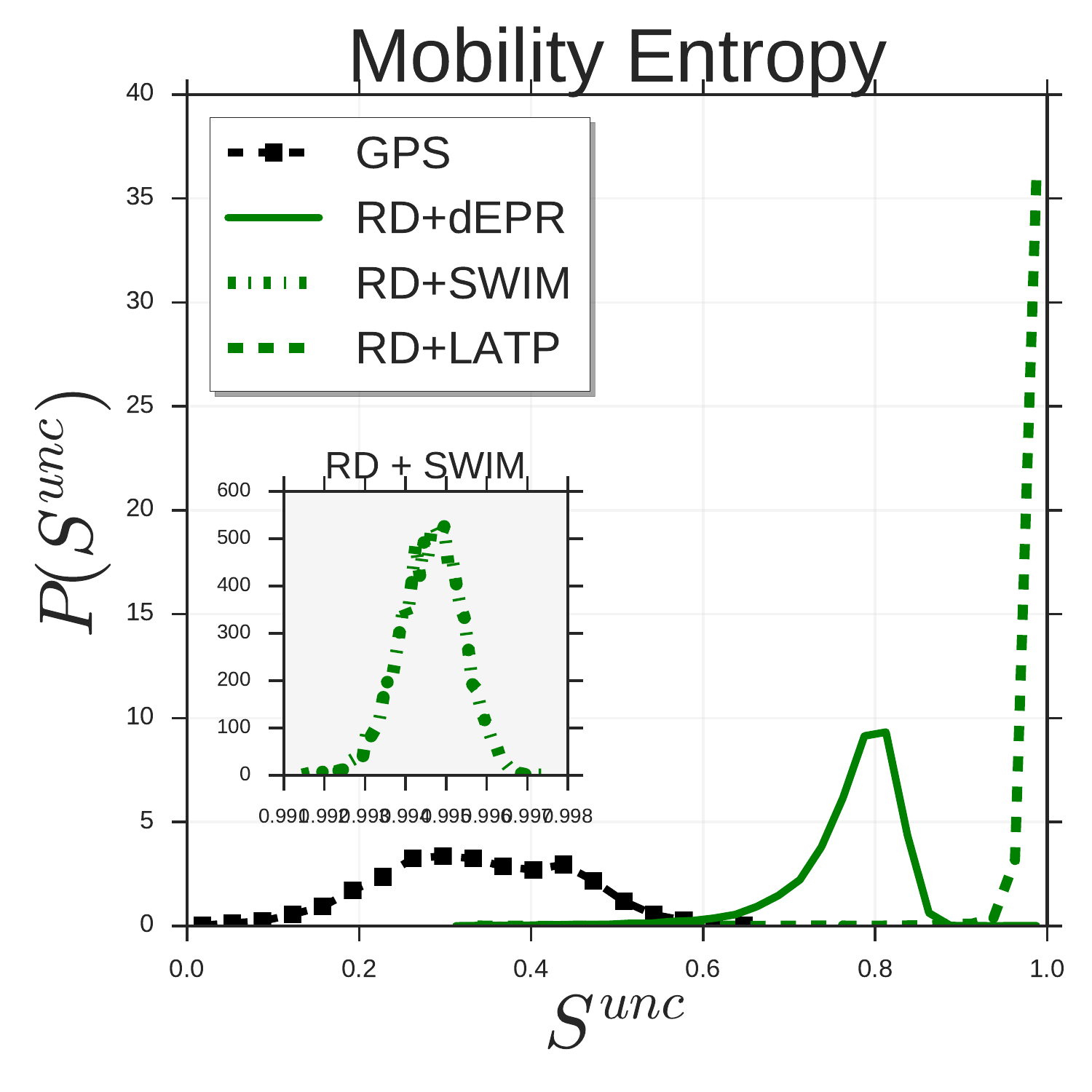}}
 \subfigure[]
   {\includegraphics[scale=0.255]{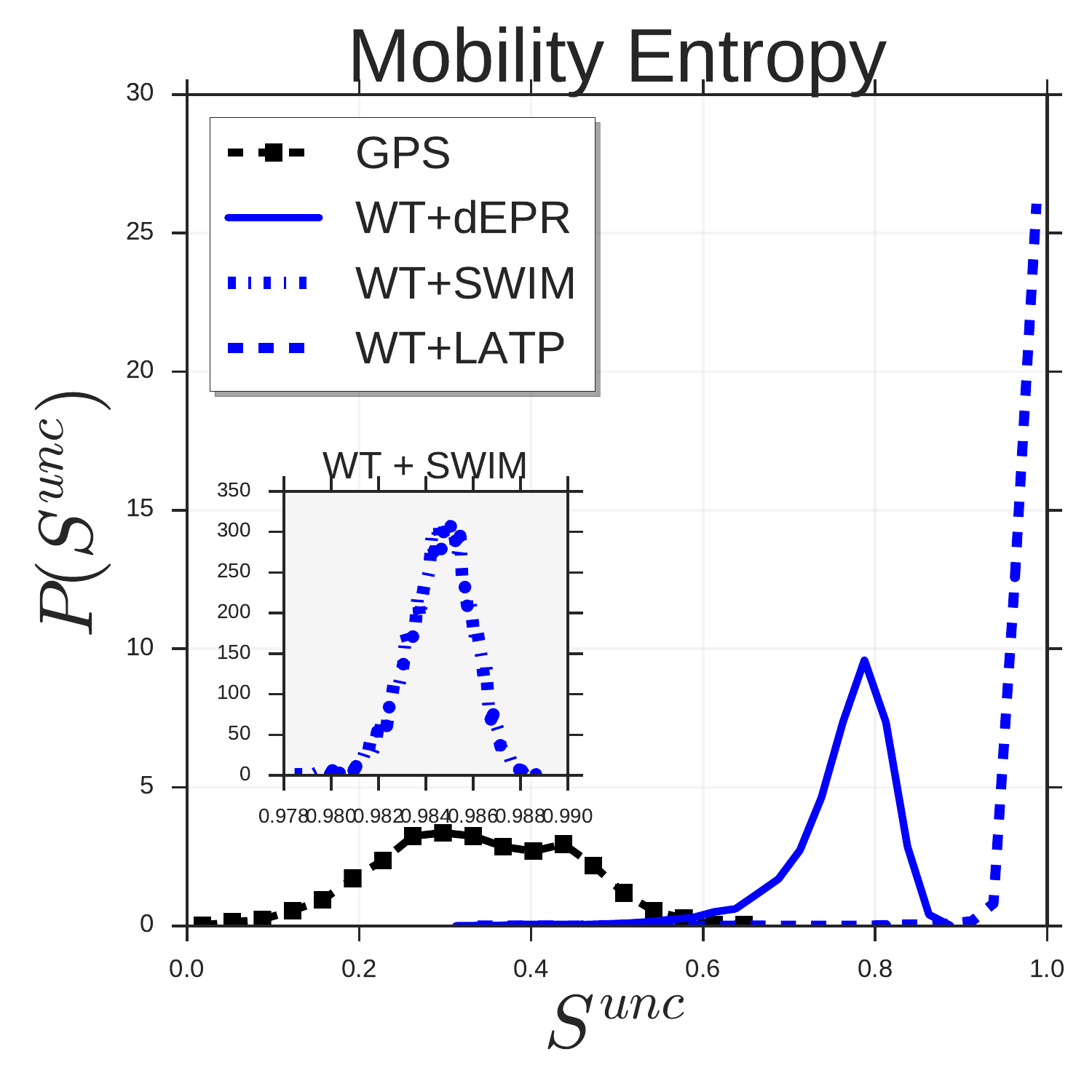}}

\caption{\textbf{Distributions of human mobility patterns (GPS).} The figure compares the models and GPS data on trip distance, radius of gyration and mobility entropy. 
}
\label{fig:plots_toscana}
\end{figure}

\paragraph{Trip distance.} The distance of a trip $\Delta r$ is the geographical distance between the trip's origin and destination locations.
We compute the trip distances for every individual and then plot the distribution $P(\Delta r)$ of trip distances in Fig.\ \ref{fig:plots}a-c (CDR data) and Fig. \ref{fig:plots_toscana}a-c (GPS data). 
Fig.\ \ref{fig:plots}a compares the distribution of trip distance of CDR data with the distributions produced by $d$-EPR$^{\mbox{\tiny (CDR)}}_{\mbox{\small MD}}$, SWIM$^{\mbox{\tiny (CDR)}}_{\mbox{\small MD}}$ and LATP$^{\mbox{\tiny (CDR)}}_{\mbox{\small MD}}$. We observe that $d$-EPR$^{\mbox{\tiny (CDR)}}_{\mbox{\small MD}}$ and LATP$^{\mbox{\tiny (CDR)}}_{\mbox{\small MD}}$ are able to reproduce the distribution of $P(\Delta r)$ although slightly overestimating long-distance trips. In contrast SWIM$^{\mbox{\tiny (CDR)}}_{\mbox{\small MD}}$ cannot reproduce the shape of the empirical distribution resulting in a RMSE(SWIM$^{\mbox{\tiny (CDR)}}_{\mbox{\small MD}}$) and KL(SWIM$^{\mbox{\tiny (CDR)}}_{\mbox{\small MD}}$) higher than the other two models (see Table \ref{tab:models}). The shape of the synthetic distributions do not vary significantly by changing the diary generator (Fig.\ \ref{fig:plots}, b-c). In other words, the choice of the diary generator does not affect the ability of the model to capture the distribution $P(\Delta r)$. This is also evident from Table \ref{tab:models} where the RMSEs and the KLs in the first column vary a little by changing the diary generator. Model $d$-EPR$^{\mbox{\tiny (CDR)}}_{\mbox{\small MD}}$ produces the best fit with CDR data, as we notef in Fig.\ \ref{fig:plots}c and Table \ref{tab:models}. This suggests that modelling preferential return and location preference is crucial to reproduce $P(\Delta r)$ as well as the preference for short-distance trips. Although SWIM embeds a preference for short-distance trips (Equation \ref{eq:swim}) the distance is chosen with respect to the home location $L_h$ leading to an underestimation of short-distance trips (Fig.\ \ref{fig:plots}a-c).
Fig.\ \ref{fig:plots_toscana}a-c compares the distribution of trip distance of GPS data with the distributions produced by the generative algorithms. Results on GPS data confirm the observations on CDRs: in contrast with SWIM, $d$-EPR and LATP are able to reproduce the distribution of $P(\Delta r)$, regardless the diary generator. Also in this case, $d$-EPR$^{\mbox{\tiny (GPS)}}_{\mbox{\small RD}}$ is the model generating the most realistic synthetic data (Table \ref{tab:models_GPS}). 

\paragraph{Radius of gyration.} The radius of gyration $r_g$ is the characteristic distance traveled by an individual during the period of observation \citep{gonzalez08,pappalardo2013,pappalardo2015}. In detail, $r_g$ characterizes the spatial spread of the locations visited by an individual $u$ from the trajectories' center of mass (i.e., the weighted mean point of the locations visited by an individual), defined as:
\begin{equation}
r_g = \sqrt{ \sum_{i \in L^{(u)}} p_i (l_i - l_{cm})^2  },
\end{equation}
where $l_i$ and $l_{cm}$ are the vectors of coordinates of location $i$ and center of mass, respectively \citep{gonzalez08,pappalardo2013,pappalardo2015}, $L^{(u)} \subseteq L$ is the set of locations visited by individual $u$, $p_i = n_i/|L^{(u)}| $ is the individual's visitation frequency of location $l_i$, equal to the number of visits to $l_i$ divided by the total number of visits to all locations. In Fig.\ \ref{fig:plots}a we observe that $d$-EPR$^{\mbox{\tiny (CDR)}}_{\mbox{\small MD}}$ is the only model capable of reproducing the shape of $P(r_g)$ of CDR data, though overestimating the presence of large radii (see Fig.\ \ref{fig:plots}d). RMSE($d$-EPR$^{\mbox{\tiny (CDR)}}_{\mbox{\small MD}}$) for $r_g$ is indeed lower than RMSE(SWIM$^{\mbox{\tiny (CDR)}}_{\mbox{\small MD}}$) and RMSE(LATP$^{\mbox{\tiny (CDR)}}_{\mbox{\small MD}}$) as shown in Table \ref{tab:models}.
SWIM$^{\mbox{\tiny (CDR)}}_{\mbox{\small MD}}$ and LATP$^{\mbox{\tiny (CDR)}}_{\mbox{\small MD}}$ cannot reproduce the shape of $P(r_g)$ because $r_g$ also depends on the preferential return mechanism \citep{SongNaturePhysics2010,pappalardo2015} which is not modeled in SWIM and LATP. In a previous work \citep{Pappalardo2016934} we also show that $P(r_g)$ depends on the preferential exploration mechanism of $d$-EPR since a version of $d$-EPR without preferential exploration -- the $s$-EPR model -- is not able to reproduce the shape of $P(r_g)$. We also observe that while $d$-EPR$^{\mbox{\tiny (CDR)}}_{\mbox{\small \{MD, RD, WT\}}}$ produce similar distributions of $r_g$, SWIM and LATP produce different distributions of $r_g$ with different choices of the diary generator (Fig.\ \ref{fig:plots}e, f). The shape of $P(r_g)$ for GPS data is slightly different from the same distribution of CDR data, since short radii are less likely in GPS due to the nature of car travels \citep{pappalardo_comparing,pappalardo2013,pappalardo2012_sebd}. Also for GPS we observe that, in contrast with LATP and SWIM, $d$-EPR is the only model that can reproduce the shape of $P(r_g)$. In particular $d$-EPR$^{\mbox{\tiny (GPS)}}_{\mbox{\small MD}}$ produces the best fitting with GPS data in terms of both RMSE and KL (Table \ref{tab:models_GPS}).

\paragraph{Mobility entropy.} The mobility entropy $S^{unc}$ of an individual $u$ is defined as the Shannon entropy of her visited locations \citep{song2010limits,pentland,pappalardo2016}: 
\begin{equation}
S^{unc}(u) = {\sum_{i \in L^{(u)}} p_i \log(p_i) \over \log |L^{(u)}|},
\end{equation}
where $p_i$ is the probability that individual $u$ visits location $i$ during the period of observation and $\log |L^{(u)}|$ is a normalization factor. The mobility entropy of an individual quantifies the possibility to predict individual's future whereabouts. Individuals having a very regular movement pattern possess a mobility entropy close to zero and their whereabouts are rather predictable. Conversely, individuals with a high mobility entropy are less predictable. 

We observe that the average $\overline{S}^{unc}$ produced by $d$-EPR$^{\mbox{\tiny (CDR)}}_{\mbox{\small MD}}$ data equals the average $\overline{S}^{unc} {=} 0.61$ in CDR data, although $d$-EPR$^{\mbox{\tiny (CDR)}}_{\mbox{\small MD}}$ underestimates the variance of distribution $P(S^{unc})$ (Fig.\ \ref{fig:plots}g). In contrast, SWIM$^{\mbox{\tiny (CDR)}}_{\mbox{\small MD}}$ and LATP$^{\mbox{\tiny (CDR)}}_{\mbox{\small MD}}$ largely overestimate $\overline{S}^{unc}$ and underestimate the variance of $P(S^{unc})$, resulting in RMSE and KL much higher than RMSE($d$-EPR$^{\mbox{\tiny (CDR)}}_{\mbox{\small MD}}$) and KL($d$-EPR$^{\mbox{\tiny (CDR)}}_{\mbox{\small MD}}$), as shown in Table \ref{tab:models}. This is because SWIM and LATP do not model the preferential return mechanism, which increases the predictability of individuals since they tend to come back to already visited locations.
$P(S^{unc})$ is not robust to the choice of diary generator: diary generator RD and WT make the models to largely overestimate $\overline{S}^{unc}$ (Figures \ref{fig:plots}h, i). In particular SWIM$^{\mbox{\tiny (CDR)}}_{\mbox{\small \{RD, WT\}}}$ and LATP$^{\mbox{\tiny (CDR)}}_{\mbox{\small \{RD, WT\}}}$ produce distributions with $\bar{S}^{unc} \approx 1$, indicating that the typical synthetic individual is much more unpredictable than a typical individual in CDR data. This makes those distributions not comparable with the distribution of MD models.
Hence, distribution $P(S^{unc})$ highly depends on both the choice of the trajectory generator and the choice of the diary generator. We observe similar results for GPS data, where only \{$d$-EPR, SWIM, LATP\}$^{\mbox{\tiny (GPS)}}_{\mbox{\small MD}}$  can reproduce $P(S^{unc})$ in reasonable agreement with real data. All the other models produce distributions that are not comparable with the entropies of private vehicles (Fig. \ref{fig:plots_toscana}g-i).

\begin{figure}[!htb]\centering
 \textbf{\LARGE CDR}\par\medskip
\subfigure[]
   {\includegraphics[scale=0.255]{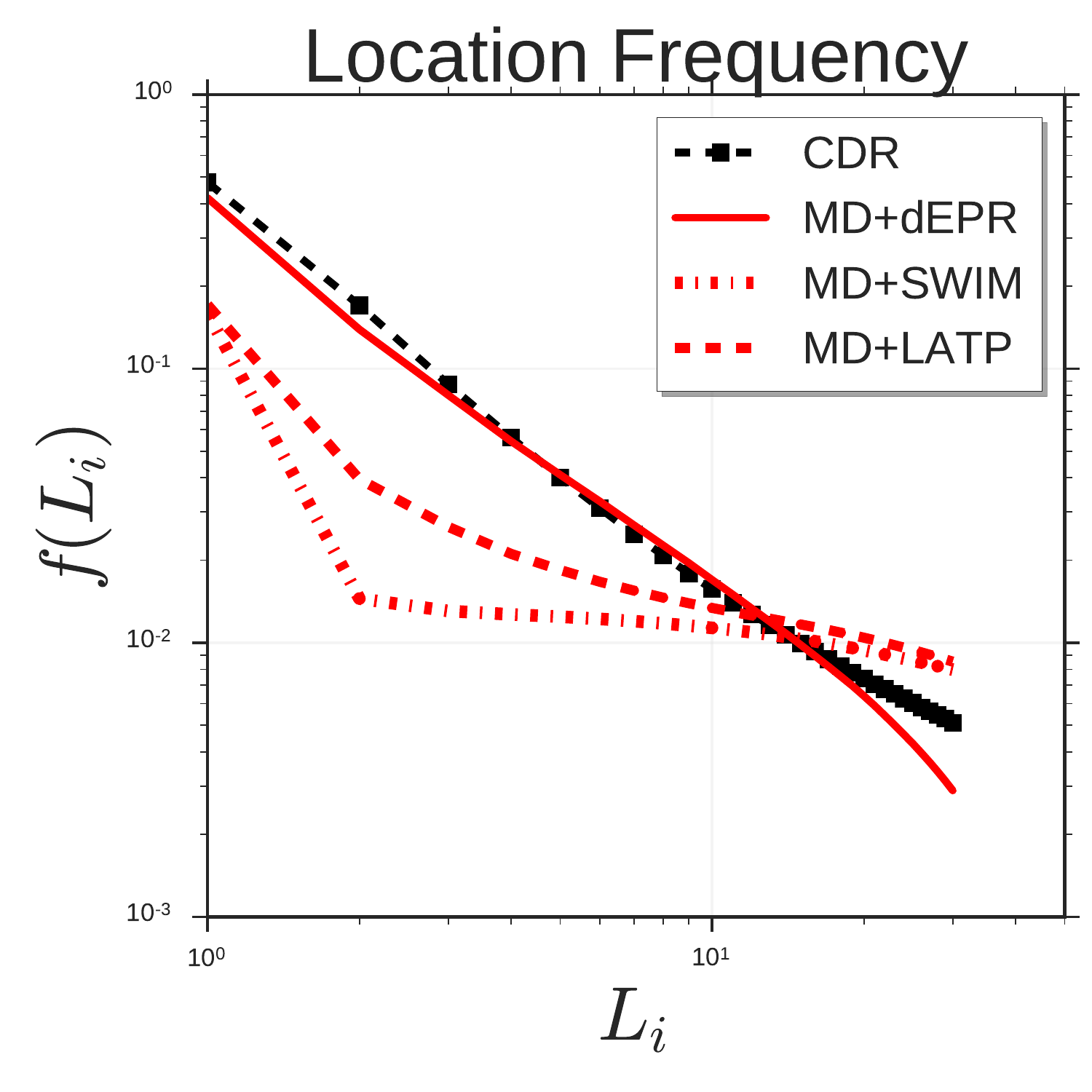}}
\subfigure[]
   {\includegraphics[scale=0.255]{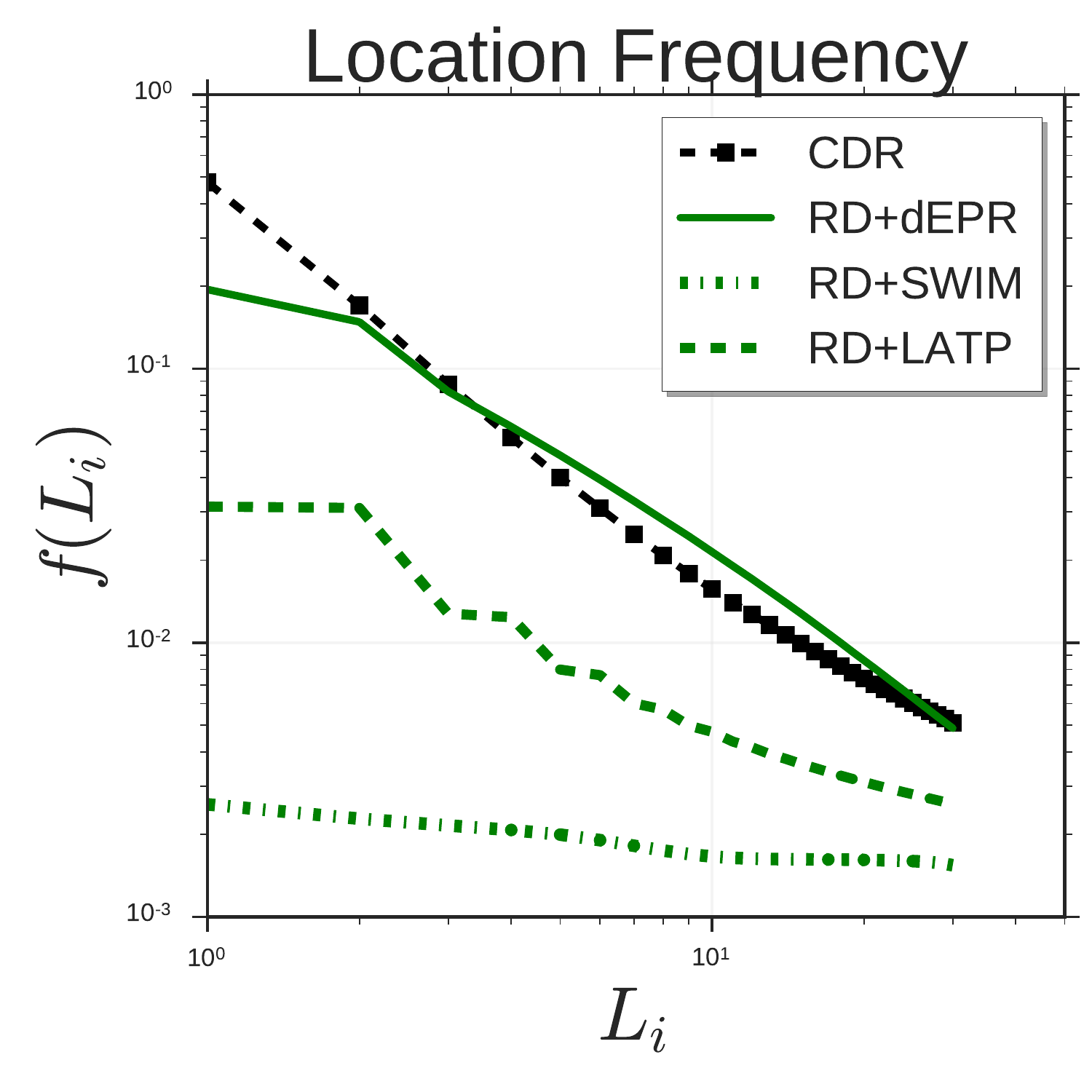}}
\subfigure[]
   {\includegraphics[scale=0.255]{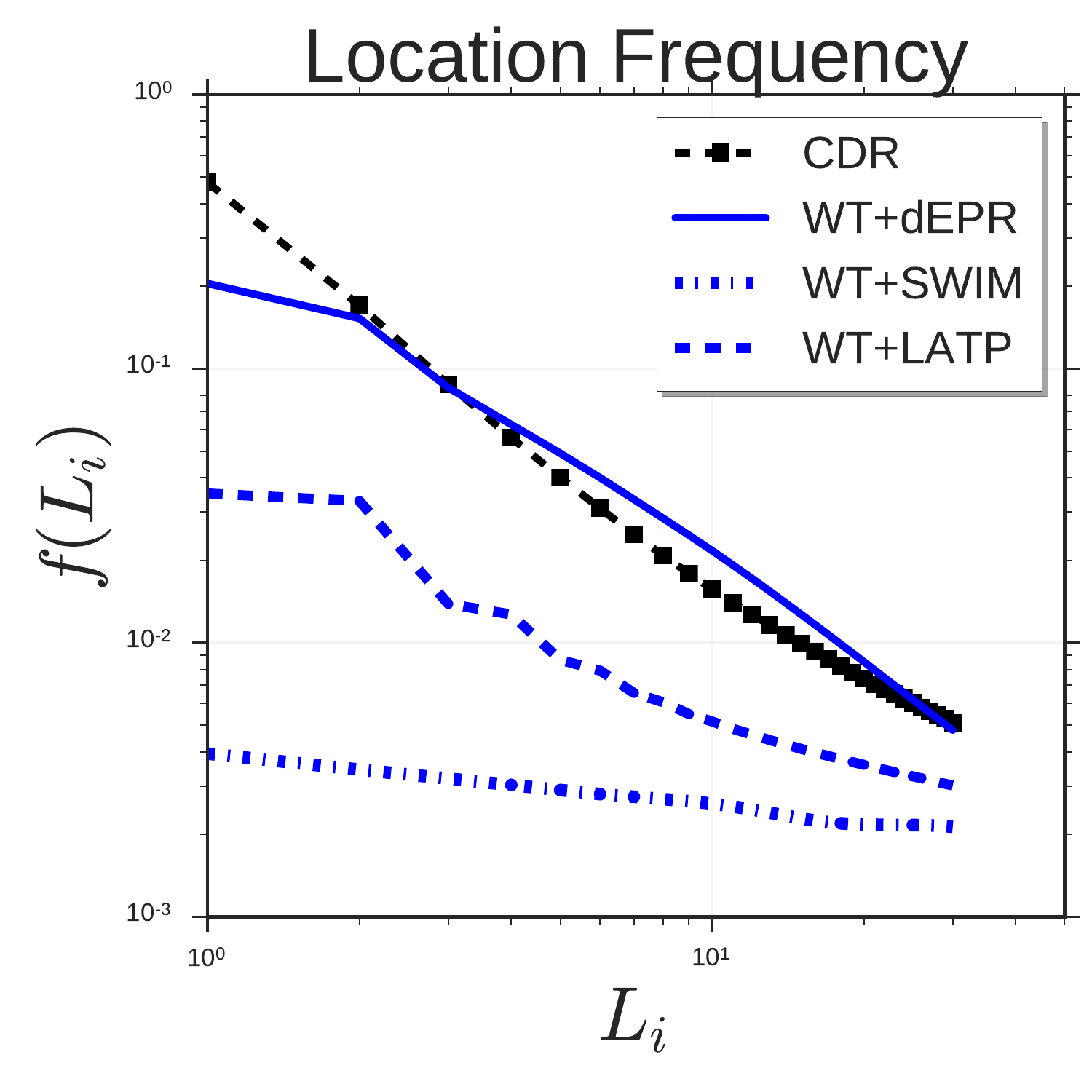}}

\subfigure[]
   {\includegraphics[scale=0.255]{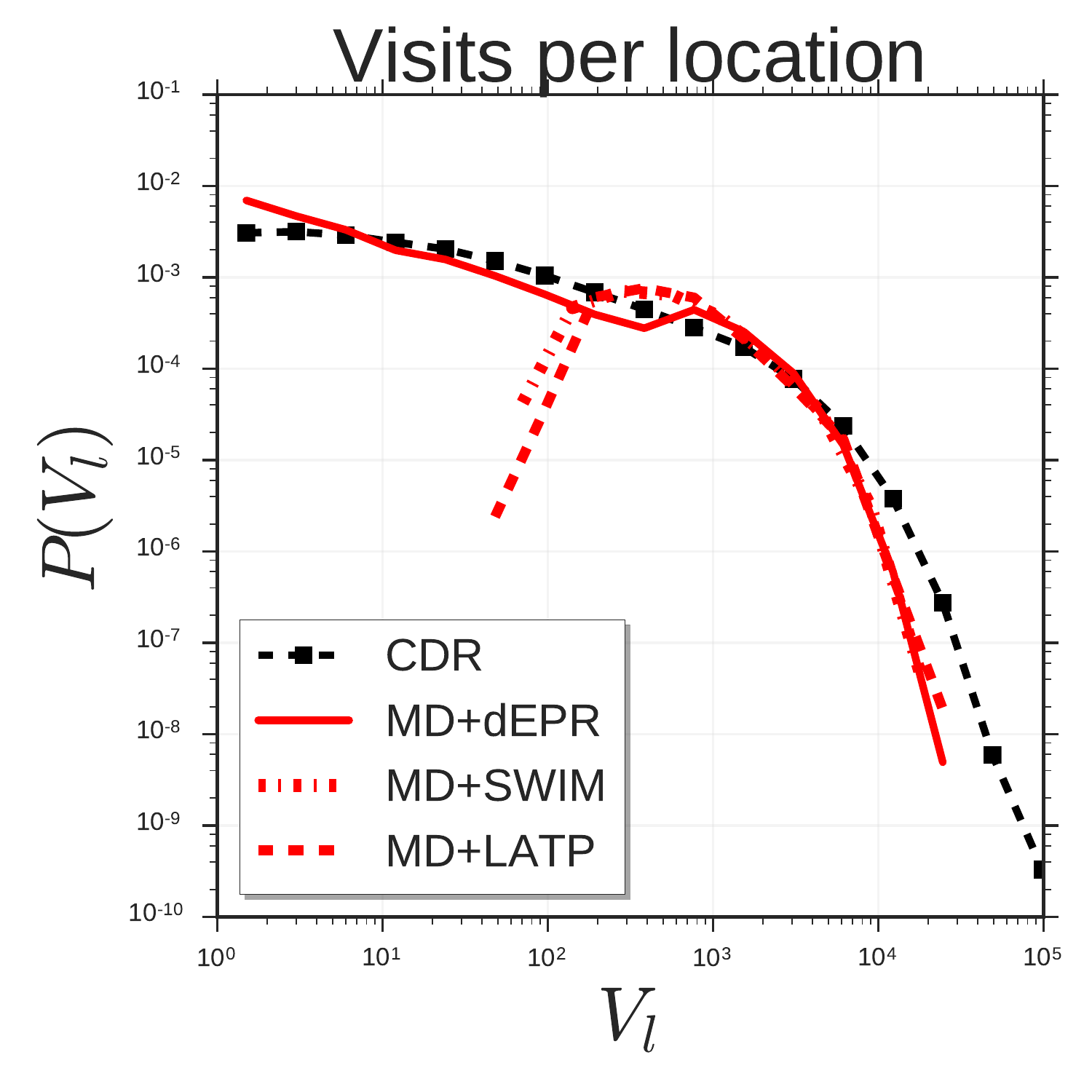}}
\subfigure[]
   {\includegraphics[scale=0.255]{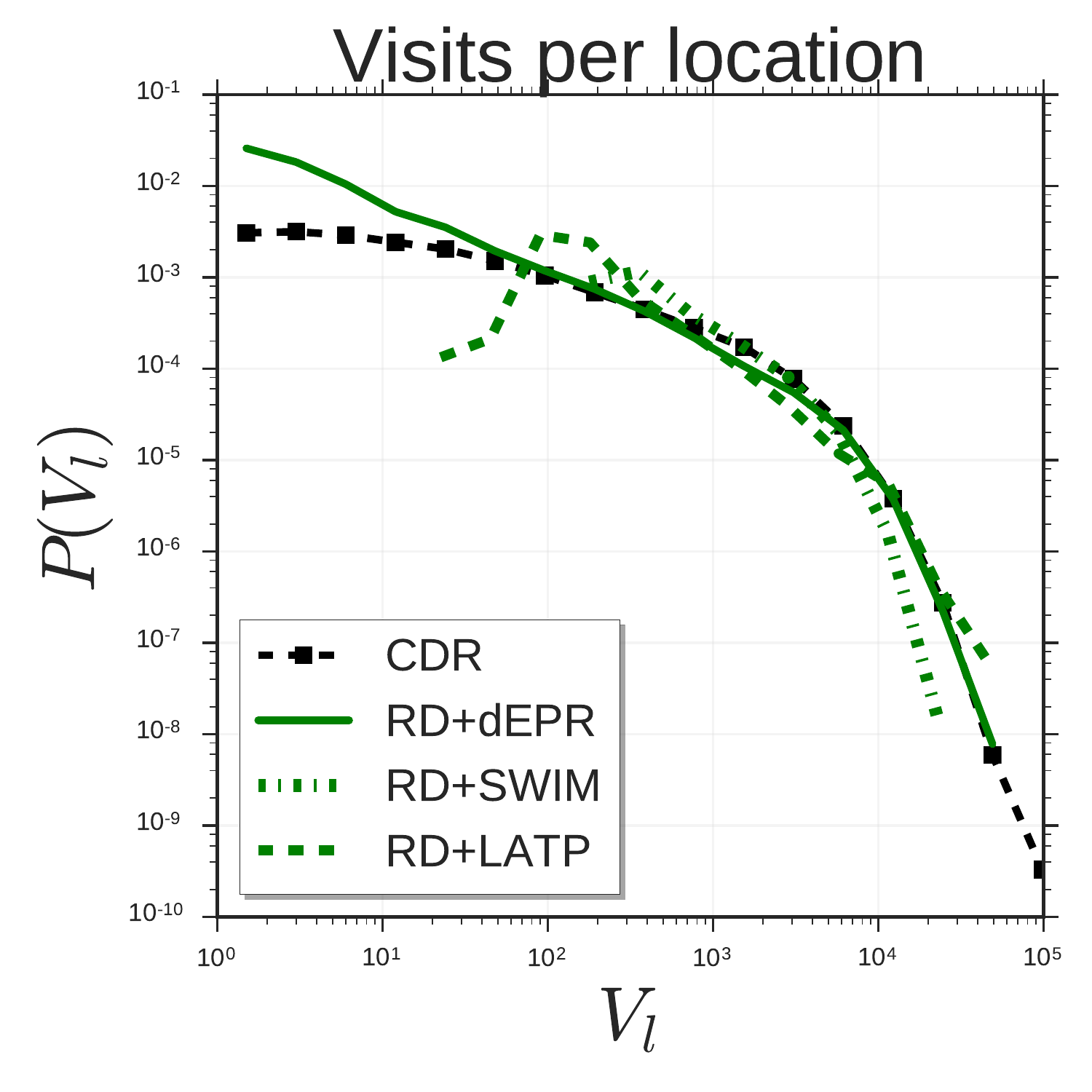}}
\subfigure[]
   {\includegraphics[scale=0.255]{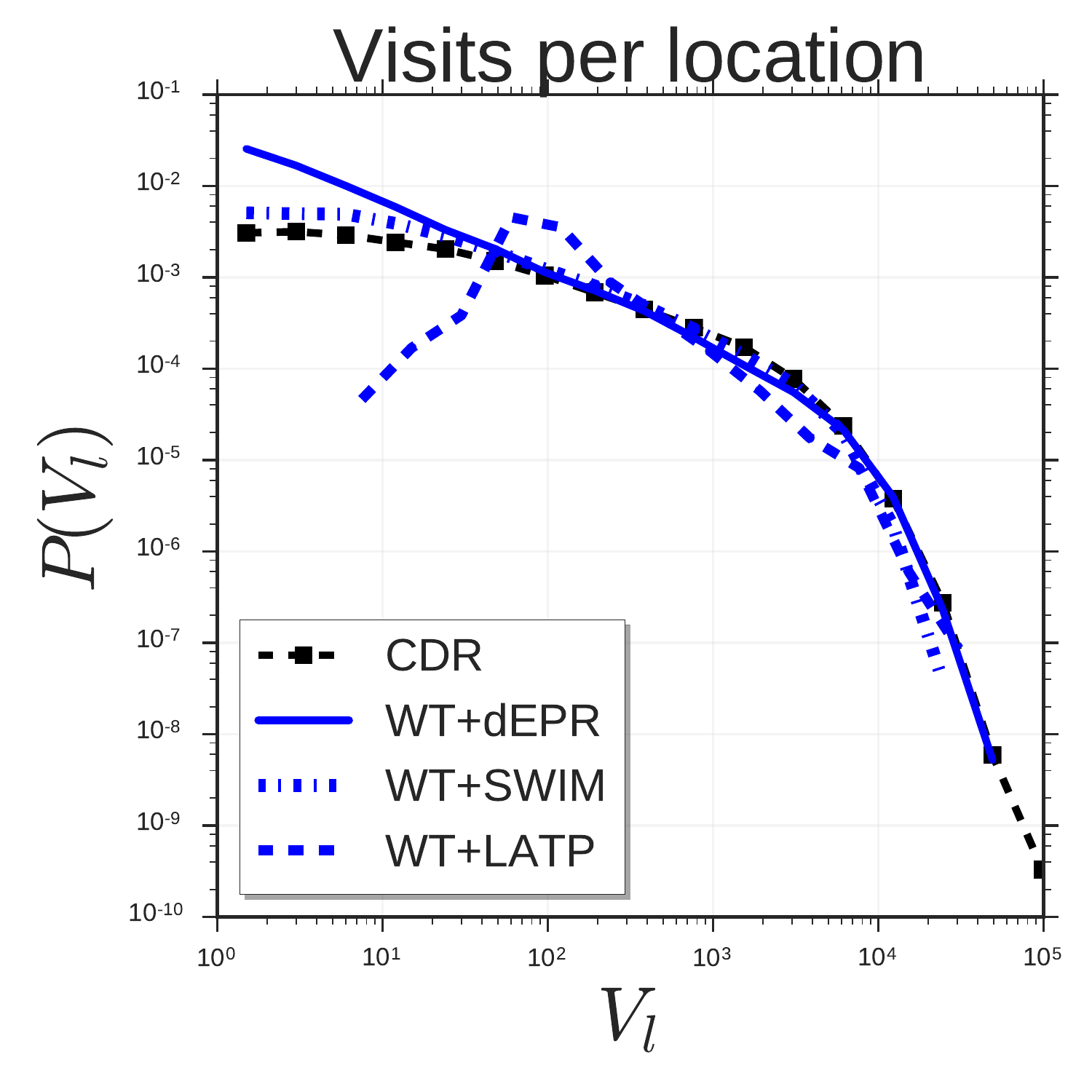}}

\subfigure[]
   {\includegraphics[scale=0.255]{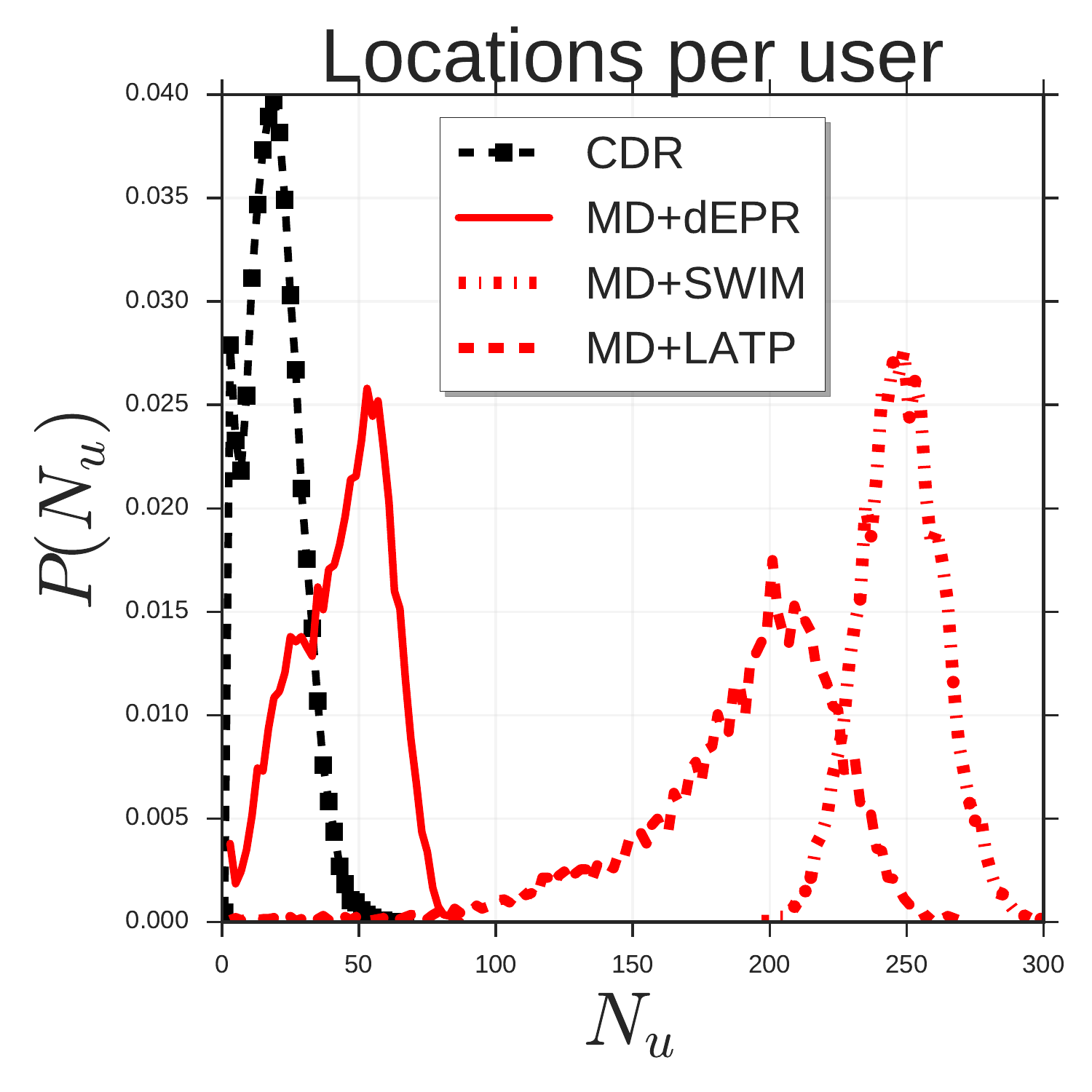}}
\subfigure[]
   {\includegraphics[scale=0.255]{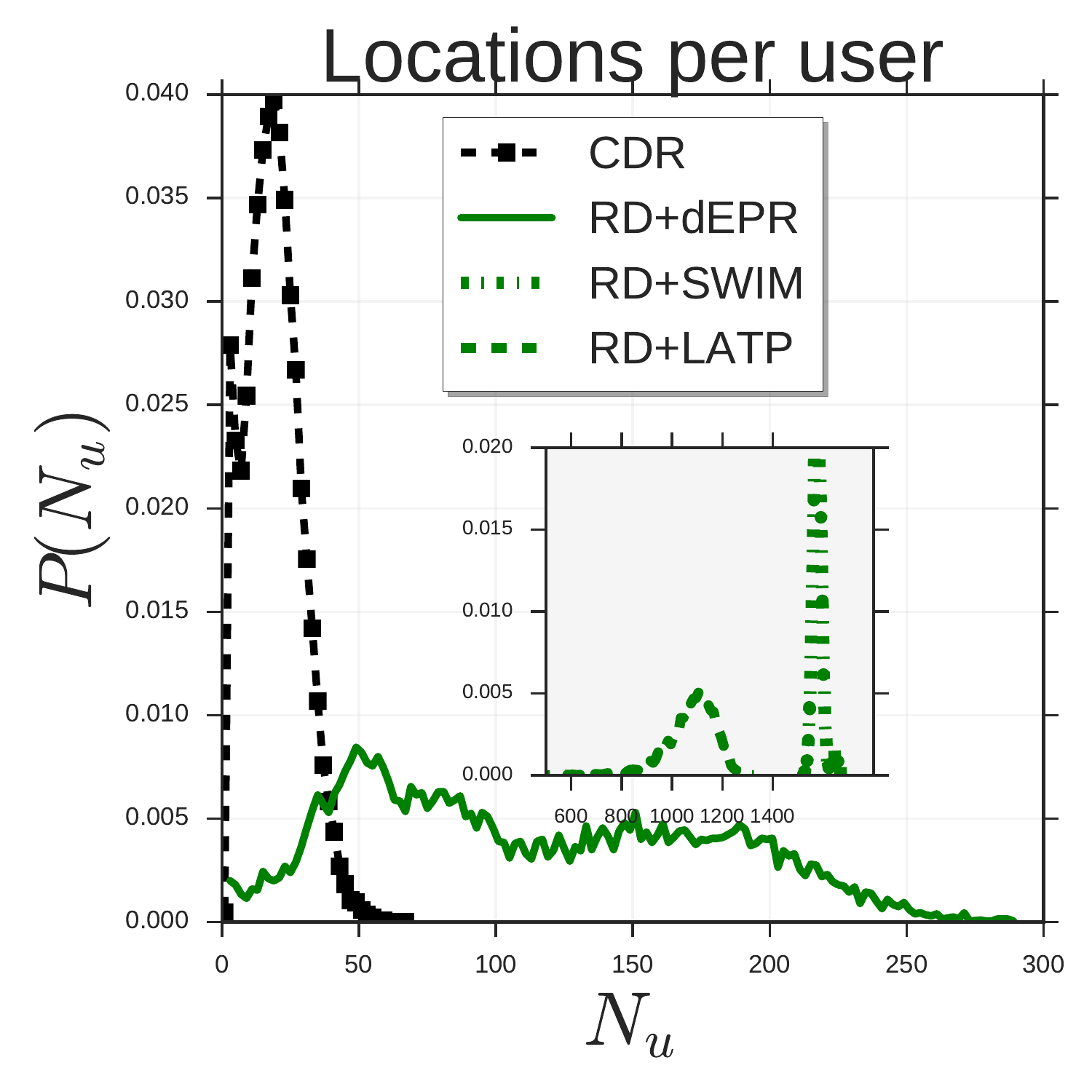}}
\subfigure[]
   {\includegraphics[scale=0.255]{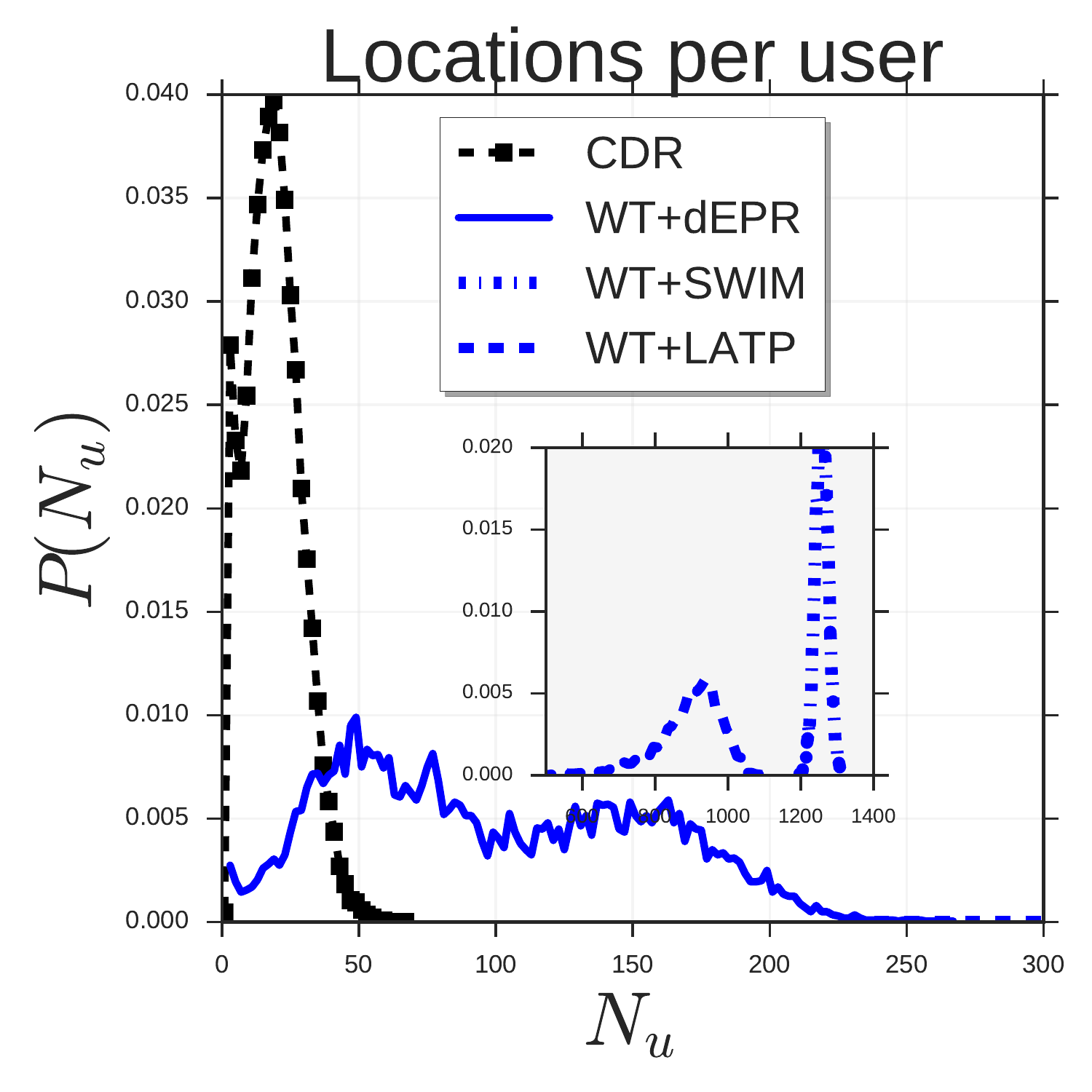}}

\caption{\textbf{Distributions of human mobility patterns (CDR)}. The figure compares the models and CDR data on location frequency, visits per location and locations per users. Plots in (a), (b) and (c) show the distribution of location frequency $f(L)$ for $d$-EPR, SWIM and LATP used in combination with MD, RD and WT respectively. Plots in (d), (e) and (f) show the distribution of the number $V$ of visits per location and plots in (g), (h), (i) show the distribution of the number $N$ of distinct visited locations per user.}
\label{fig:temporal}
\end{figure}

\begin{figure}[!htb]\centering
 \textbf{\LARGE GPS}\par\medskip
\subfigure[]
   {\includegraphics[scale=0.255]{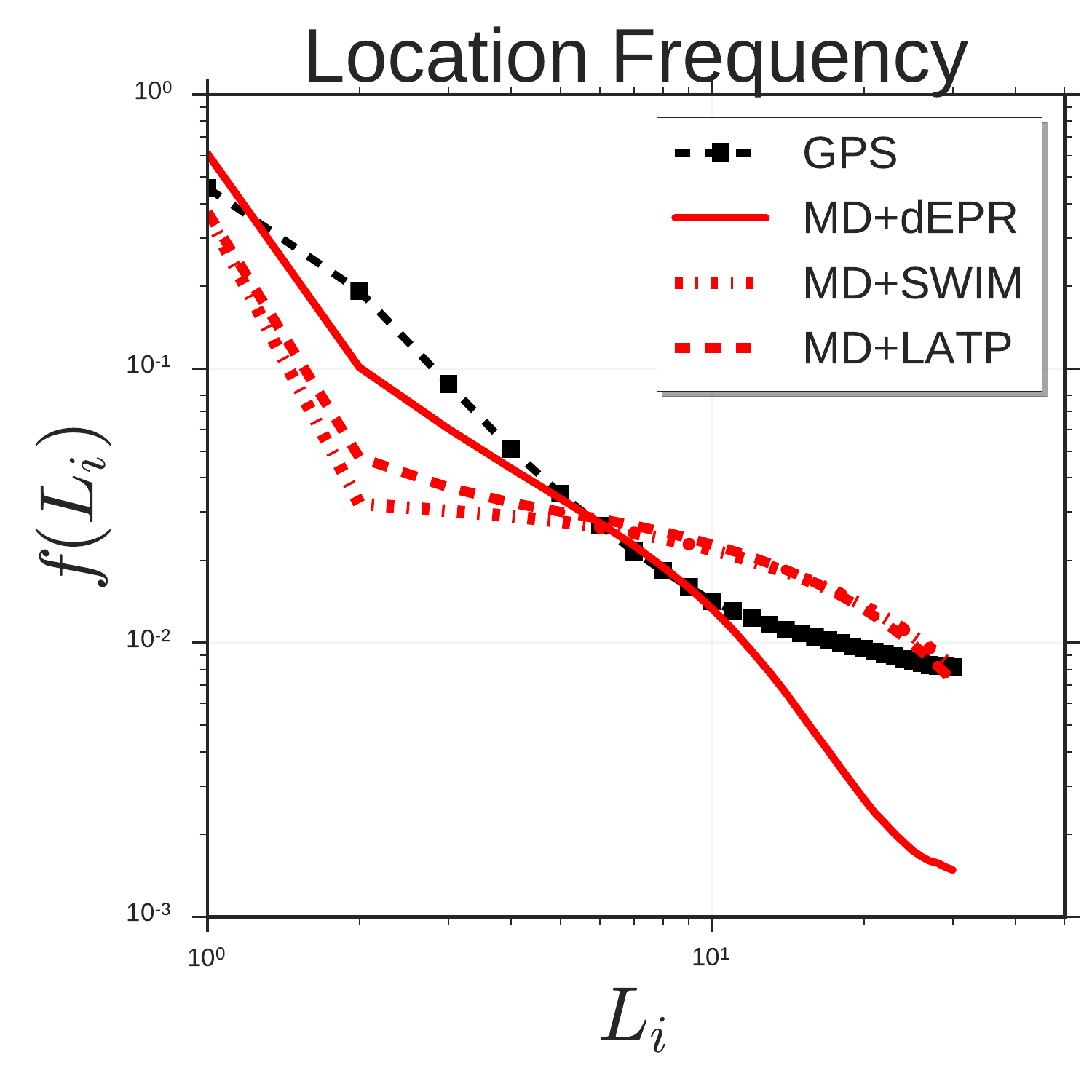}}
\subfigure[]
   {\includegraphics[scale=0.255]{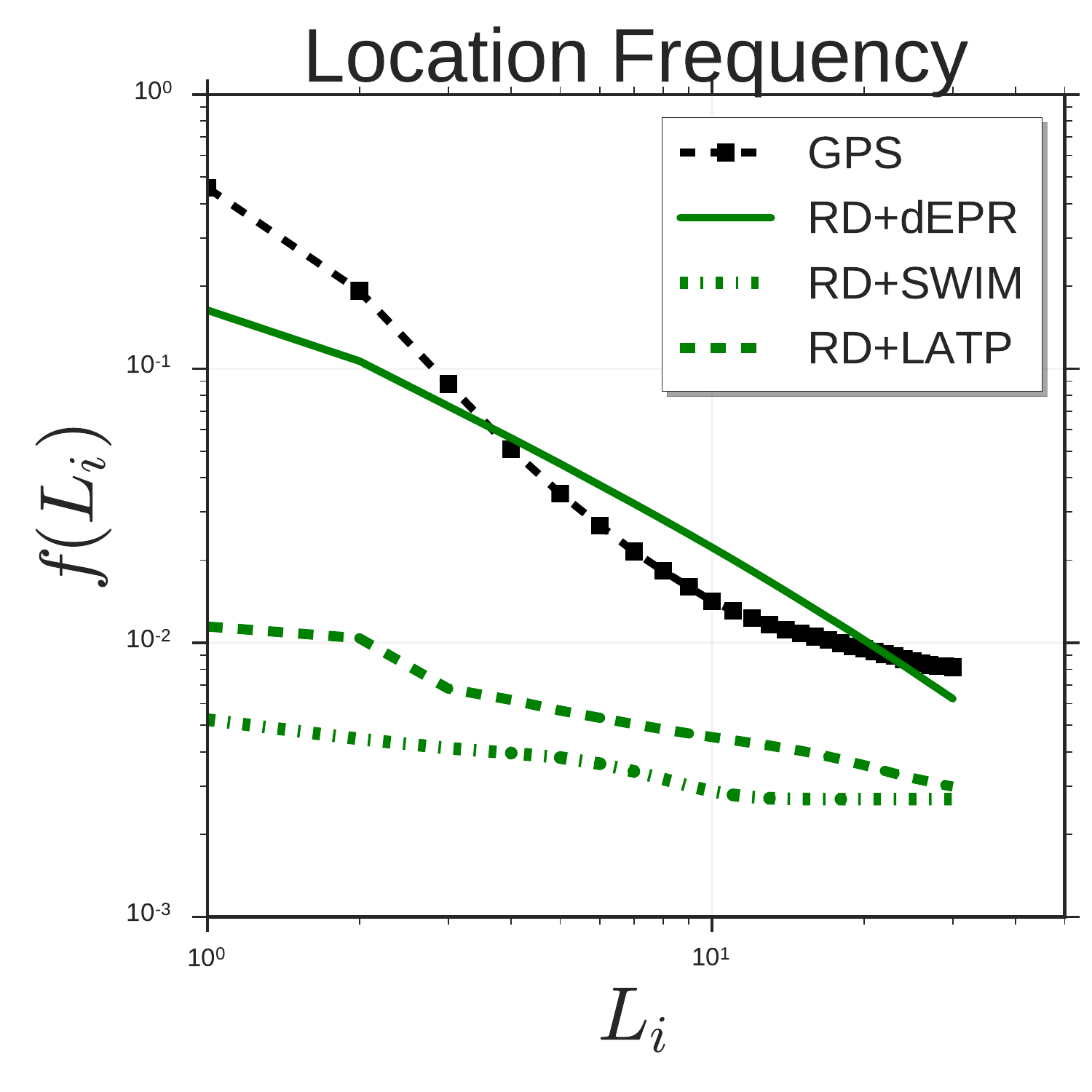}}
\subfigure[]
   {\includegraphics[scale=0.255]{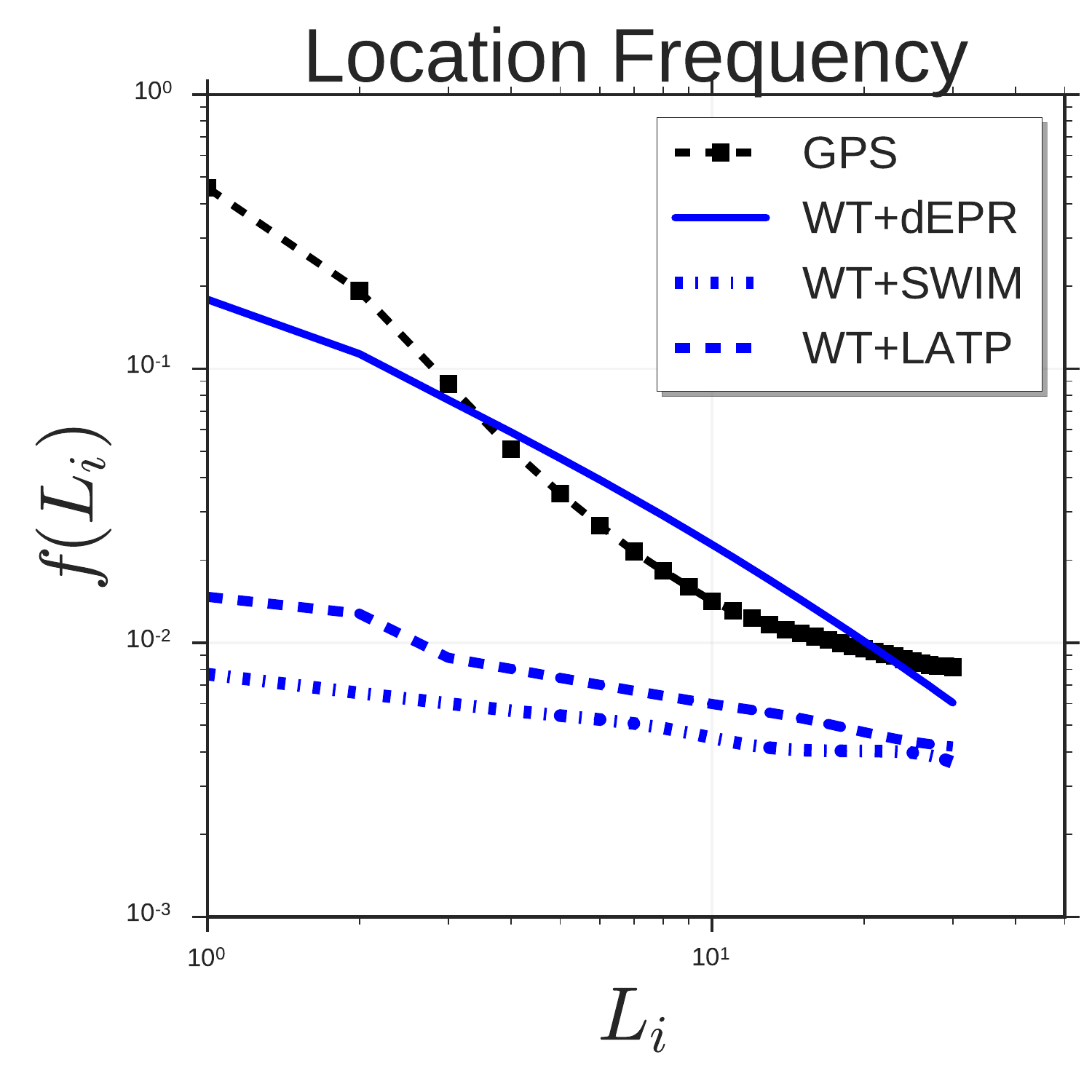}}

\subfigure[]
   {\includegraphics[scale=0.255]{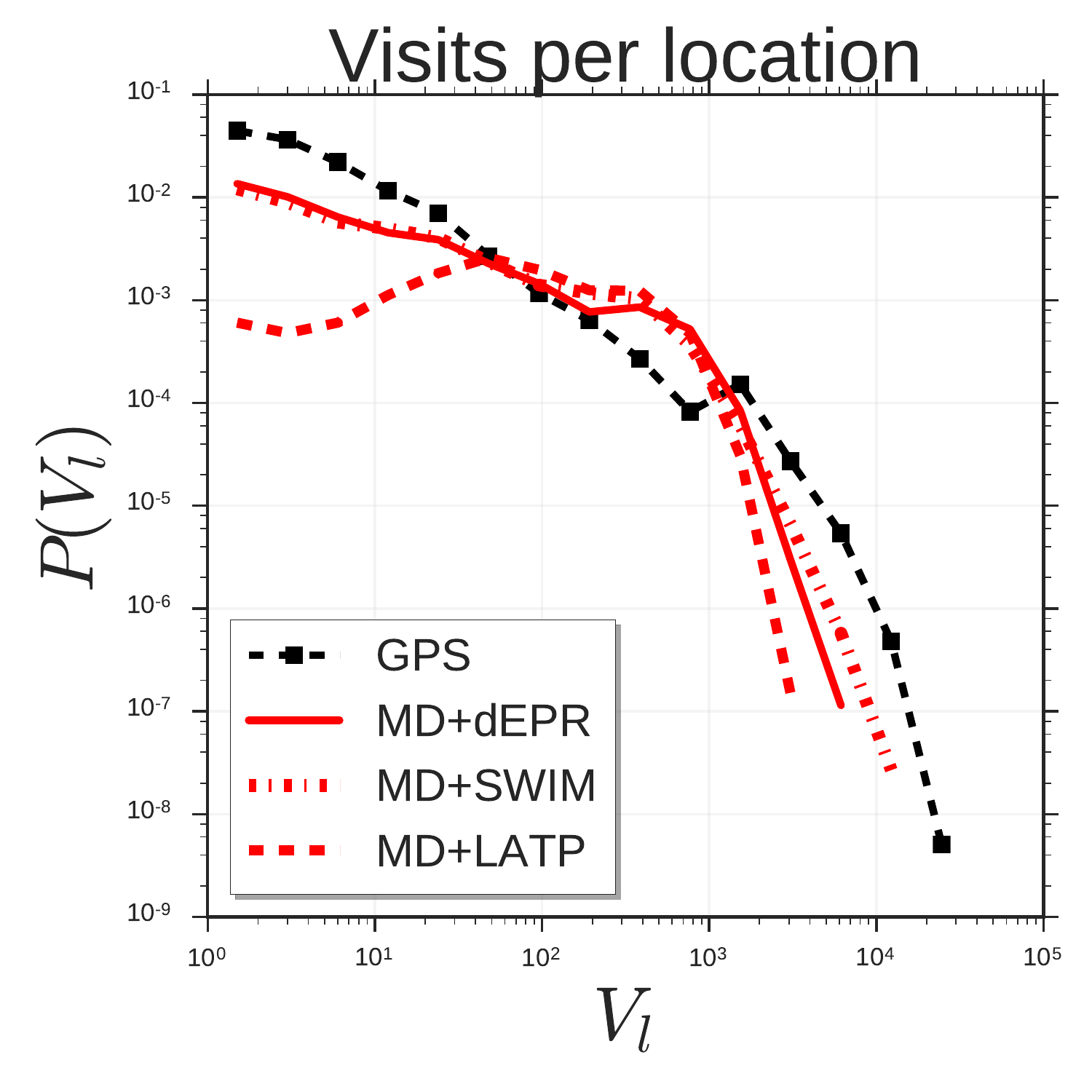}}
\subfigure[]
   {\includegraphics[scale=0.255]{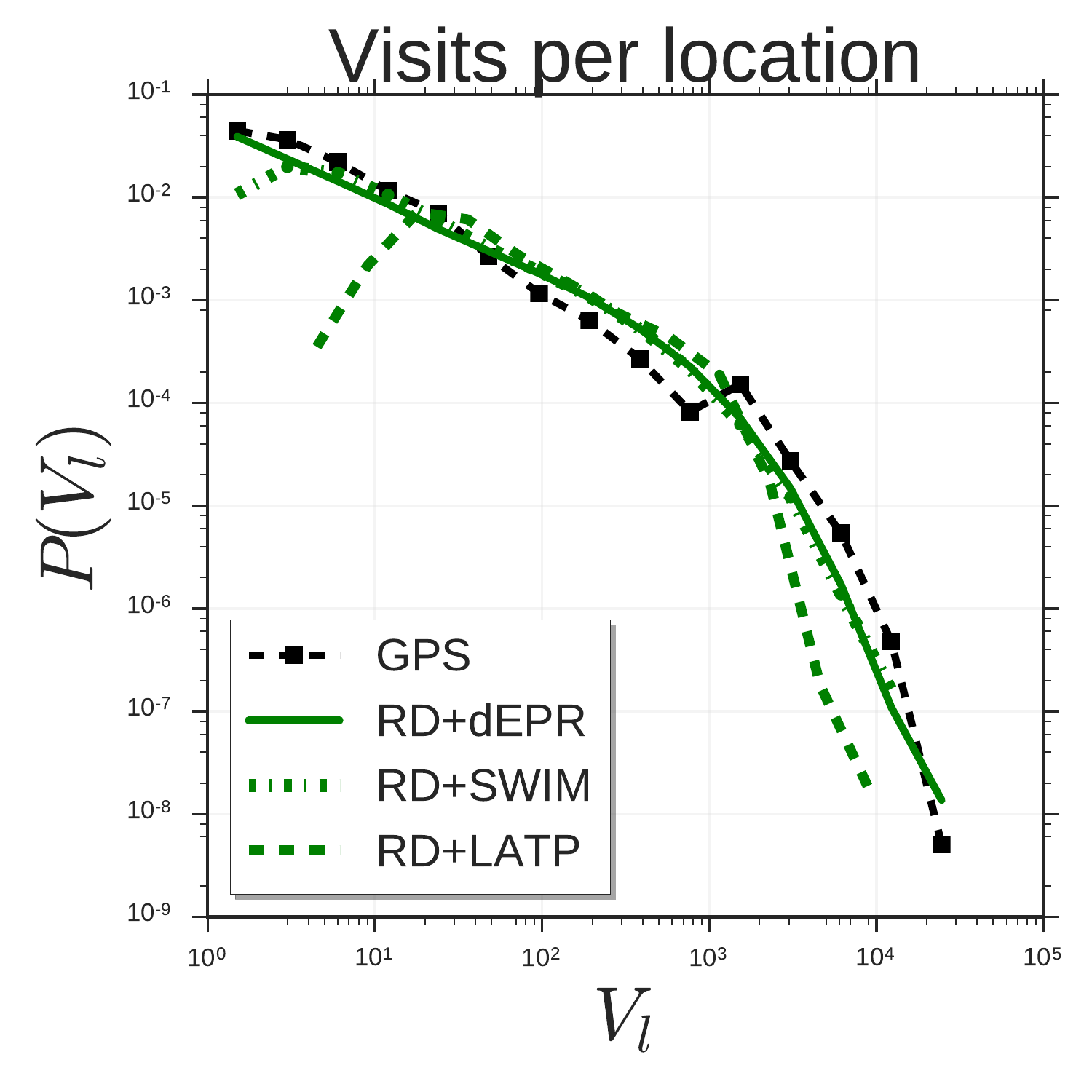}}
\subfigure[]
   {\includegraphics[scale=0.255]{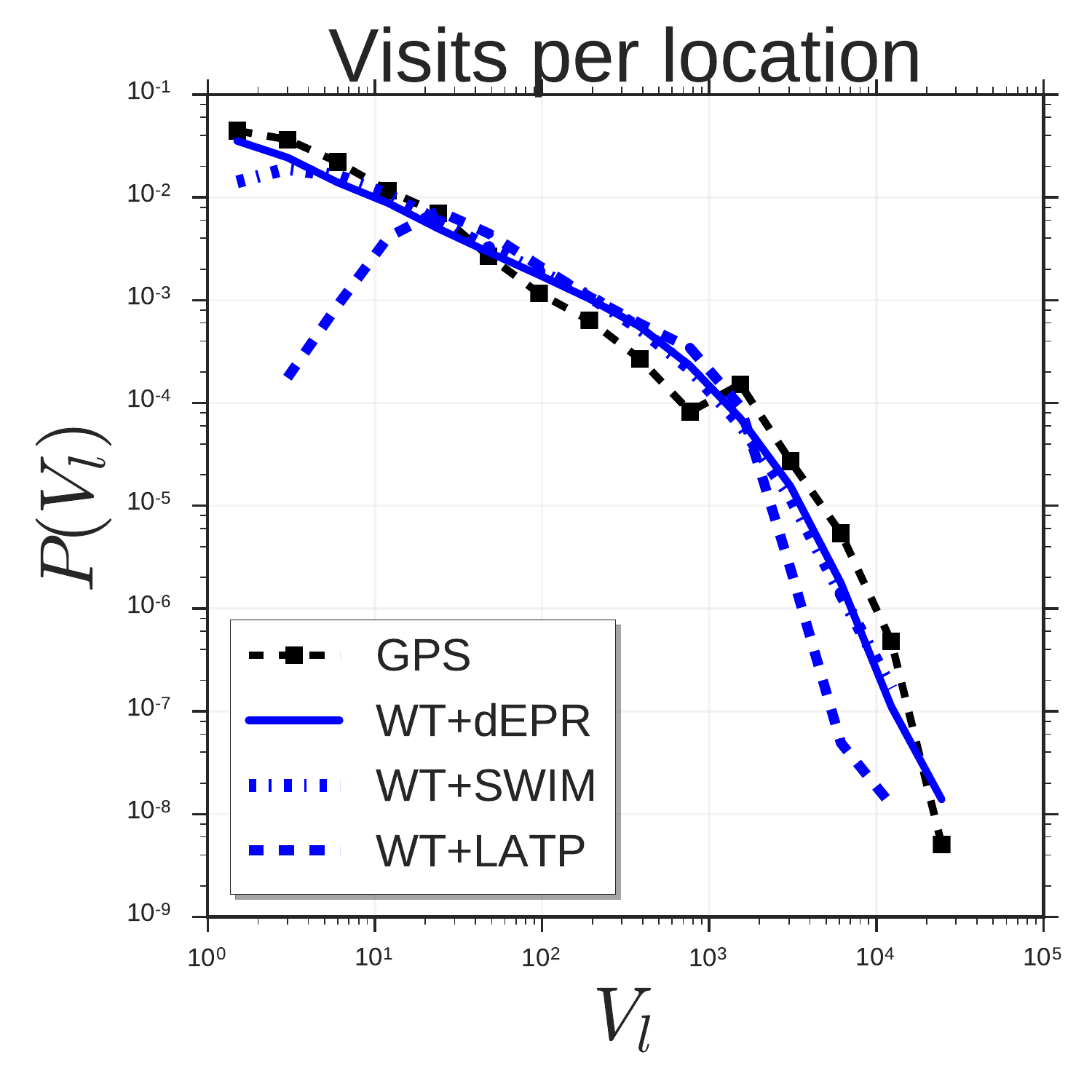}}

\subfigure[]
   {\includegraphics[scale=0.255]{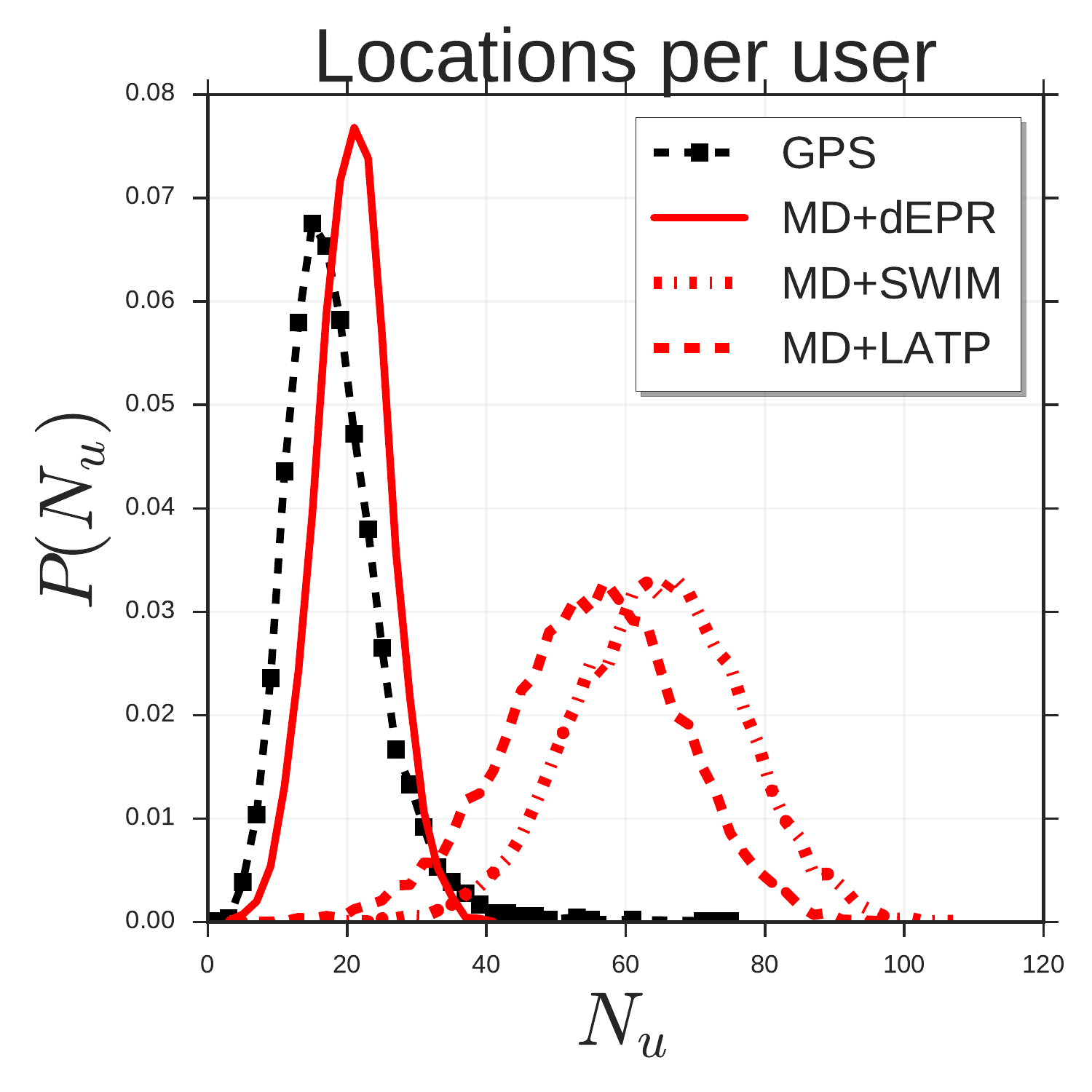}}
\subfigure[]
   {\includegraphics[scale=0.255]{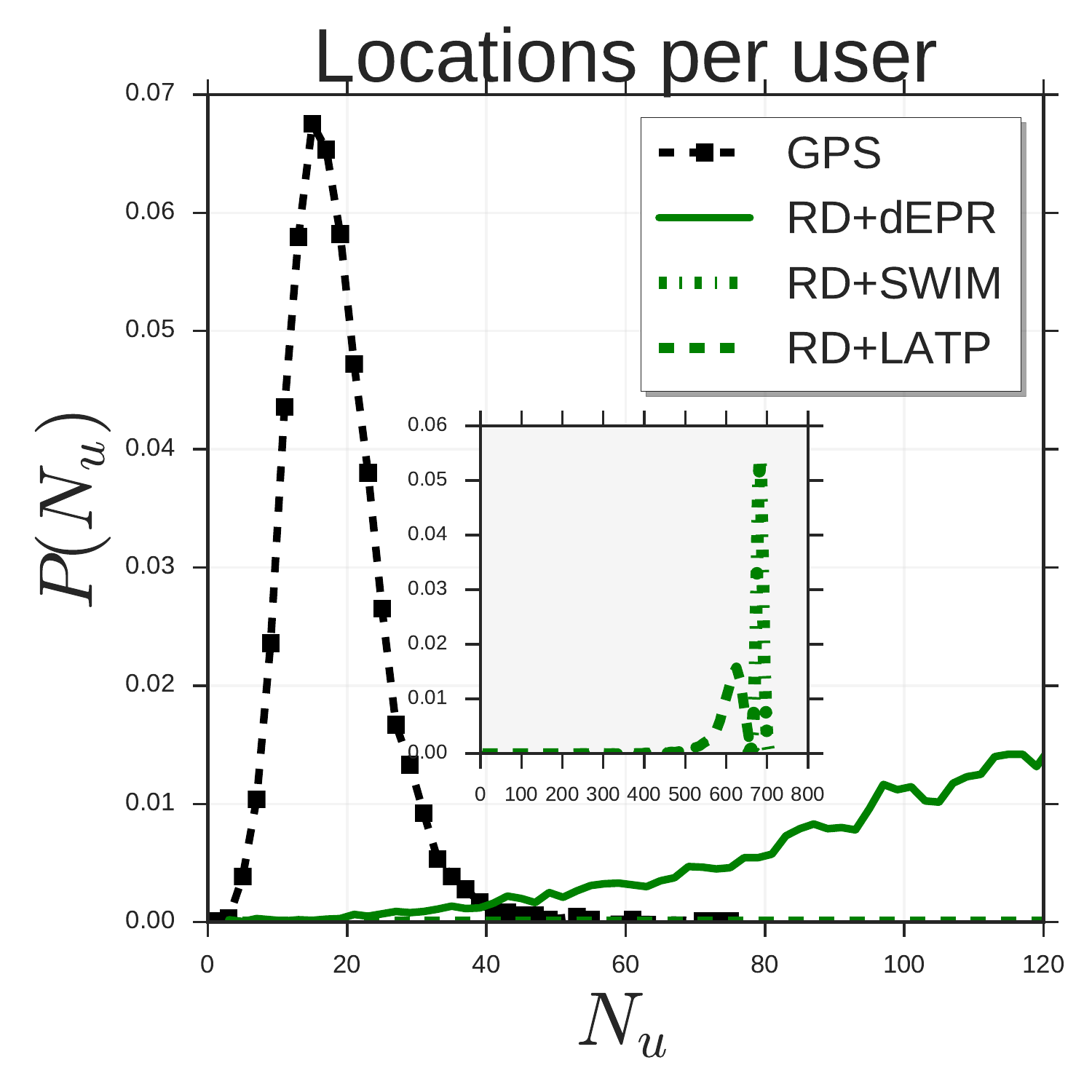}}
\subfigure[]
   {\includegraphics[scale=0.255]{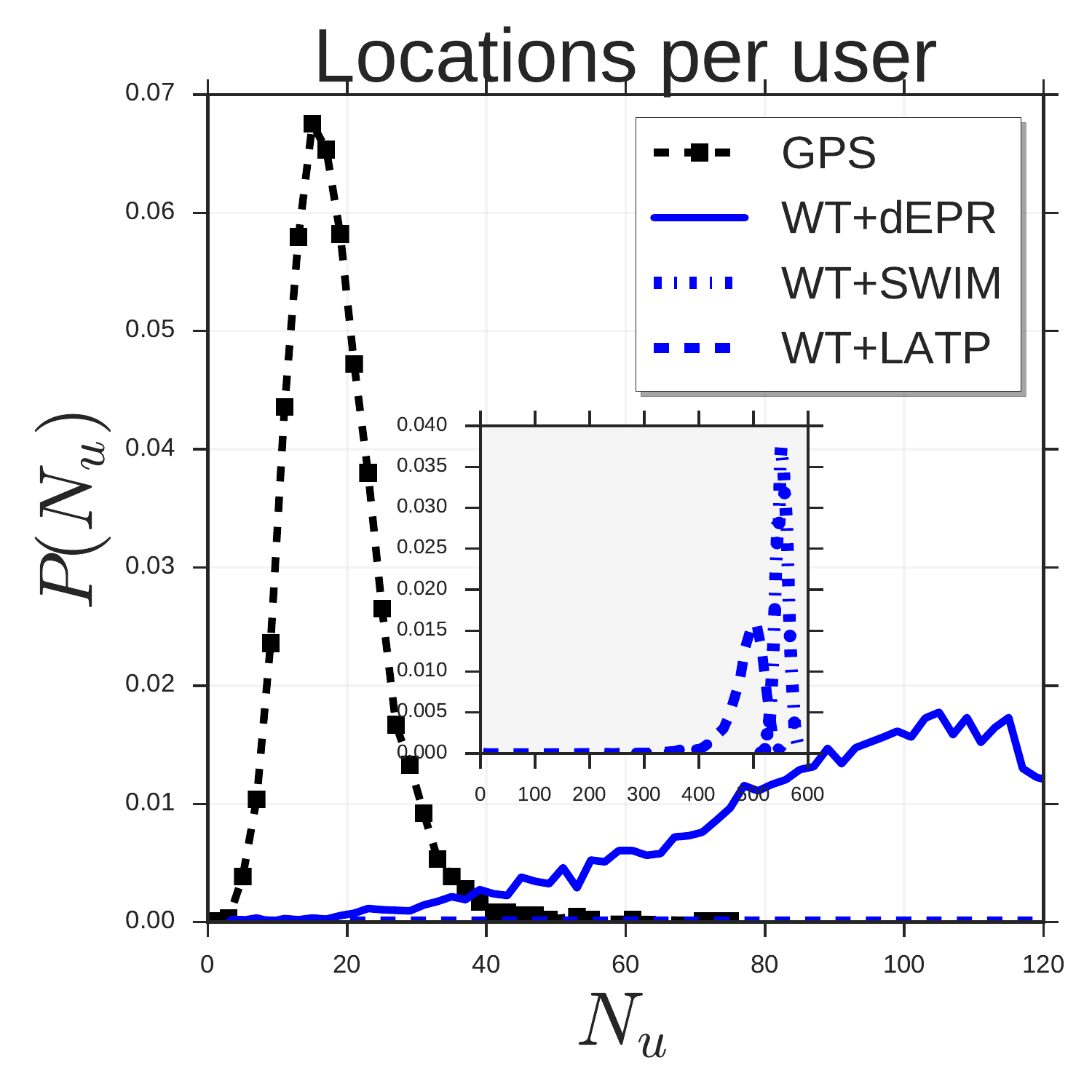}}

\caption{\textbf{Distributions of human mobility patterns (GPS)}. The figure compares the models and GPS data on location frequency, visits per location and locations per users. 
}
\label{fig:temporal_toscana}
\end{figure}

\paragraph{Location frequency.} Another important characteristic of an individual's mobility is the probability of visiting a location given the location's rank. The rank of a location depends on the number of times the individual visits the locations over the period of observation. For instance, rank 1 represents the most visited location (generally home place); rank 2 the second most visited location (e.g., work place) and so on. We compute the frequency of each of these ranked locations for every individual and plot the distribution of frequencies $f(L_i)$ in Fig.\ \ref{fig:temporal}a-c (CDR) and Fig.\ \ref{fig:temporal_toscana}a-c (GPS). For CDR data, we observe that $d$-EPR$^{\mbox{\tiny (CDR)}}_{\mbox{\small MD}}$ reproduces the shape of $f(L_i)$ (with RMSE=0.0122 and KL=0.12) better than SWIM$^{\mbox{\tiny (CDR)}}_{\mbox{\small MD}}$ and LATP$^{\mbox{\tiny (CDR)}}_{\mbox{\small MD}}$ (which have RMSE=0.0669, KL=1.2892 and RMSE=0.0626, KL=0.9353 respectively). If we change the diary generator in the model, $d$-EPR$^{\mbox{\tiny (CDR)}}_{\mbox{\small \{RD, WT\}}}$ underestimate the frequency of the top-ranked location and slightly overestimate the frequency of the less visited locations with respect to CDR data (Fig.\ \ref{fig:temporal}b-c). A reason for this discrepancy is that RD and WT do not take into account the circadian rhythm of individuals, hence underestimating the number of returns to the most frequent location (usually the home place). In SWIM$^{\mbox{\tiny (CDR)}}_{\mbox{\small MD}}$ and LATP$^{\mbox{\tiny (CDR)}}_{\mbox{\small MD}}$, the absence of a preferential return mechanism produce a more uniform distribution of location frequencies (Fig.\ \ref{fig:temporal}b-c), which is further exacerbated for SWIM$^{\mbox{\tiny (CDR)}}_{\mbox{\small \{RD, WT\}}}$ and LATP$^{\mbox{\tiny (CDR)}}_{\mbox{\small \{RD, WT\}}}$. Location frequency $f(L_i)$ is another case where the choice of the diary generator and the choice of the trajectory generator are both crucial to reproduce the shape of the distribution in an accurate way. Experiments on GPS data confirm results observed on CDRs (Fig.\ \ref{fig:temporal_toscana}a-c): model $d$-EPR$^{\mbox{\tiny (GPS)}}_{\mbox{\small MD}}$ produces the best fit with real data, while changing either the diary or the trajectory generators produces worse fits.

\begin{figure}[!htb]\centering
\textbf{\LARGE CDR}\par\medskip
\subfigure[]
   {\includegraphics[scale=0.255]{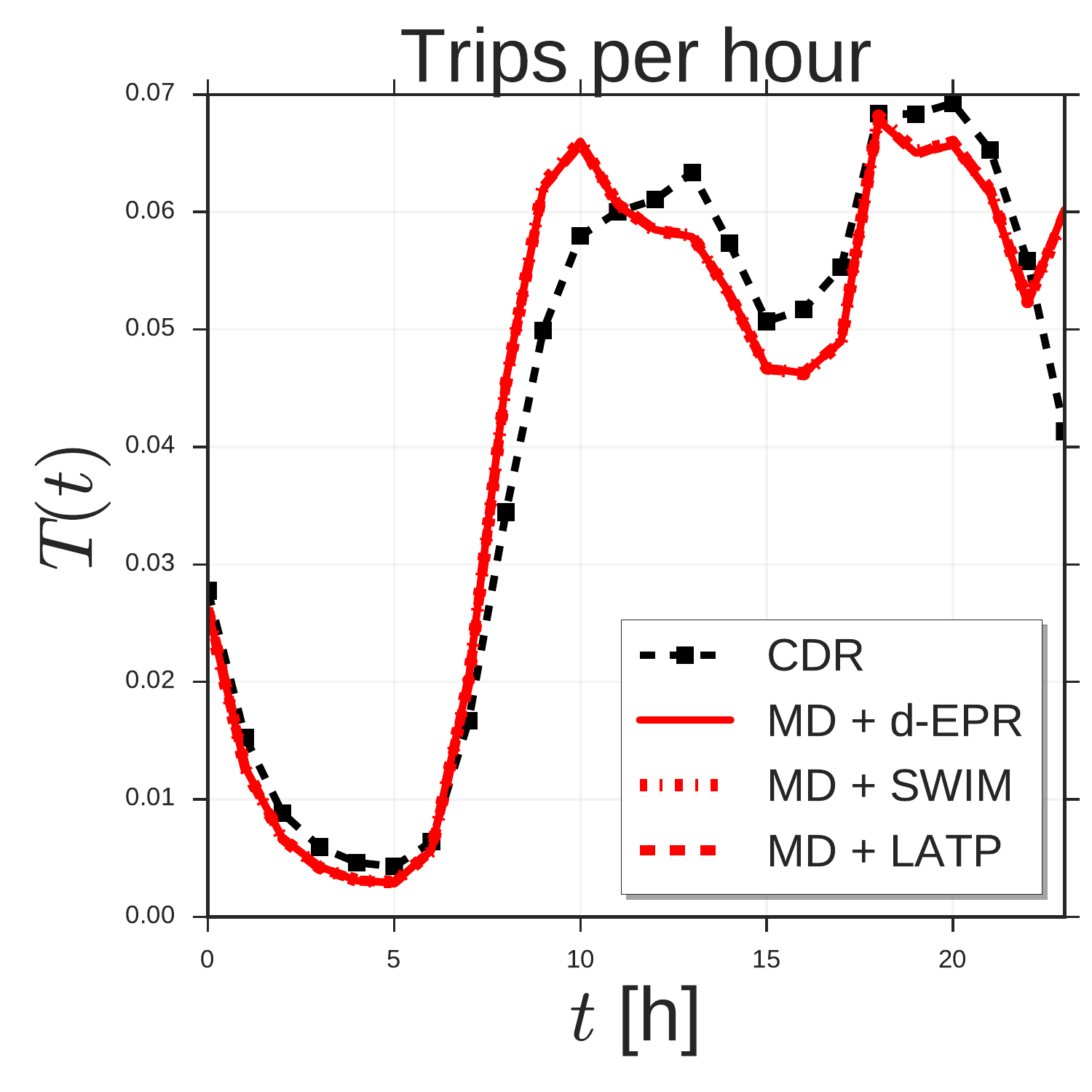}}
\subfigure[]
   {\includegraphics[scale=0.255]{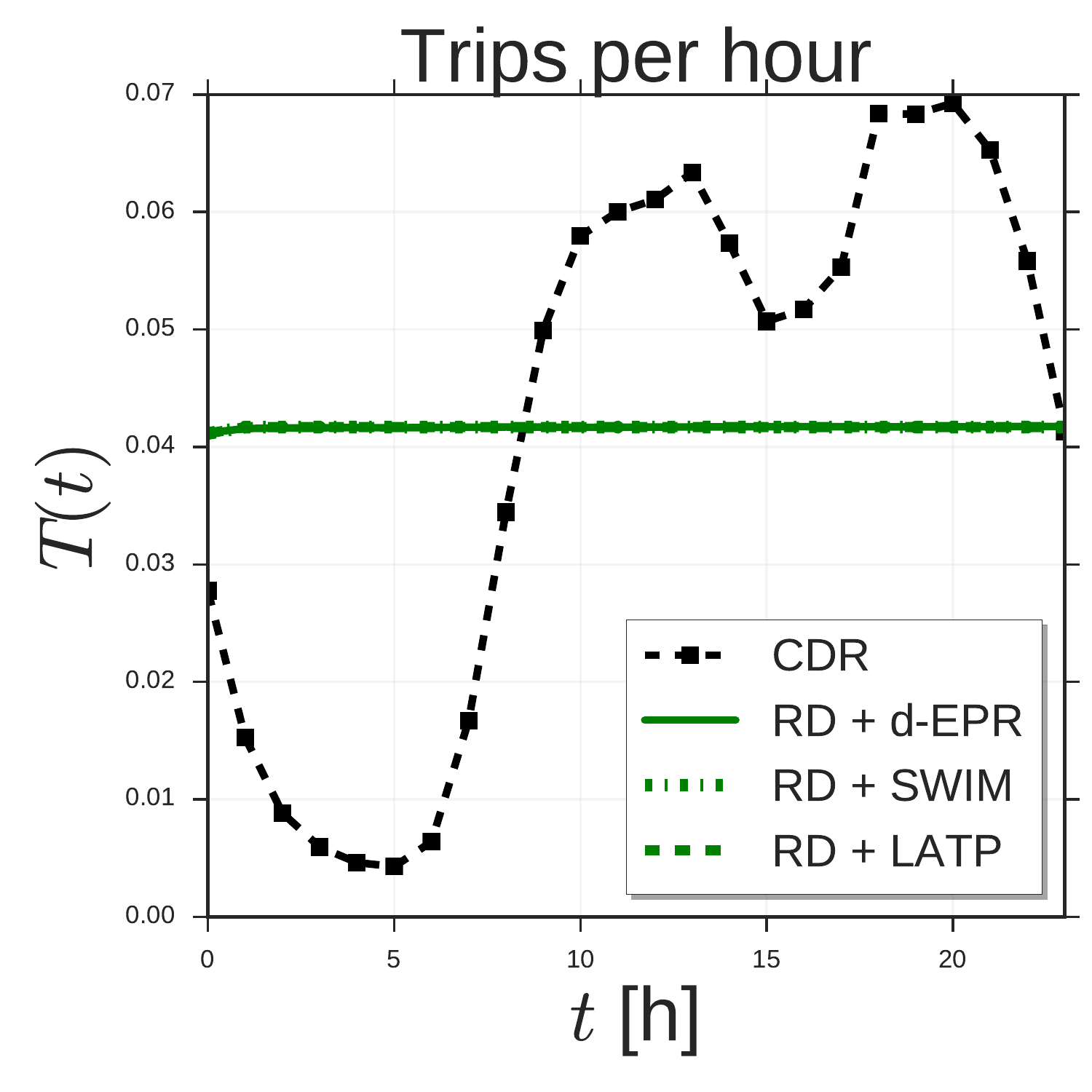}}
\subfigure[]
   {\includegraphics[scale=0.255]{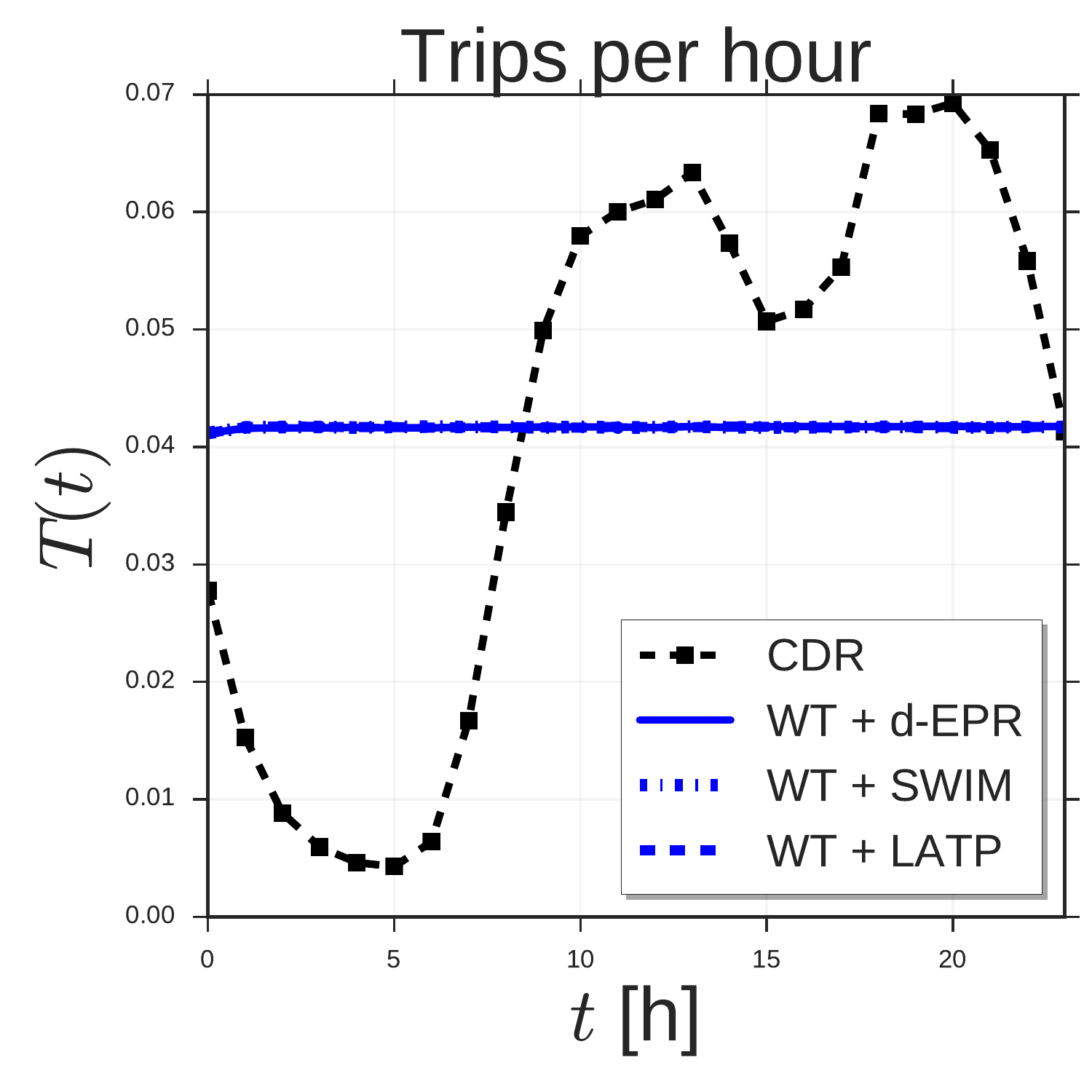}}
   
  \subfigure[]
   {\includegraphics[scale=0.255]{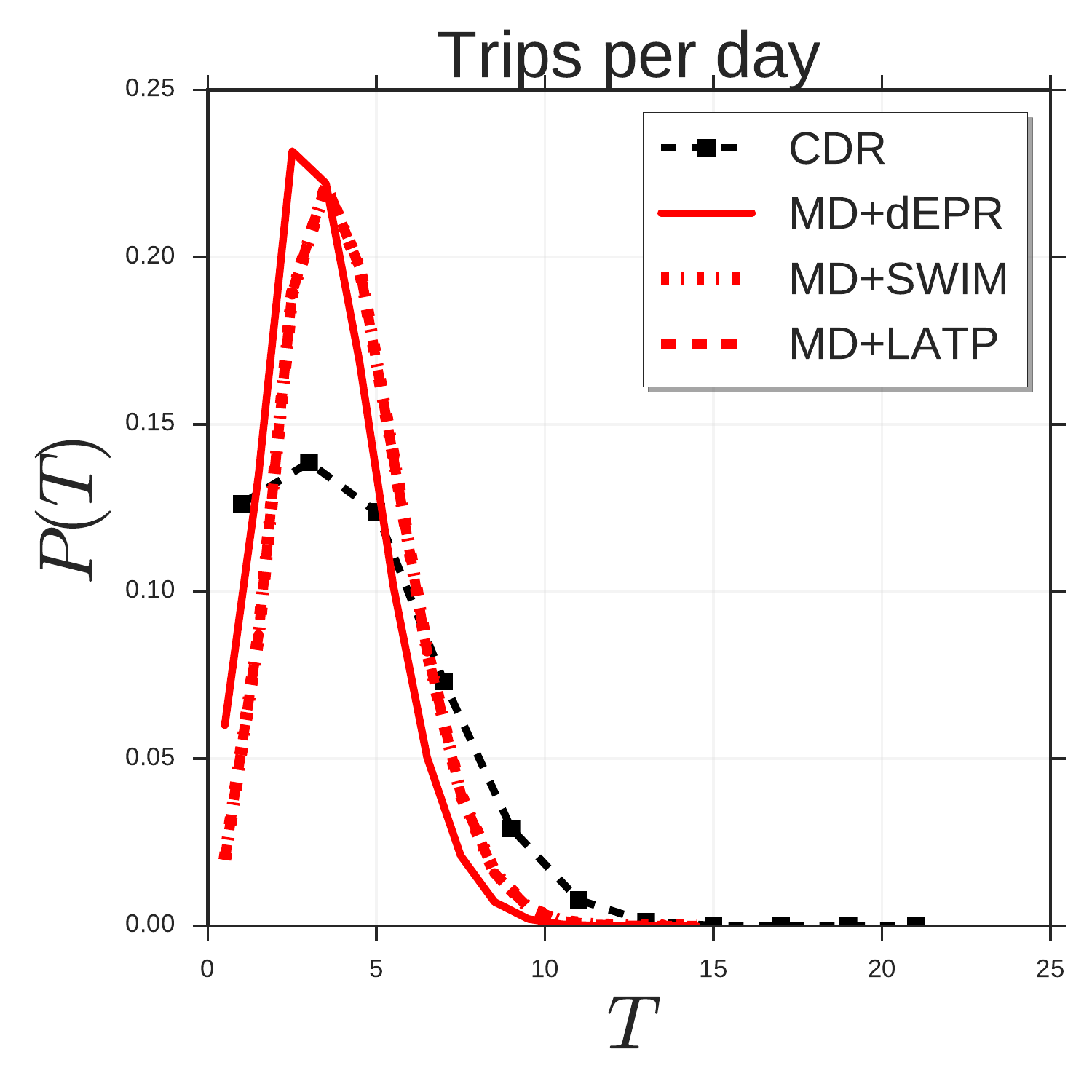}}
\subfigure[]
   {\includegraphics[scale=0.255]{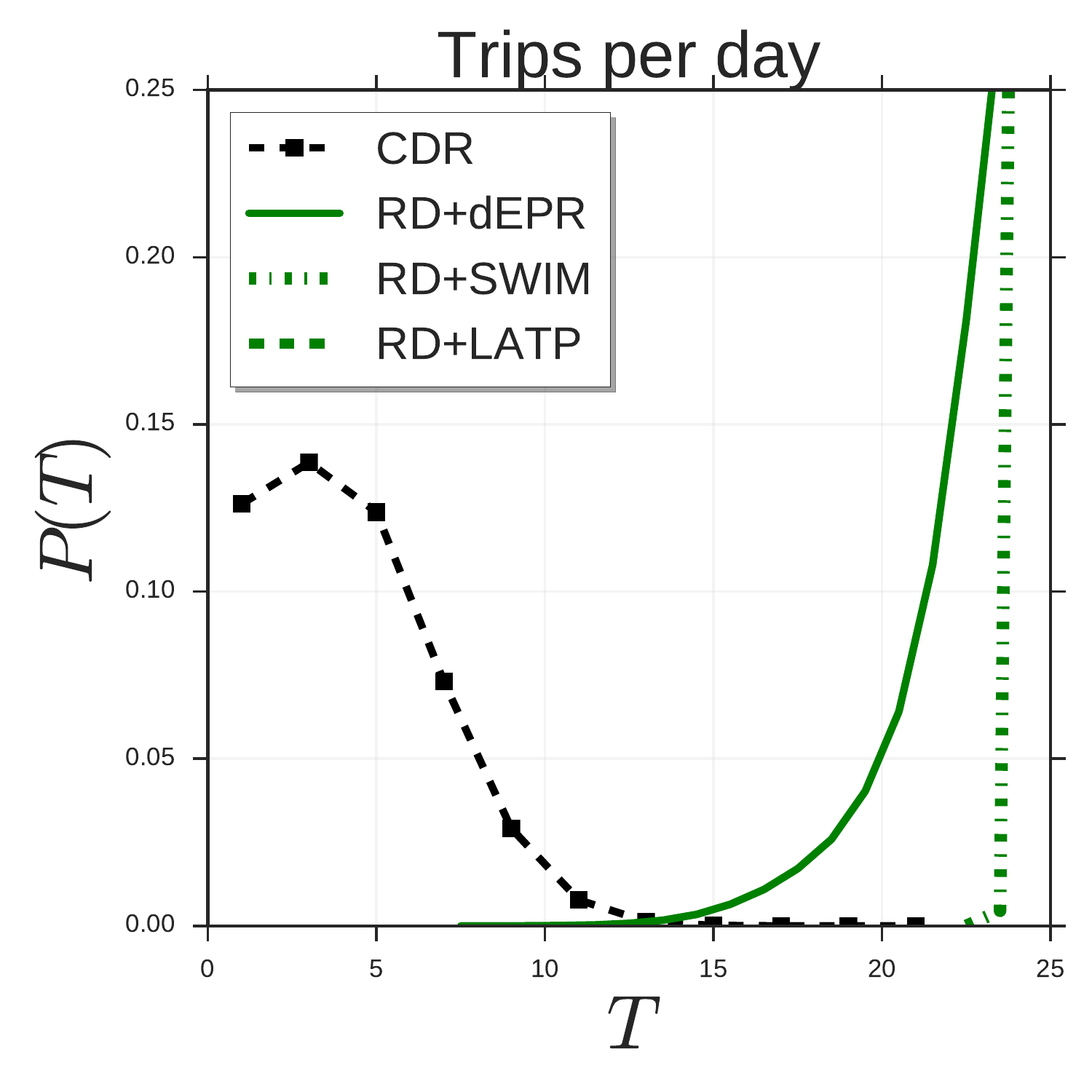}}
\subfigure[]
   {\includegraphics[scale=0.255]{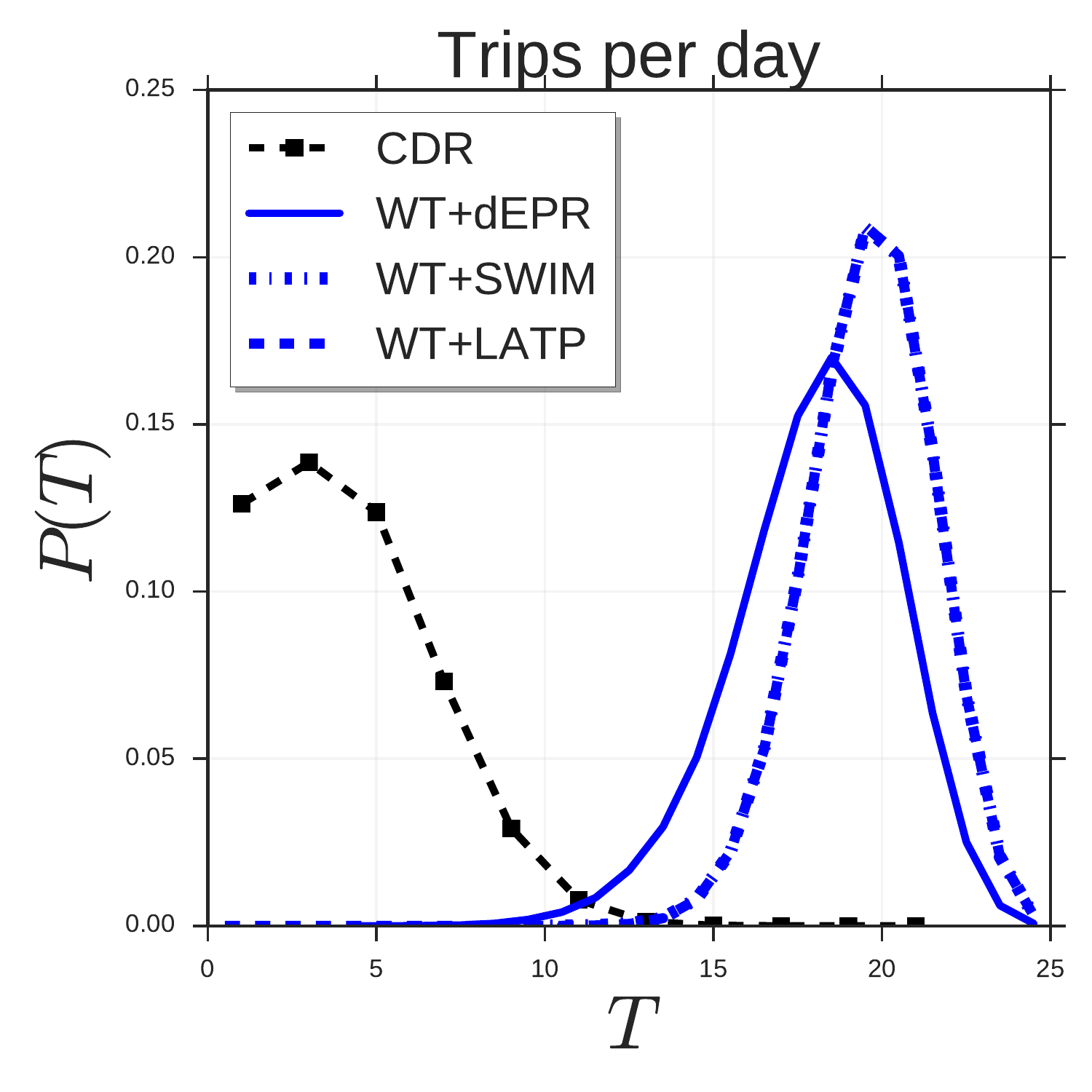}}

\subfigure[]
   {\includegraphics[scale=0.255]{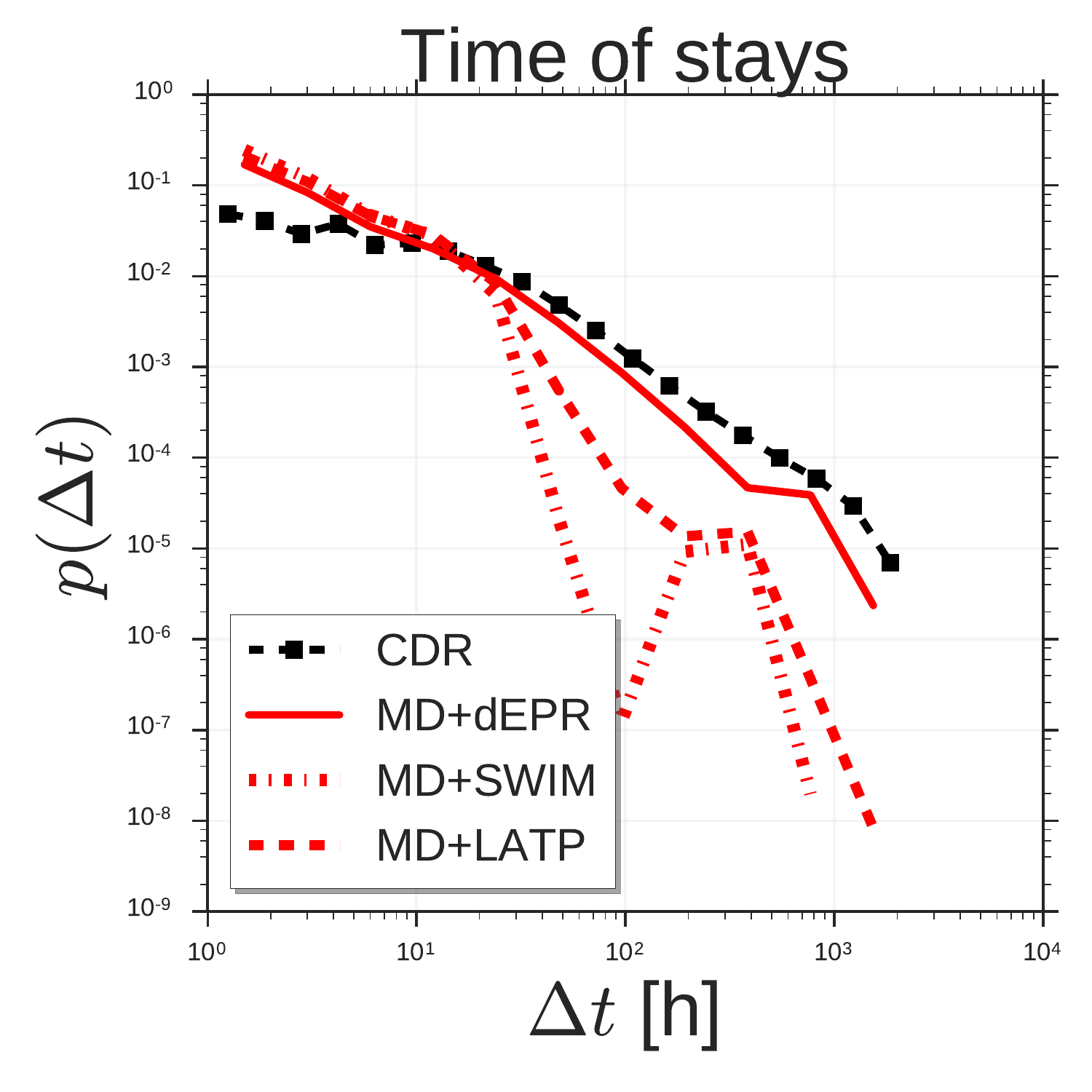}}
\subfigure[]
   {\includegraphics[scale=0.255]{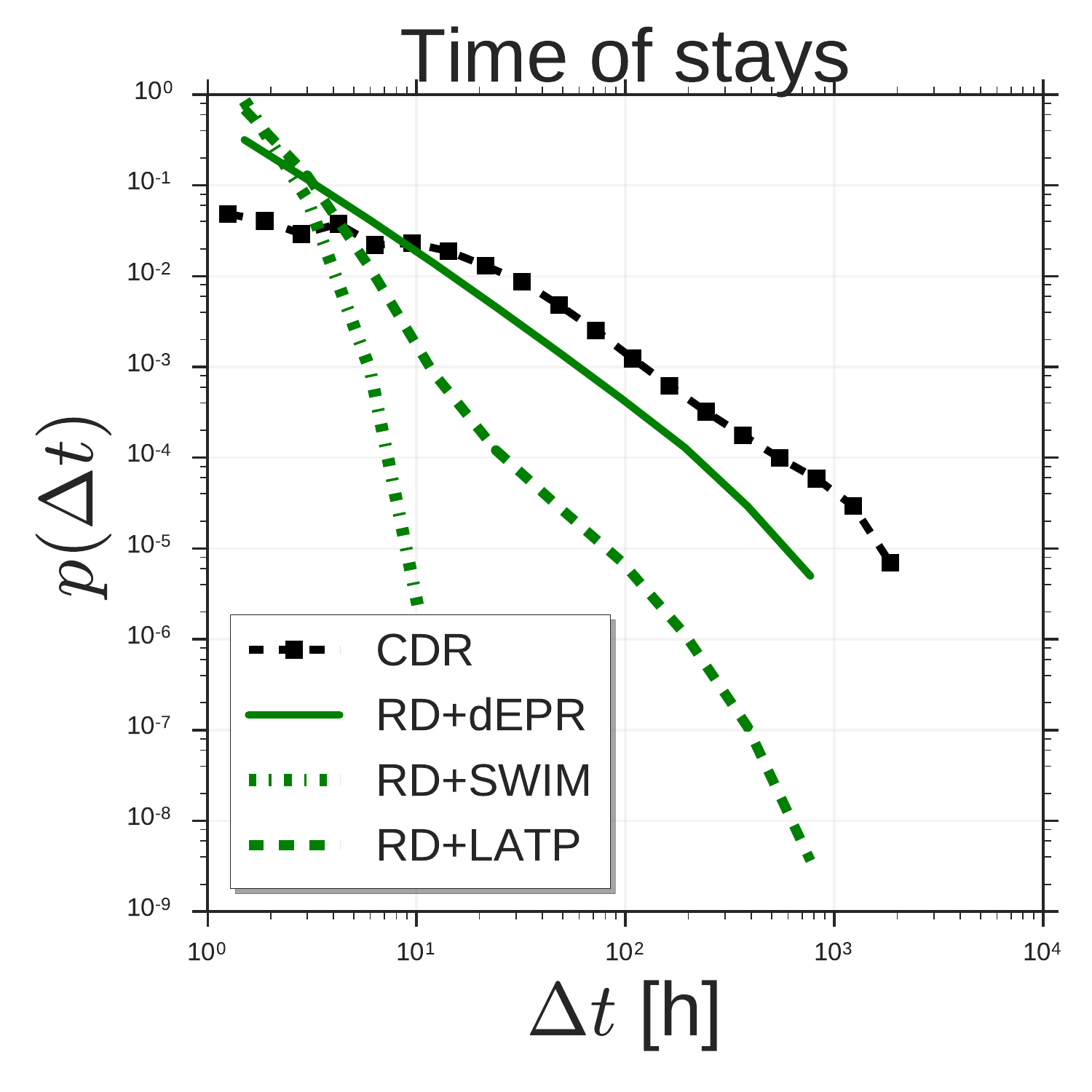}}
\subfigure[]
   {\includegraphics[scale=0.255]{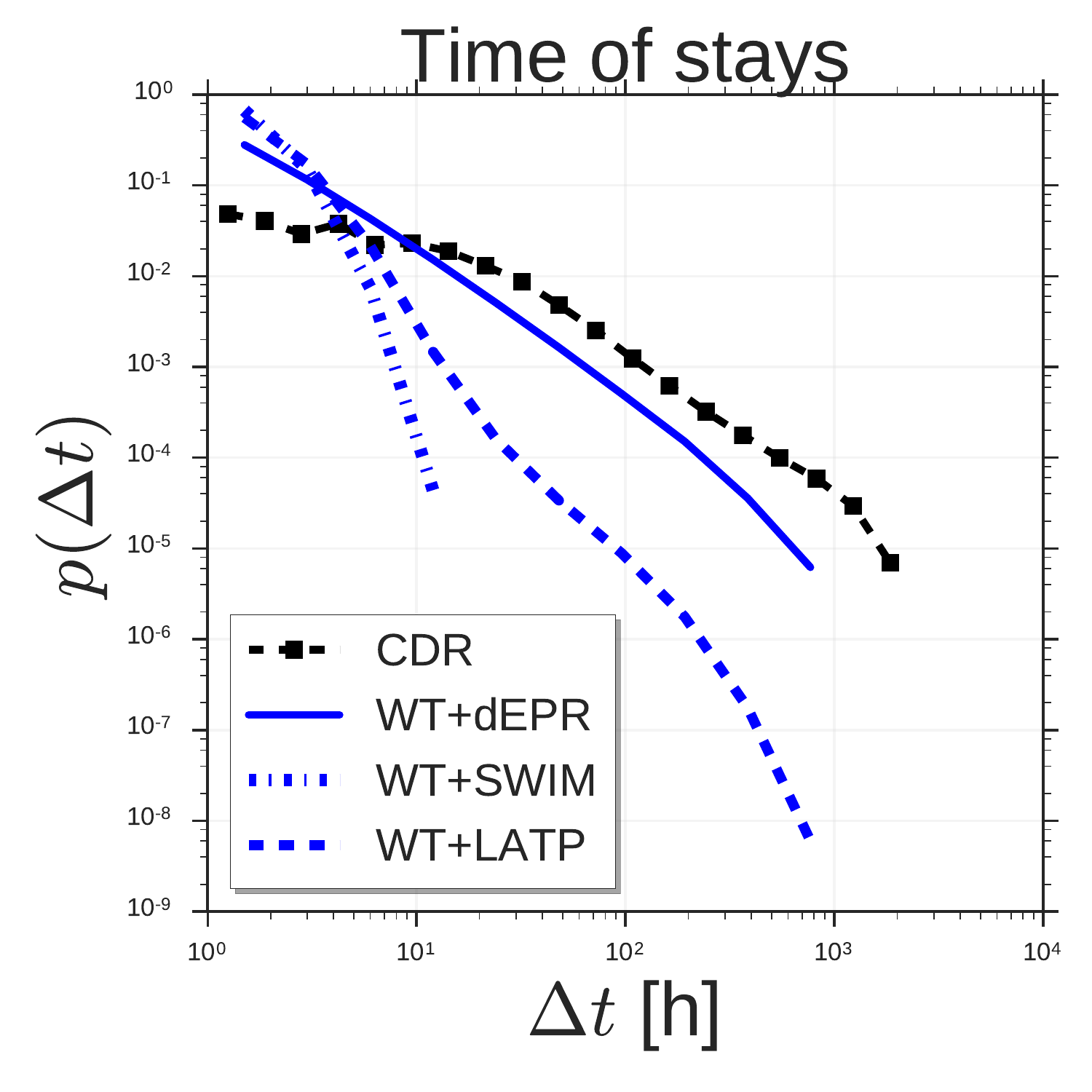}}
   
\caption{\textbf{Distributions of human mobility patterns (CDR)}. The figure compares the models and CDR data on trips per hour, trips per day and time of stays. Plots in (a), (b) and (c) show the distribution of the number $T$ of trips per hour of the day for $d$-EPR, SWIM and LATP used in combination with MD, RD and WT respectively. Plots in (d), (e) and (f) show the distribution of the number $D$ of trips per day, plots in (g), (h), (i) show the distribution of time of stays $\Delta t$.}
\label{fig:others}
\end{figure}

\begin{figure}[!htb]\centering
\textbf{\LARGE GPS}\par\medskip
\subfigure[]
   {\includegraphics[scale=0.255]{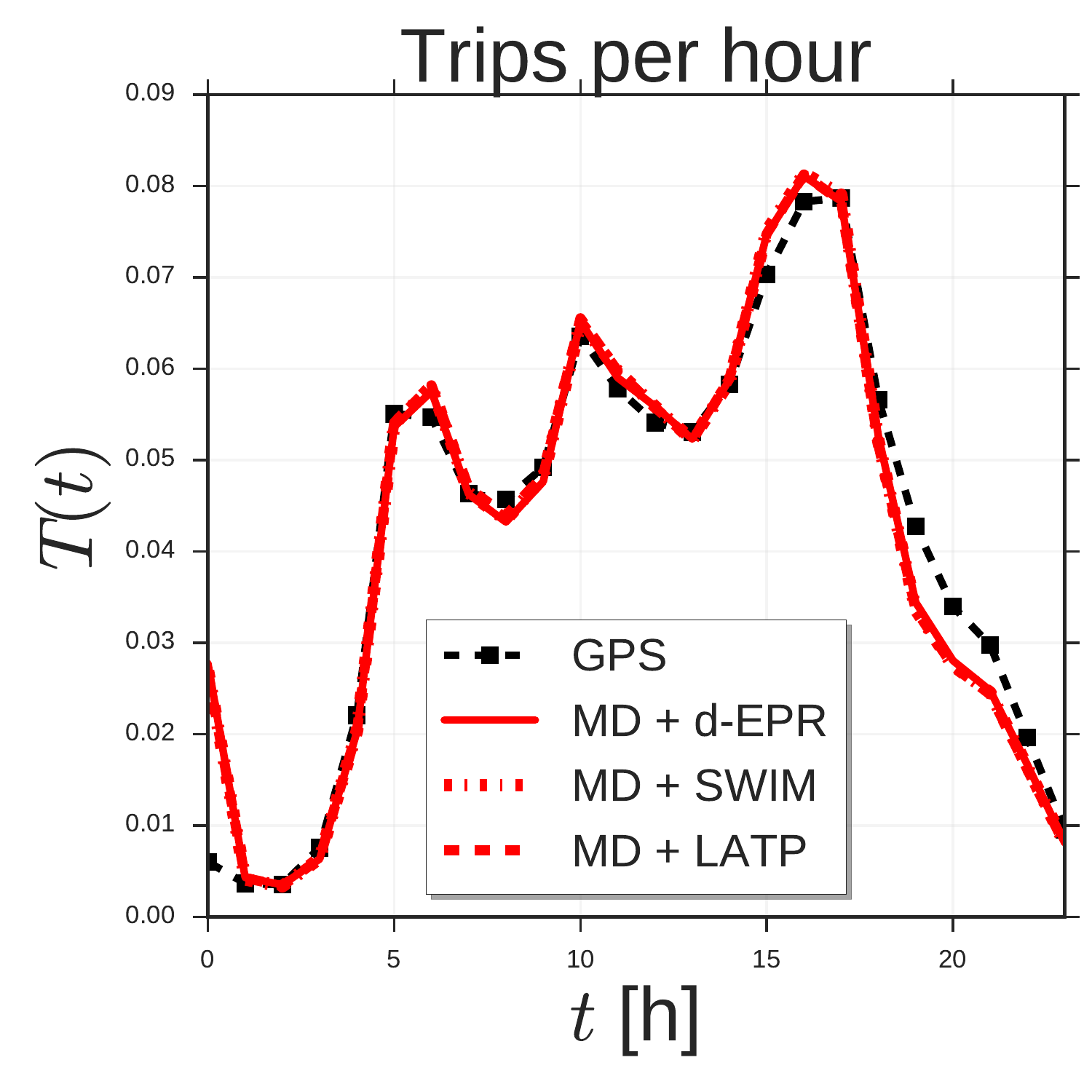}}
\subfigure[]
   {\includegraphics[scale=0.255]{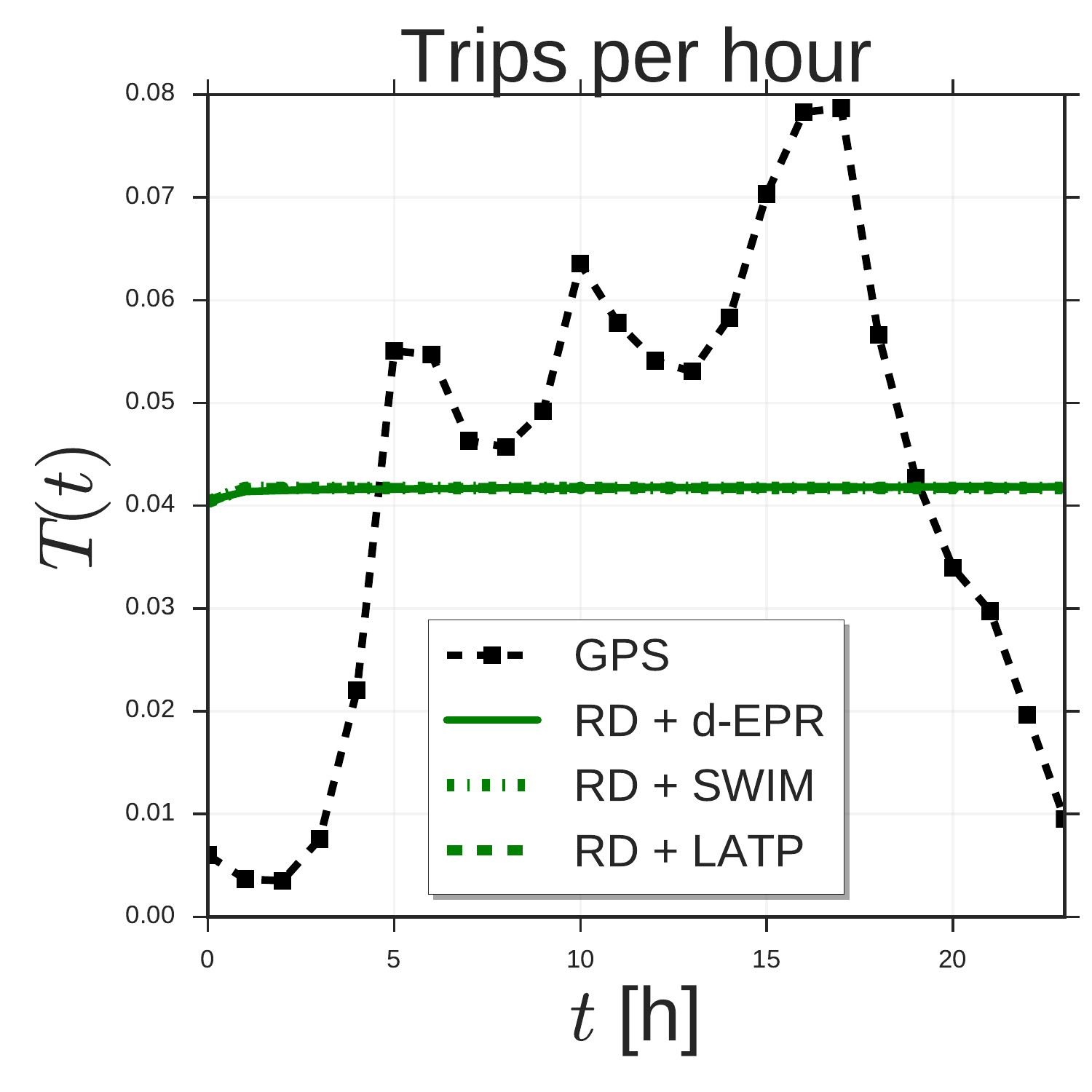}}
\subfigure[]
   {\includegraphics[scale=0.255]{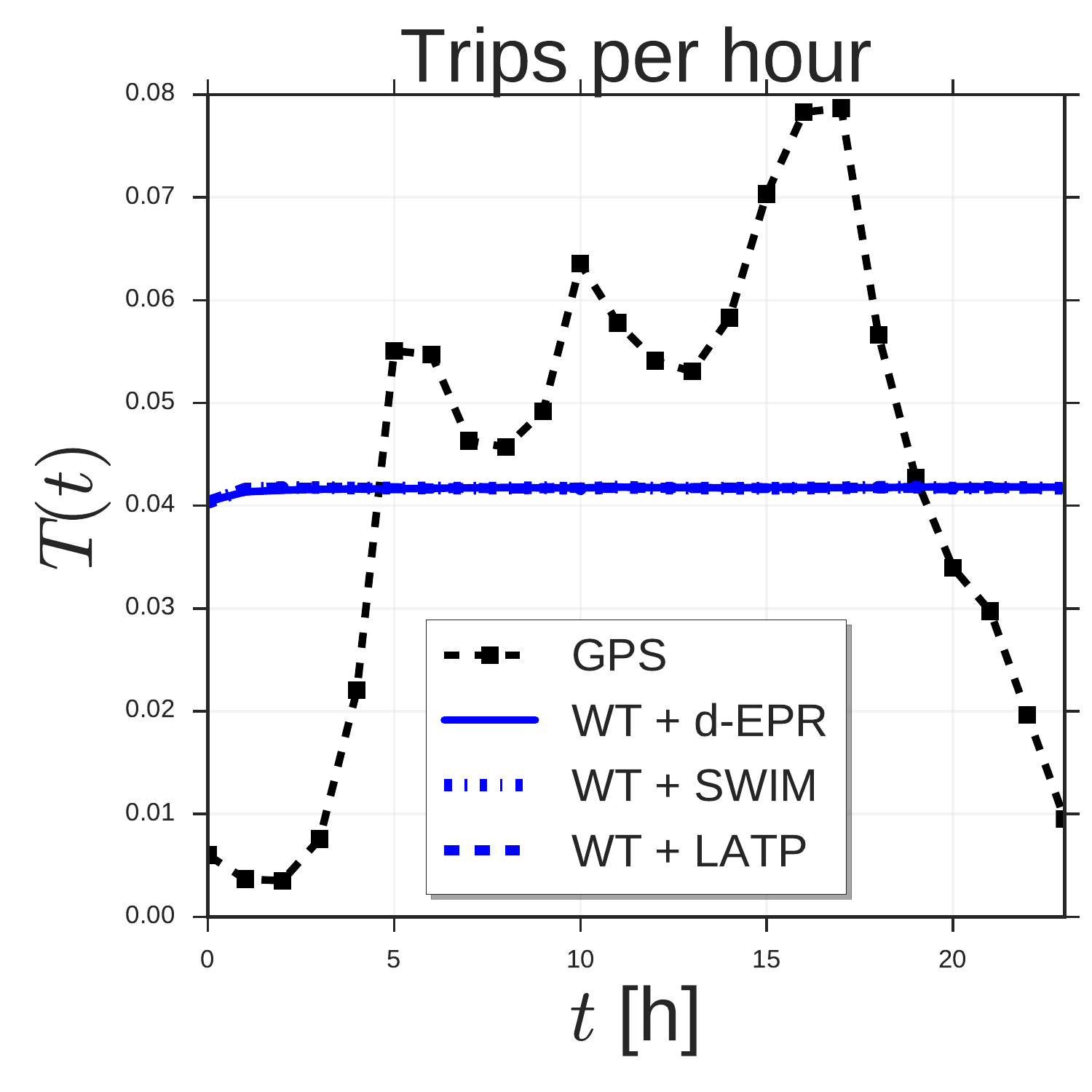}}
   
  \subfigure[]
   {\includegraphics[scale=0.255]{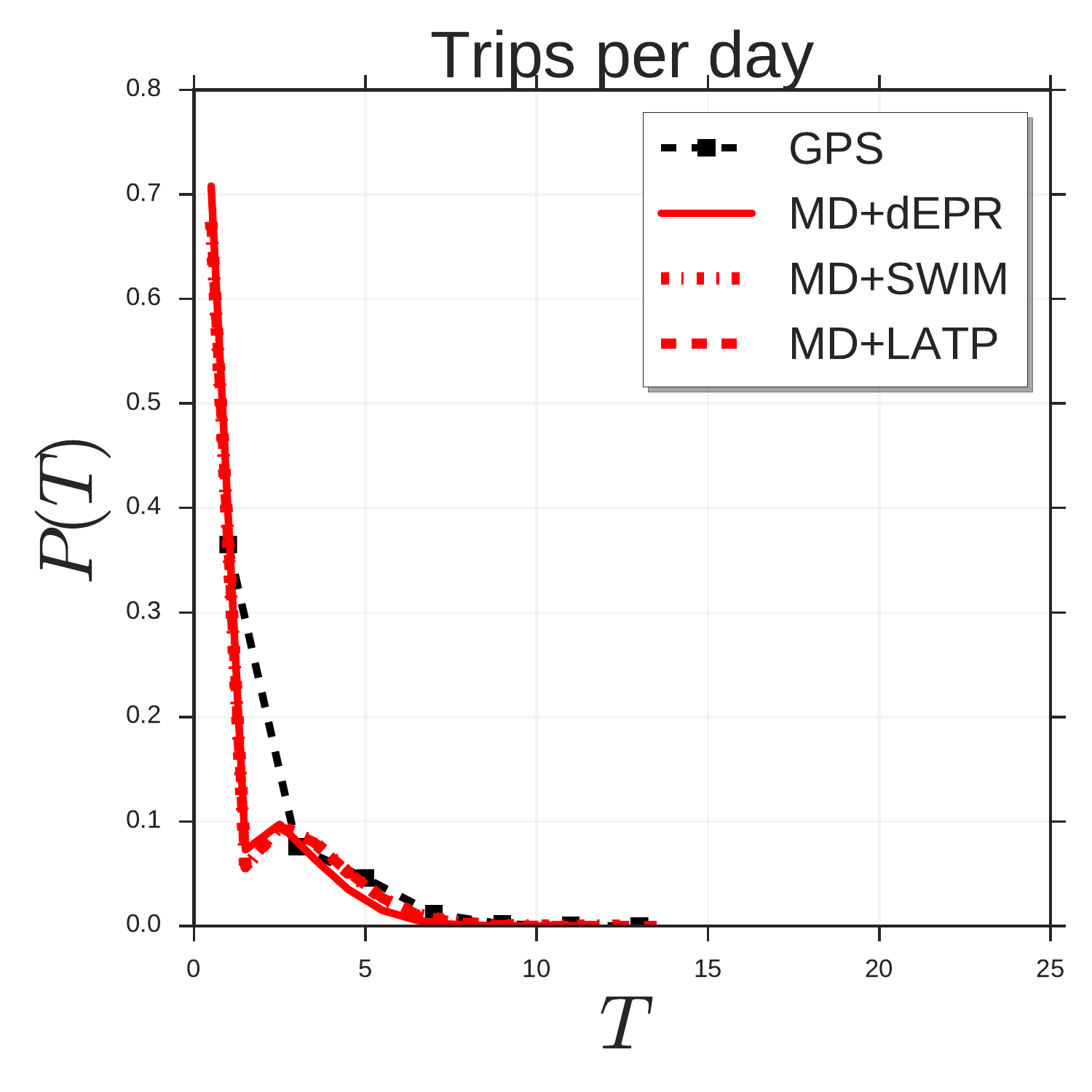}}
\subfigure[]
   {\includegraphics[scale=0.255]{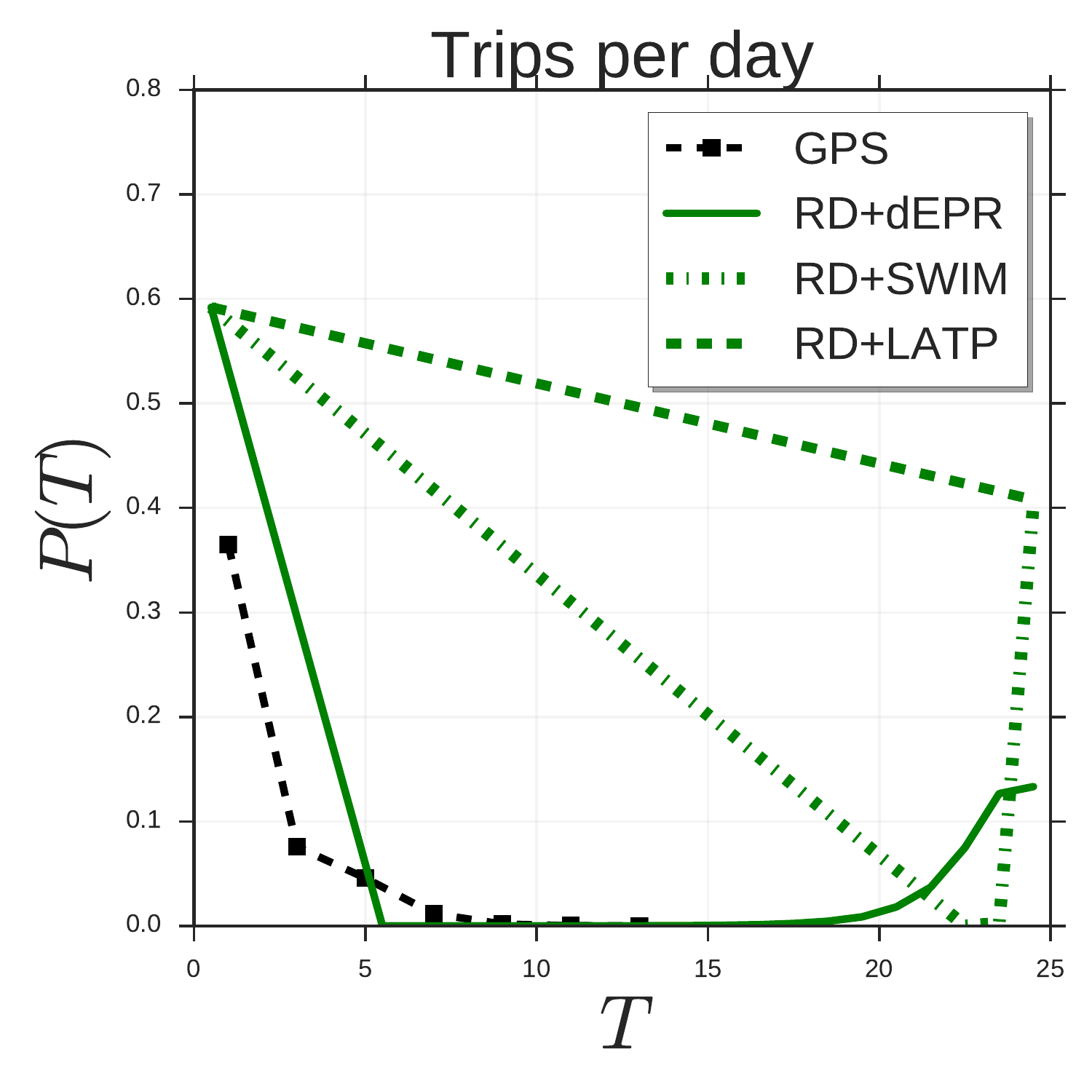}}
\subfigure[]
   {\includegraphics[scale=0.255]{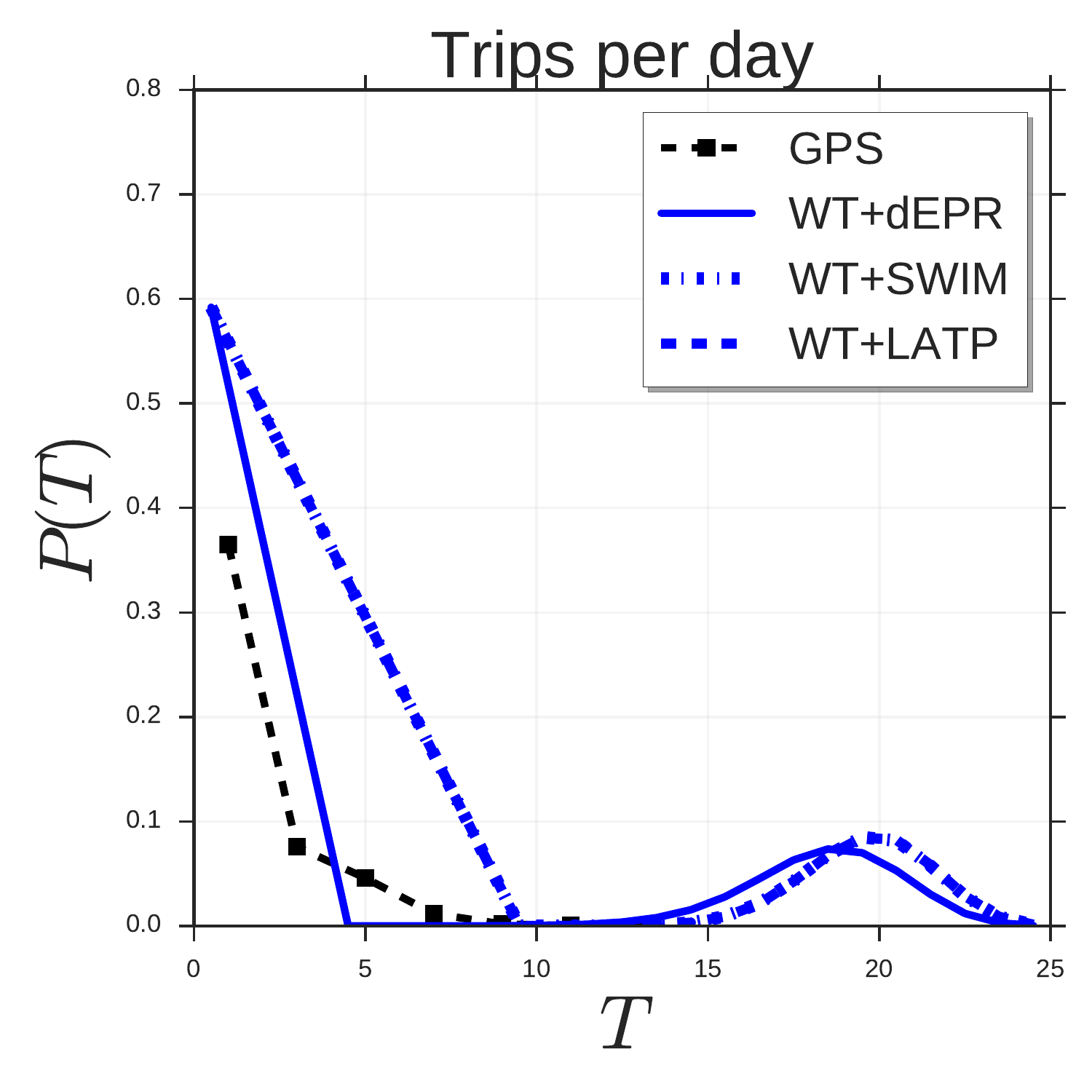}}
\subfigure[]
   {\includegraphics[scale=0.255]{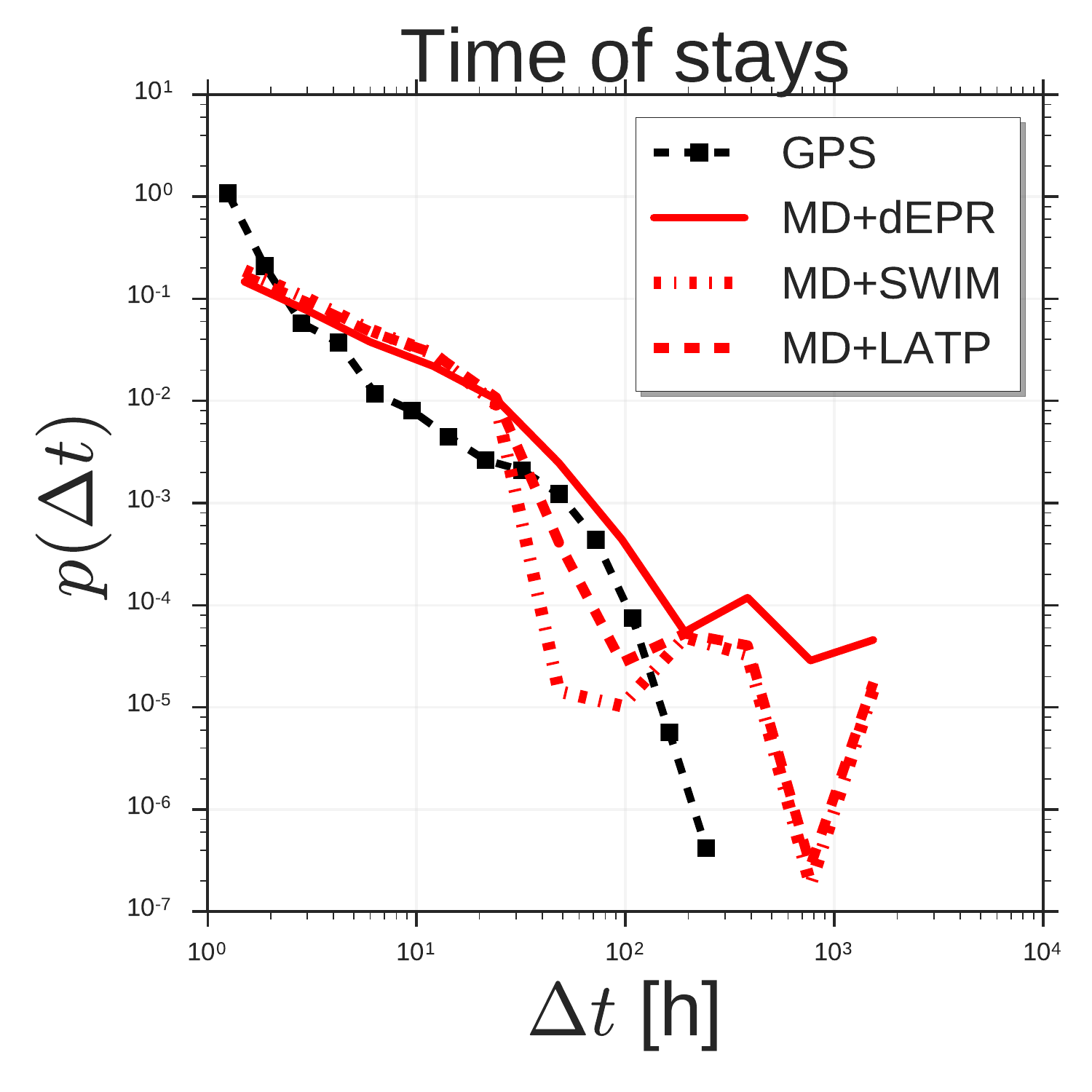}}
\subfigure[]
   {\includegraphics[scale=0.255]{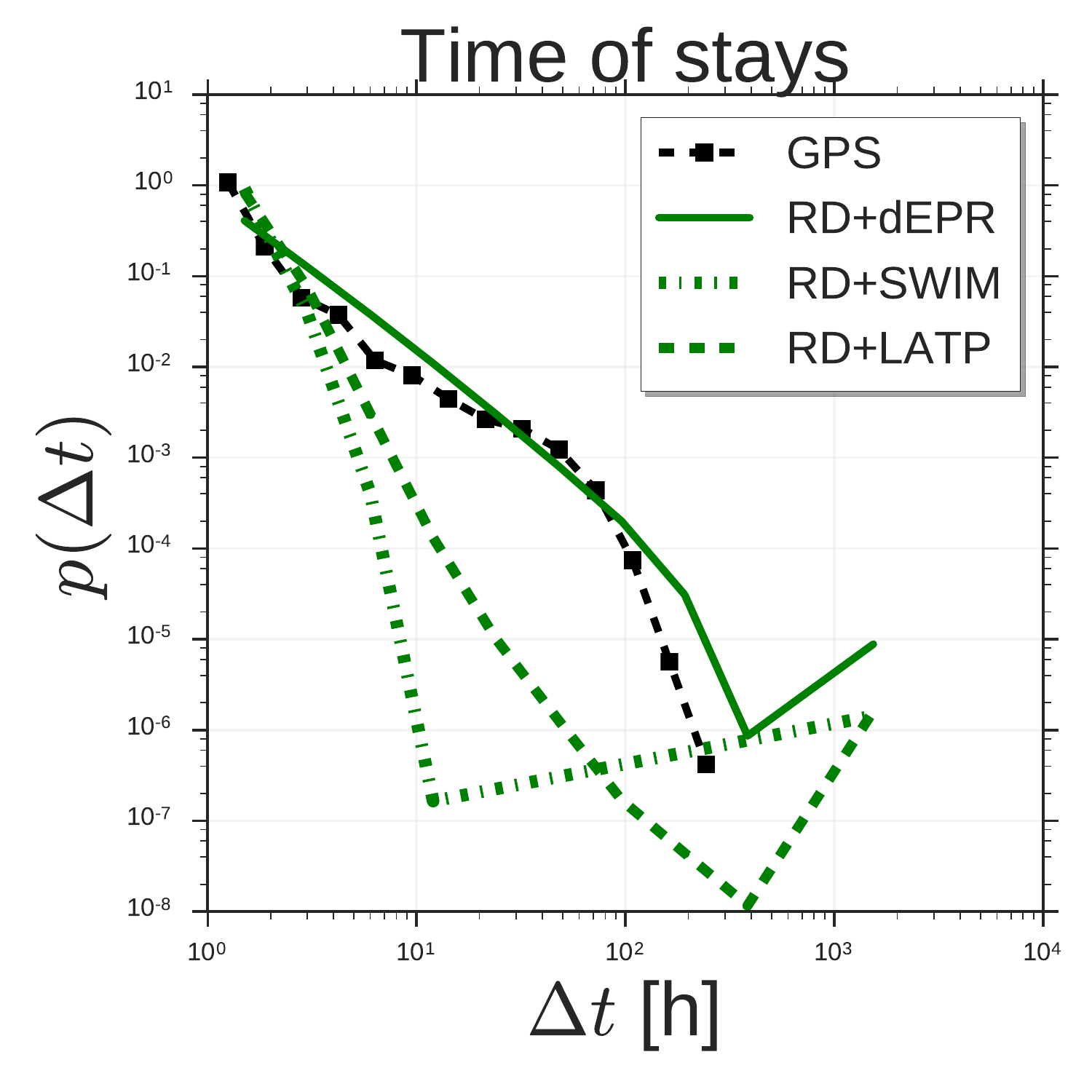}}
\subfigure[]
   {\includegraphics[scale=0.255]{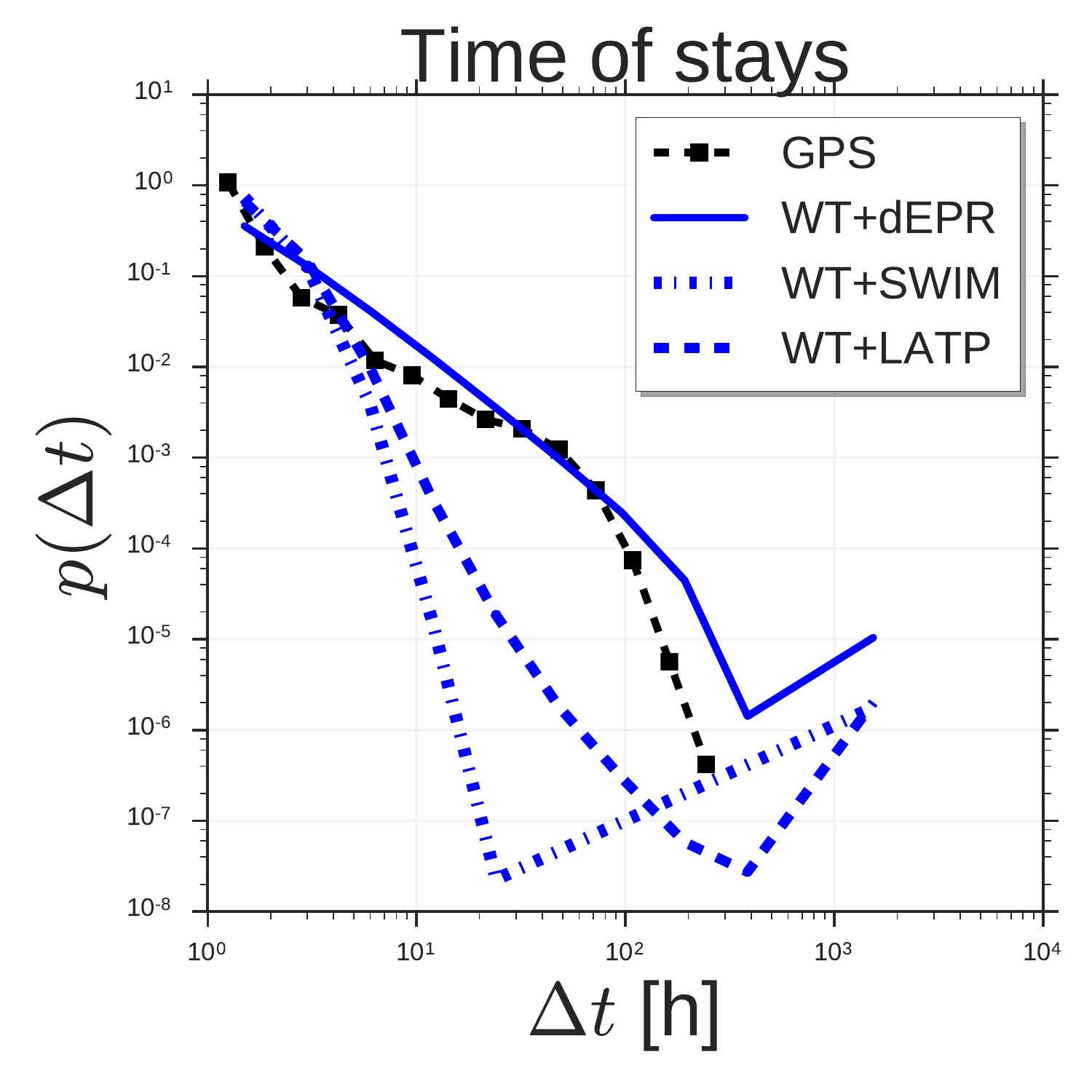}}
   
\caption{\textbf{Distributions of human mobility patterns (GPS)}. The figure compares the models and GPS data on trips per hour, trips per day and time of stays.  
}
\label{fig:others_toscana}
\end{figure}

\paragraph{Visits per location.} A useful measure to understand how a set of individuals exploit the mobility space is the number $V$ of overall visits per location, i.e., the total number of visits by all the individuals in every location during the period of observation. For every dataset, we compute the number of visits for every location of the weighted spatial tessellation and plot the distribution $P(V)$ in Fig.\ \ref{fig:others}d-f (CDR) and Fig.\ \ref{fig:others_toscana}d-f (GPS). As for CDR data, $d$-EPR$^{\mbox{\tiny (CDR)}}_{\mbox{\small MD}}$ produces a $P(V)$ which follows a heavy tail distribution: the majority of locations have just one visit while a minority of locations have up to several thousands visits during the 11 weeks. The value of $V$ of a location depends on two factors: (i) its relevance in the weighted spatial tessellation; (ii) its position in the weighted spatial tessellation. The higher the relevance of a location in the weighted spatial tessellation, the higher is the probability for the location to be visited in the exploration mechanisms of $d$-EPR and SWIM. Indeed, from Fig.\ \ref{fig:others}e-f we observe that $d$-EPR and SWIM are the models which better fit $P(V)$. In contrast LATP does not take into account the relevance of a location during the exploration being unable to capture the shape of $P(V)$. Experiments on GPS data substantially confirm these results (Fig.\ \ref{fig:others_toscana}d-f): $d$-EPR and SWIM generates the most realistic distributions of $P(V)$.

\paragraph{Locations per user.} The number $N_u$ of distinct locations visited by an individual during the period of observation describes the degree of exploration of an individual, i.e., how the single individuals exploit the mobility space. In Fig.\ \ref{fig:temporal}g we observe that the MD models do not capture the shape of $P(N_u)$ in CDR data: the average number of distinct locations $\overline{N}$ according to $d$-EPR$^{\mbox{\tiny (CDR)}}_{\mbox{\small MD}}$ is about twice $\overline{N}$ in CDR data, while SWIM$^{\mbox{\tiny (CDR)}}_{\mbox{\small MD}}$ and LATP$^{\mbox{\tiny (CDR)}}_{\mbox{\small MD}}$ produce distributions whose $\overline{N}$ is more than ten times $\overline{N}$ in CDR data. By changing diary generator (Fig.\ \ref{fig:temporal}h-i) the difference with CDR data becomes even larger: $d$-EPR$^{\mbox{\tiny (CDR)}}_{\mbox{\small \{RD, WT\}}}$ produce a much broader variance of $P(N_u)$, SWIM$^{\mbox{\tiny (CDR)}}_{\mbox{\small \{RD, WT\}}}$ and LATP$^{\mbox{\tiny (CDR)}}_{\mbox{\small \{RD, WT\}}}$ predict a number of distinct visited locations very far from CDR data. These results suggest that the considered models overestimate the degree of exploration of individuals. In the case of $d$-EPR$^{\mbox{\tiny (CDR)}}_{\mbox{\small MD}}$ the overestimation may depend on the distribution of time of stays, as the distribution of time stays $P(\Delta t)$ produced by $d$-EPR$^{\mbox{\tiny (CDR)}}_{\mbox{\small MD}}$ overestimates the number of short stay times, leading to a larger total number of visited locations (Fig.\ \ref{fig:others}g). For GPS data, model $d$-EPR$^{\mbox{\tiny (GPS)}}_{\mbox{\small MD}}$ produces a $P(N)$ that is more realistic than the other models, as it is evident from Fig.\ \ref{fig:others_toscana}g and from Table \ref{tab:models_GPS}.

\paragraph{Trips per hour.} Human movements follow the circadian rhythm, i.e., they are prevalently stationary during the night and move preferably at specific times of the day \citep{gonzalez08,pappalardo2013}. To verify whether the considered models are able to capture this characteristic of human mobility, we compute the number of trips $T$ made by the individuals at every hour of the period of observation. Fig.\ \ref{fig:others}a-c and Fig.\ \ref{fig:others_toscana}a-c show how $T$ distribute across the 24 hours of the day, for CDRs and GPS data respectively. We observe that, regardless the trajectory generator used, diary generator MD produces a distribution of trips per hour very similar to real data (Fig.\ \ref{fig:others}a and Fig.\ \ref{fig:others_toscana}a). The mobility diary generator MD proposed in Section \ref{sec:mobility_diary} is hence able to create mobility diaries which reproduce the circadian rhythm of individuals in an accurate way. In contrast, diary generators RD and WT are not able to capture this distribution, regardless the trajectory generator used (Fig.\ \ref{fig:others}b-c and Fig.\ \ref{fig:others_toscana}b-c). This is because: (i) in RD individuals are always in motion; (ii) WT takes into account the waiting times but not the preference of individuals to move at specific times of the day.

\paragraph{Trips per day.}
The number of trips per day $D$ indicates the tendency of individuals to travel in their every-day life. For every dataset, we compute the number of trips per day made by each individual during the period of observation and plot the distribution $P(D)$ in Fig.\ \ref{fig:others}d-f (CDR) and Fig.\ \ref{fig:others_toscana}d-f (GPS).
We observe that $d$-EPR$^{\mbox{\tiny (CDR, GPS)}}_{\mbox{\small MD}}$, SWIM$^{\mbox{\tiny (CDR, GPS)}}_{\mbox{\small MD}}$ and LATP$^{\mbox{\tiny (CDR, GPS)}}_{\mbox{\small MD}}$ are able to capture the shape of $P(D)$ but overestimate the variance of the distribution (Fig.\ \ref{fig:others}d). The other diary generators, RD and WT, are not able to reproduce the CDR distribution since the average number $\overline{D}$ of trips per day is much higher than CDR data (Fig.\ \ref{fig:others}e-f). Again, this is because in RD individuals are always in motion and because WT does not take into account the circadian rhythm of individuals.

\paragraph{Time of stays.} The distribution of stay times $\Delta t$ is another important temporal features observed in human mobility. Stay time is the amount of time an individual spends at a particular location. In our experiments we compute the stay time as the number of hours every individual spends in her visited locations and plot the distribution $P(\Delta t)$ in Fig.\ \ref{fig:temporal}g-i (CDR) and Fig.\ \ref{fig:temporal_toscana}g-i (GPS). We observe that, for both CDRs and GPS data, $d$-EPR$^{\mbox{\tiny (CDR, GPS)}}_{\mbox{\small \{MD, RD, WT\}}}$ capture the shape of the distribution while the other models do not, though overestimating the presence of short time stays.

\subsection{Discussion of results}
\label{sec:discussion}
Two main results emerge from our experiments. First, model {$d$-EPR$_{\mbox{\small MD}}$} produces sampled mobility trajectories having in general the best fit to both CDR data and GPS data (i.e., having the lowest RMSE and KL for most of the measures), as evident in Table \ref{tab:models} and Table \ref{tab:models_GPS}. Diary generator MD, indeed, simulates in a realistic way temporal human mobility patterns such as the distribution of location frequency (Fig.\  \ref{fig:temporal}a) and the distribution of trips per hour (Fig.\ \ref{fig:others}a and Fig.\ \ref{fig:others_toscana}a). This is mainly because MD reproduces the circadian rhythm of individuals, while RD and WT do not. Moreover, trajectory generator $d$-EPR embeds two mobility mechanisms: preferential return and preferential exploration. The preferential return mechanism -- absent in SWIM and LATP -- allows for a realistic simulation of, for example, the distribution of radius of gyration (Fig.\ \ref{fig:plots}d and Fig.\ \ref{fig:plots_toscana}d) and the distribution of stay times (Fig.\ \ref{fig:others}g). The preferential exploration mechanism, which is modeled by both $d$-EPR and SWIM but it is absent in LATP, allows for a realistic description of the territory exploitation by individuals, in terms of the distribution of the number of visits per location (Fig.\ \ref{fig:temporal}d and Fig.\ \ref{fig:temporal_toscana}d). Also, model $d$-EPR$_{\mbox{\small MD}}$ produces realistic distributions for both CDR and GPS data, suggesting that it can be used in different simulation scenarios where its parameters are fitted on different types of data and different spatio-temporal resolutions.

Second interesting result is that the temporal and the spatial mechanisms have different roles in shaping the distribution of standard mobility measures. Some measures, such as trip distance (Fig.\ \ref{fig:plots}a-c and Fig.\ \ref{fig:plots_toscana}a-c), radius of gyration (Fig.\ \ref{fig:plots}d-f and Fig.\ \ref{fig:plots_toscana}d-f), visits per location (Fig.\ \ref{fig:temporal}d-f and Fig.\ \ref{fig:temporal_toscana}d-f) and time of stays (Fig.\ \ref{fig:plots}g-i) mainly depend on the choice of the trajectory generator, i.e., on the spatial mechanism of the model. Indeed, by changing the underlying diary generator the shape of these distribution, the RMSE and the KL divergence w.r.t.\ real data do not change in a significant way. Other measures, such as trips per hour (Fig.\ \ref{fig:others}a-c and Fig.\ \ref{fig:others_toscana}a-c) and trips per day (Fig.\ \ref{fig:others}d-f) mainly depend on the choice of the diary generator, i.e., on the temporal mechanism of the model. Conversely, both the spatial and the temporal mechanism are determinant in reproducing the distribution of some other measures like mobility entropy (Fig.\ \ref{fig:plots}g-i and Fig.\ \ref{fig:plots_toscana}g-i) and locations per user (Fig.\ \ref{fig:temporal}g-i and Fig.\ \ref{fig:temporal_toscana}g-i). Moreover the right combination of diary and trajectory generator, $d$-EPR$_{\mbox{\small MD}}$, leads to more accurate fits w.r.t.\ both CDR data and GPS data for the majority of measures (Table \ref{tab:models} and Table \ref{tab:models_GPS}). Human mobility patterns depend on both where people go and when people move: our results show that to reproduce them in an accurate way we need proper choices for the spatial and the temporal generative models to use in the {\scshape Ditras} framework. 

\section{Conclusion and future work}
\label{sec:conclusion}
In this paper we propose {\scshape Ditras}, a framework for the generation of individual human mobility trajectories with realistic spatio-temporal patterns. The framework consists of two steps: (i) the generation of a mobility diary by using a diary generator; (ii) the generation of a mobility trajectory by using a trajectory generator. In the paper we propose a novel diary generator MD together with MDL, a data-driven algorithm to build it from real mobility data. 

We instantiate {\scshape Ditras} by using MD and the state-of-the-art trajectory generator $d$-EPR and obtain a novel generative algorithm, $d$-EPR$_{\mbox{\small MD}}$. We use it to generate the spatio-temporal trajectories of thousands of agents visiting the locations on a large European country and a region in Italy. The generated sampled mobility trajectories are compared with CDR data, GPS vehicular data, and the trajectories produced by other generative algorithms, each obtained by using a different combination of diary generator and trajectory generator in the {\scshape Ditras} framework. Among the considered algorithms, $d$-EPR$_{\mbox{\small MD}}$ produces the best fit with respect to both CDR data and GPS data. We also observe that different combinations of diary and trajectory generators show different abilities to reproduce the distribution of standard mobility measures. This result highlights the importance of considering both the spatial and temporal dimensions in human mobility modelling. 

The proposed model $d$-EPR$_{\mbox{\small MD}}$ has a limited number of parameters to fit. The generation of the mobility diary is parameter-free as the Markov chain is a non-parametric model where each element of the transition matrix MD is estimated using the empirical frequencies observed in the data. The generation of the mobility trajectory is based on the $d$-EPR model. The details on how to fit the $d$-EPR parameters are explained in detail in \citep{pappalardo2015,Pappalardo2016934}. Here, for the two parameters of the exploration probability $p_{new}$, we choose the values $\rho = 0.6$ and $\gamma = 0.21$ that have been estimated in previous work \citep{SongNaturePhysics2010}. For the gravity model used in the exploration phase, we use a power law deterrence function of the distance with exponent $-2$, although other types of gravity or intervening opportunities models can be used. 
Given that the model is non-parametric or depends on a very small number of parameters, it does not suffer from training/test issues and its calibration is quite robust to changes in the size of the training set. 

\paragraph{Applications.}
Given its flexibility, {\scshape Ditras} can be used in a wide range of applications. Here we provide three examples where {\scshape Ditras} and $d$-EPR$_{\mbox{\small MD}}$ can be particularly useful and profitably applied.

In urban science, the generation of what-if scenarios to imagine the new mobility that could emerge from the construction of new infrastructures requires the generation of realistic mobility data and hence the presence of an accurate generative algorithm \citep{barbosa2017survey,koppetal2014}. $d$-EPR$_{\mbox{\small MD}}$ could be used to generate synthetic data given the tessellation of the territory that emerges from the construction of the new infrastructure, allowing urban planners and managers to quantify changes in urban mobility and visualize preferred path that could emerge from the simulation.

Computational epidemiology has attracted particular attention in the last decade, as the arrival of the 2009 flu pandemic prompted scientists to develop realistic mobility models to simulate the spread of viruses on a territory \citep{merler2013containing, ajelli2010comparing, venkatramanan2017using}. The possibility to use {\scshape Ditras} to combine different temporal and spatial mechanisms is particularly valuable for this type of studies, as generative algorithms for individual human mobility are the basic mechanism used in computational epidemiology to generate synthetic population mimicking at an individual level the realistic aspects related to disease propagation.

Opportunistic Networks (OppNets) enable communications in disconnected environments in the absence of an end-to-end path between the sender and the receiver. In OppNets, the mobility of nodes (e.g., mobile devices such as smartphones and tables) help the delivery of messages by connecting, asynchronously in time, otherwise disconnected subnetworks. This means that the network protocols responsible for finding a route between two disconnected devices must embed patterns in human movements and make prediction of future encounters. Realistic generative algorithms for human mobility are fundamental for testing the efficiency of OppNets protocol, as real data about the functioning of the network is obviously not available during the protocol design \citep{tomasini2017effect}. {\scshape Ditras} can be used to instantiate many generative algorithms and then generate realistic mobility routines to test the efficiency of a given network protocol for OppNets. Given its accuracy in reproducing human mobility patterns, $d$-EPR$_{\mbox{\small MD}}$ can be used to uncover the characteristics of the network protocol in real-life, such as the speed of message delivery.

A possible application of {\scshape Ditras} and $d$-EPR$_{\mbox{\small MD}}$ in data mining is anomaly detection. The proposed model can be used to detect individuals with an anomalous mobility behavior with respect to the typical mobility patterns of the majority of the individuals. In particular, within our framework an individual is anomalous if her trajectory is not a likely outcome of the model, i.e., if the probability that the model would generate such trajectory is below a given threshold. To this end, the log-likelihood of each individual's trajectory can be computed and the individuals can be ranked according to their log-likelihood values: individuals with a low rank and a very high log-likelihood values would be the most typical, whereas individuals with the highest ranks and low log-likelihood values would be the most anomalous.

\paragraph{Improvements.} 
The instantiation of {\scshape Ditras} we propose, $d$-EPR$_{\mbox{\small MD}}$, can be further improved in several directions. 
First, in this work the construction of the diary generator MD$^{(t)}$ through the mobility diary learner MDL is based on the simplest possible typical diary $W^{(t)}$, where the most likely location where a synthetic individual can be found at any time is her home location. More complex typical diaries can be used specifying, for example, the typical times where an individual can be found at work, school, friends' home and so on. Such a composition of $W^{(t)}$ can be constructed by using surveys or generative algorithms describing the daily schedule of human activities \citep{rinzi14,gonzalez_clustering,Liao2007} as a way to enrich an individual's trajectory with information about the type of activity associated to a location.

Second, in $d$-EPR the preference for short-distance trips is embedded in the preferential exploration phase only. A preference for short-distance trips can be introduced during the preferential return mechanisms as well, in order to eliminate the overestimation of long-distance trips and long-distance radii observed in Figures \ref{fig:plots}a and \ref{fig:plots}d. 

Third, in $d$-EPR$_{\mbox{\small MD}}$ we make the simplifying assumption that the travel time is of negligible duration. This may not be a good assumption especially when the duration of the time slot is one hour or less. 
The proposed algorithm can be modified to explicitly include realistic information on the travel time between locations, which imposes constraints on the locations that are reachable in a given time window and on the time that can be spent in a location given the travel time needed to reach the next location in the mobility diary.
Moreover, another interesting improvement can be to map the sampled mobility trajectories to a road network specifying specific road routes with specific velocities. This mapping would be of great help, for example, in what-if analysis where we want to study how human mobility changes with the construction of a new infrastructure in an urban context. 

Finally, there is a large number of studies that demonstrate the connection between human mobility and social networks \citep{brown2013place,hristova2016multilayer,wang2011,volkovich2012length,brown2013social,hossmann2011complex,hossmann2011putting}, as well as several approaches that include information on social connections in human mobility models \citep{borrel2009simps,yang2010using,fischer2010gesomo,boldrini2010hcmm,musolesi2007designing}. A mechanism to account for the influence of social connections on human mobility can be introduced in {\scshape DITRAS} as a third phase, between the mobility diary generation and the sampled trajectory construction. 

We leave these improvements of {\scshape DITRAS} for future work.

\appendix
\section{Homogeneity of typical mobility diaries}
\label{app:clustering}
\begin{figure}[htb]\centering
\subfigure[]{\includegraphics[scale=0.35]{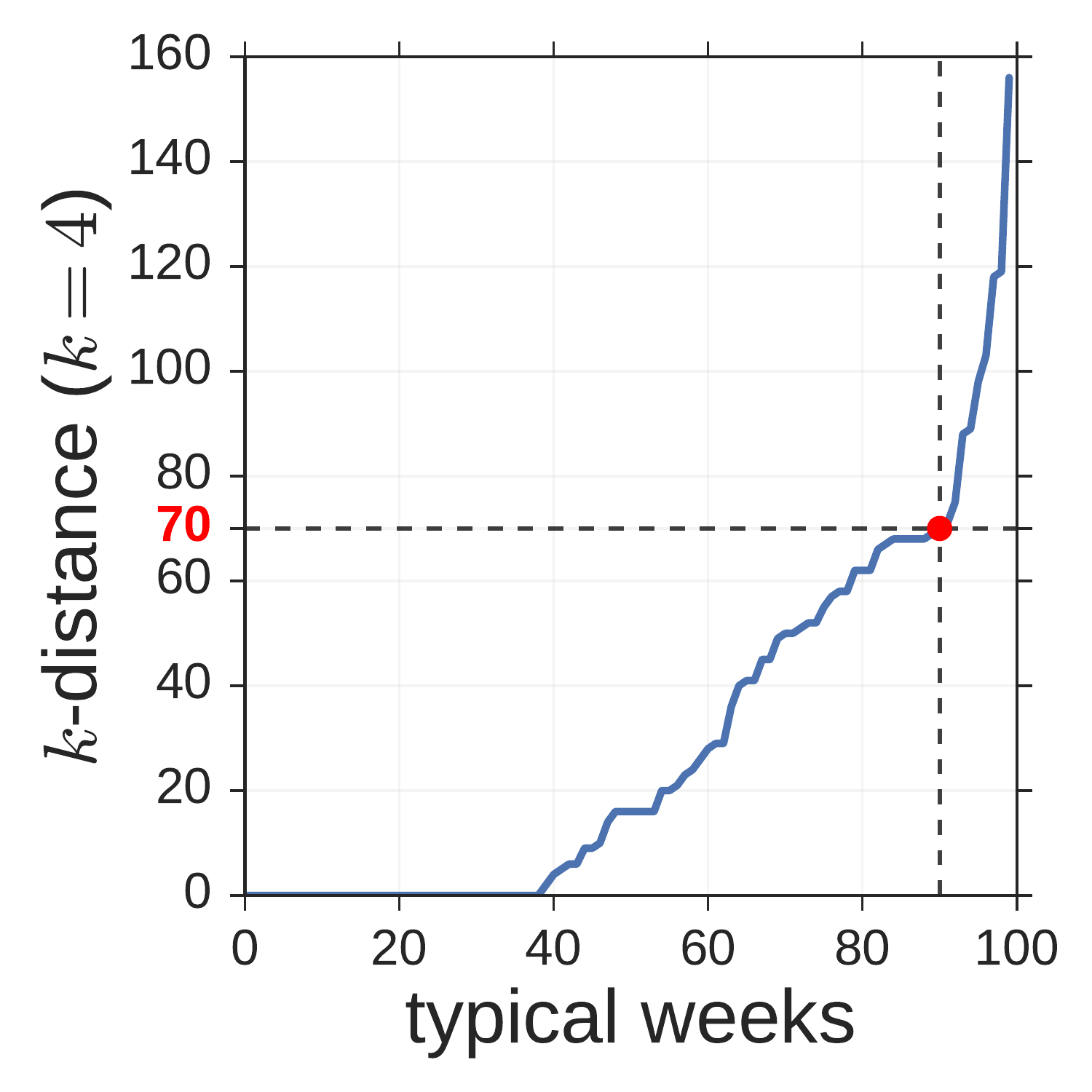}}
\subfigure[]{\includegraphics[scale=0.35]{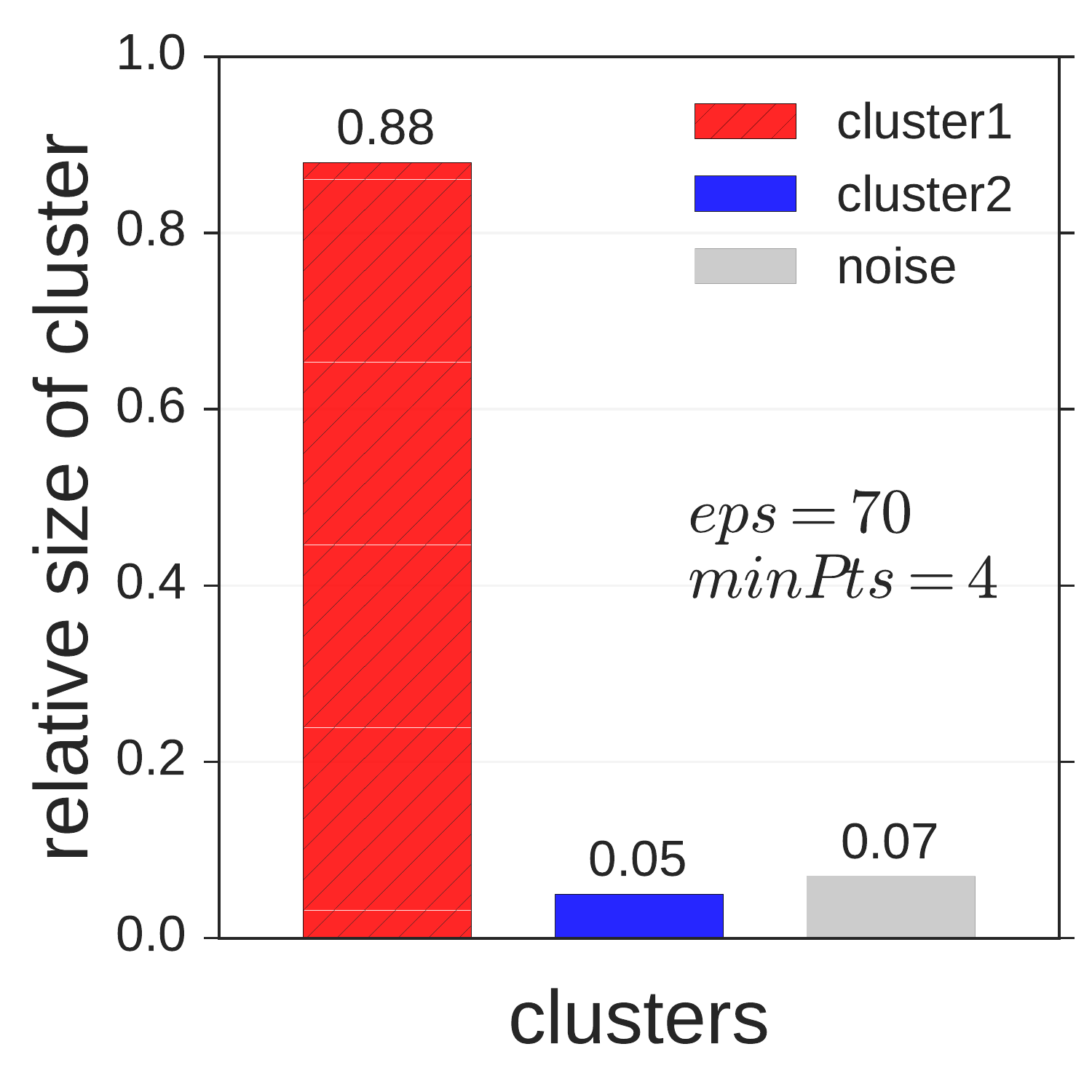}}
\subfigure[]{\includegraphics[scale=0.35]{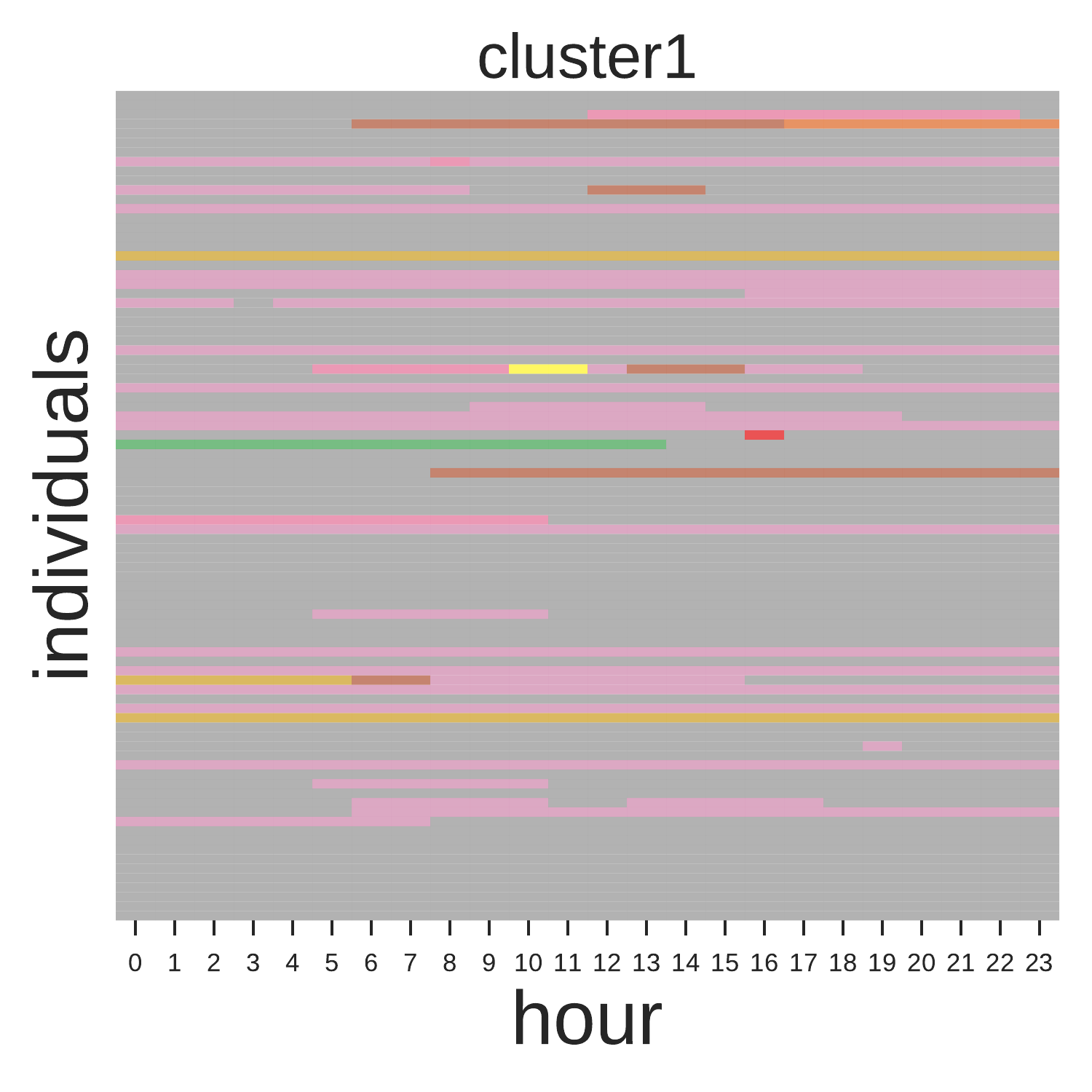}}
\subfigure[]{\includegraphics[scale=0.35]{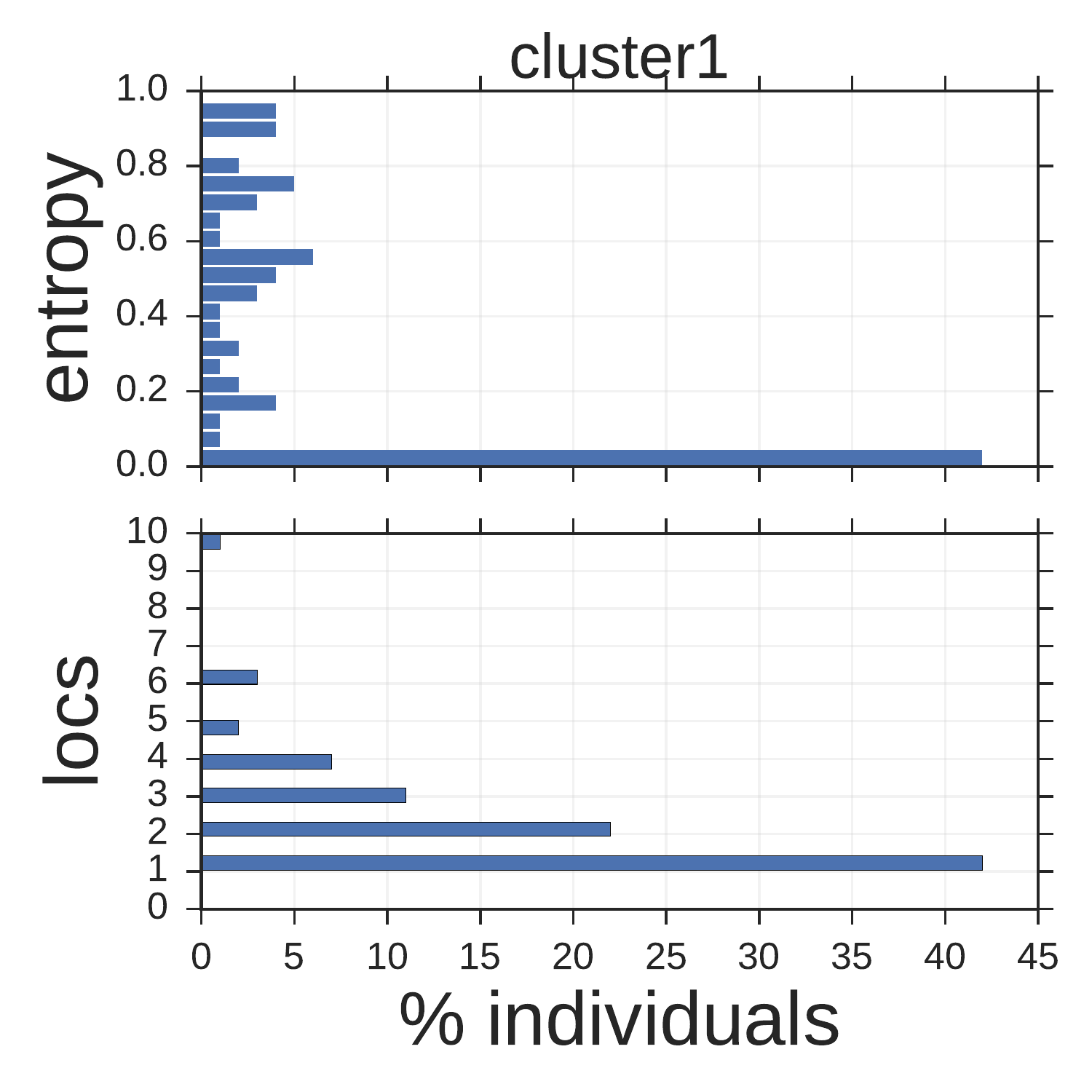}}
\caption{\textbf{First Row}: (a) Typical weeks sorted by distance to the 4th nearest neighbor, the elbow suggests to use $eps=70$; (b) Relative size of the clusters resulting from DBSCAN algorithm with $minPts=4$ and $eps=70$ and their relative size. \textbf{Second Row}: (c) Visualization of a day of the typical weeks of 100 individuals in the GPS dataset for the first cluster. Every color represents a different abstract location in the typical diary. (d) Distribution of abstract location entropy and number of distinct abstract locations of time series of individuals in cluster 1.}
\label{fig:clustering}
\end{figure}

We investigate to what extent the typical mobility diaries of real individuals are homogeneous by performing a clustering experiment. For every individual in the GPS dataset we compute her typical week, i.e. a time series of length 168 hours. Every time slot is the most frequent location of the individual in that hour of the week. We then apply the DBSCAN clustering algorithm \citep{dbscan} to group the typical weeks in dense clusters. We use the Levenshtein metric \citep{navarro2001guided} to measure the similarity between two typical weeks. DBSCAN takes two input parameters: $minPts$ and $eps$ \citep{dbscan}. We set $minPts=4$ and $eps=70$. We estimate the value of these parameters using the procedure suggested in \citep{tan2005introduction}: (i) we fix $minPts=4$ and  compute for every typical week the distance $d$ to its 4th nearest neighbor; (ii) we sort the typical weeks in increasing order with respect to $d$ and set $eps$ to the distance corresponding to an elbow in the curve of Figure \ref{fig:clustering}a. We observe no significant differences in the clustering results by varying $minPts$ in the range $[2, 5]$.

DBSCAN produces two clusters, one of them consisting of $\approx$90\% of the typical weeks (Figure \ref{fig:clustering}b). The silhouette coefficient of the clustering \citep{rousseeuw1987silhouettes}, a measure of how similar a typical diary is to its own cluster compared to other clusters, is $s = 0.50$ (in general, $s \in [-1, 1]$). 
The typical weeks in the biggest cluster have typically one or two locations, while the representative typical week (i.e., the medoid of the cluster) consists of just one location, the most frequent location of the individual (Figure \ref{fig:clustering}c-d). This result supports the validity of the simplifying assumption to consider one typical diary with a single location for all agents.

\section{Computational Complexity of $d$-EPR$_{\mbox{\small MD}}$}
\paragraph{Learning phase.} In the learning phase, two main tasks are performed: 
\begin{enumerate}
\item[(1)] the construction of the MD model by the MDL algorithm (Algorithm \ref{alg:diary_generator_builder}). The procedure \texttt{UpdateMarkovChain} has computational complexity $\mathcal{O}(N)$, where $N$ is the number of slots in the period of observation. As we repeat the procedure for all the $n$ individuals in the dataset, the computational complexity of Algorithm \ref{alg:diary_generator_builder} is $\mathcal{O}(Nn)$. When $n \gg N$, (e.g., when the period of observation is short and the dataset contains hundreds of thousands of individuals), the factor $N$ is negligible and the computational complexity of Algorithm 2 can be approximated to $\mathcal{O}(n)$.
\item[(2)] the construction of the probability matrix $P$ in the $d$-EPR model, which has complexity $\mathcal{O}(L^2)$ where $L$ is the number of locations in the spatial tessellation.
\end{enumerate}

\paragraph{Generation phase.} In the generation phase, the generation of the mobility diary with MD has complexity $\mathcal{O}(N)$. The generation of the trajectory from the mobility diary has complexity $\mathcal{O}(LNn)$ (Algorithm \ref{alg:dEPR}): in the worst case, for each individual we assign a spatial location in each time slot, and the assignment of a spatial location requires a call to procedure \texttt{weightedRandom} which has complexity $\mathcal{O}(L)$. When $n \gg N$, the computational complexity can be approximated to $\mathcal{O}(Ln)$. 

The total complexity of the generation phase is hence $\mathcal{O}(L^2 + Ln)$ when the probability matrix has to be constructed for the first time. In this case, when $L \sim n$ the computational complexity can be approximated to $\mathcal{O}(L^2)$. If the probability matrix is already available or has been already computed, the computational complexity of the generation phase is $\mathcal{O}(LNn)$, which can be approximated to $\mathcal{O}(Ln)$ when $n \gg N$.

\begin{acknowledgements}
We thank Paolo Cintia, Gianni Barlacchi and Salvatore Rinzivillo for their invaluable suggestions. This work has been partially funded by the EU under the H2020 Program by project Cimplex grant n.\ 641191. Filippo Simini has been supported by EPSRC First Grant EP/P012906/1.
\end{acknowledgements}

\bibliographystyle{plainnat}
\bibliography{biblio.bib}   

\end{document}